\documentclass[11pt]{article}
\usepackage[margin=1.1in,a4paper]{geometry}

\usepackage{mathtools}
\usepackage{amsmath, amssymb, natbib, graphicx, url, amsthm, subcaption, float}
\usepackage{trimclip}
\usepackage{algorithm, algpseudocode}
\usepackage{tabularx, array}
\usepackage[mathcal]{eucal}
\usepackage{tikz}
\usepackage{verbatim}
\usepackage{tcolorbox}
\newtcolorbox{assbox}{colback=black!5!white,colframe=black!75!black}
\newtcolorbox{thmbox}{colback=red!5!white,colframe=red!75!black}

\usepackage{bm, dsfont}

\usepackage{booktabs, multirow}
\usepackage{amsfonts}
\usepackage[inline]{enumitem}
\usepackage{microtype}
\usepackage{xcolor}
\usepackage[colorlinks=true,citecolor=blue]{hyperref}
\usepackage{diagbox}   
\usepackage{tcolorbox}
\usepackage{nicefrac}

\usepackage{pgf}
\usepackage{pgfplots}
\usepackage{anyfontsize}
\pgfplotsset{compat=1.18}

\newcommand{\ind}{\mathds{1}}

\newcommand{\RR}{\mathbb{R}}
\newcommand{\NN}{\mathbb{N}}

\newcommand{\EE}{\mathbb{E}}

\newcommand{\PP}{\mathbb{P}}

\newcommand{\Bb}{\mathcal{B}}

\newcommand{\Gg}{\mathcal{G}}

\newcommand{\Nn}{\mathcal{N}}

\newcommand{\Rr}{\mathcal{R}}
\newcommand{\Ss}{\mathcal{S}}

\newcommand{\iid}{\overset{\mathrm{iid}}{\sim}}

\newcommand{\indeps}{\perp\!\!\!\!\perp}

\DeclareMathOperator*{\argmin}{arg\,min}
\DeclareMathOperator*{\argmax}{arg\,max}

\theoremstyle{plain}
\newtheorem{theorem}{Theorem}[section]
\newtheorem{proposition}[theorem]{Proposition}
\newtheorem{lemma}[theorem]{Lemma}
\newtheorem{corollary}[theorem]{Corollary}
\theoremstyle{definition}
\newtheorem{example}[theorem]{Example}
\newtheorem{definition}[theorem]{Definition}

\newtheorem{remark}[theorem]{Remark}

\graphicspath{ {./images/} }

\title{\fontsize{17}{18.5}\selectfont A Local False Discovery Rate Framework for Compound p-values}
\author{
Jierui Zhu\thanks{Department of Statistics, The University of Chicago \texttt{jerryzhu@uchicago.edu}}
, Rina Foygel Barber\thanks{Department of Statistics, The University of Chicago \texttt{rina@uchicago.edu}}
, Nikolaos Ignatiadis\thanks{Department of Statistics and Data Science Institute, The University of Chicago \texttt{ignat@uchicago.edu}}
}

\begin{document}

\maketitle
\begin{abstract}
Compound p-values relax individual p-value validity to an average condition
across the null hypotheses.
This flexibility, however, need not protect particular rejections, especially
when paired with an average error criterion such as the false discovery rate (FDR).
In this work, we introduce ``density-compound p-values,'' 
a concept that connects compound p-values and the frequentist local FDR paradigm. 
Under independence, we establish boundary FDR (bFDR) guarantees for the support-line (SL) procedure, with nominal-level control for weighted and compound-e-weighted p-values. For general density-compound p-values, our bounds depend on a tractable null-distribution parameter, and we show that without further restrictions on this parameter, the bFDR can substantially exceed its nominal level.
Our theory extends to discrete compound p-values constructed by pooling
permutation tests.
Applications illustrate the method in low-resolution
testing problems, where individually valid permutation p-values are
too coarse to yield discoveries after multiplicity correction.
\end{abstract}

\section{Introduction}
Multiple testing aims to identify signals while controlling type-I errors among the reported discoveries.
Let $\{H_{0,i}\}_{i=1}^m$ denote a family of null hypotheses and let $\Nn_0$ be the set of true null indices.
Standard procedures take as input p-values $P_1,\ldots,P_m$ that are superuniform under the null:
\begin{equation} \label{eq:superuniform}
    \mathbb{P}(P_i \le t) \le t \text{ for all } t \in [0,1] \text{ and } i \in \Nn_0.
\end{equation}
For example, the Benjamini-Hochberg (BH) procedure
\citep{888cd474-50a6-33fd-a789-415b80e67e78}
uses such p-values to control the false discovery rate (FDR), the expected proportion of false discoveries among all rejections.

Requiring individually valid p-values can be restrictive when null distributions are
difficult to compute or when the resulting p-values have low power.
Compound p-values
\citep{armstrong2025falsediscoveryrateadjustments,
ignatiadis2025asymptoticcompoundevaluesmultiple}
weaken this requirement by imposing superuniformity only on average across
the true nulls:

\begin{definition}
    \label{def:compound-p-values}
    We say $P_1,\ldots,P_m\in[0,1]$ form compound p-values if
    \begin{equation}\label{eq:cp}
        \frac{1}{m}\sum_{i\in\Nn_0}\PP(P_i\leq t)\leq t
        \qquad\text{for all }t\in[0,1].
    \end{equation}
\end{definition}
Under compound validity \eqref{eq:cp} rather than individual validity \eqref{eq:superuniform}, anti-conservative behavior for some nulls may be
offset by conservative behavior for others.
This flexibility can yield substantially more discoveries in certain practically relevant settings
\citep{10.1214/24-AOS2431,10.1093/jrsssb/qkag116} at the cost of weaker inferential guarantees. 
When compound validity is paired with an average error criterion such as FDR, neither condition guarantees the reliability of any particular rejected hypothesis. 

\begin{figure}[t]
    \centering
    \includegraphics[width=\textwidth]{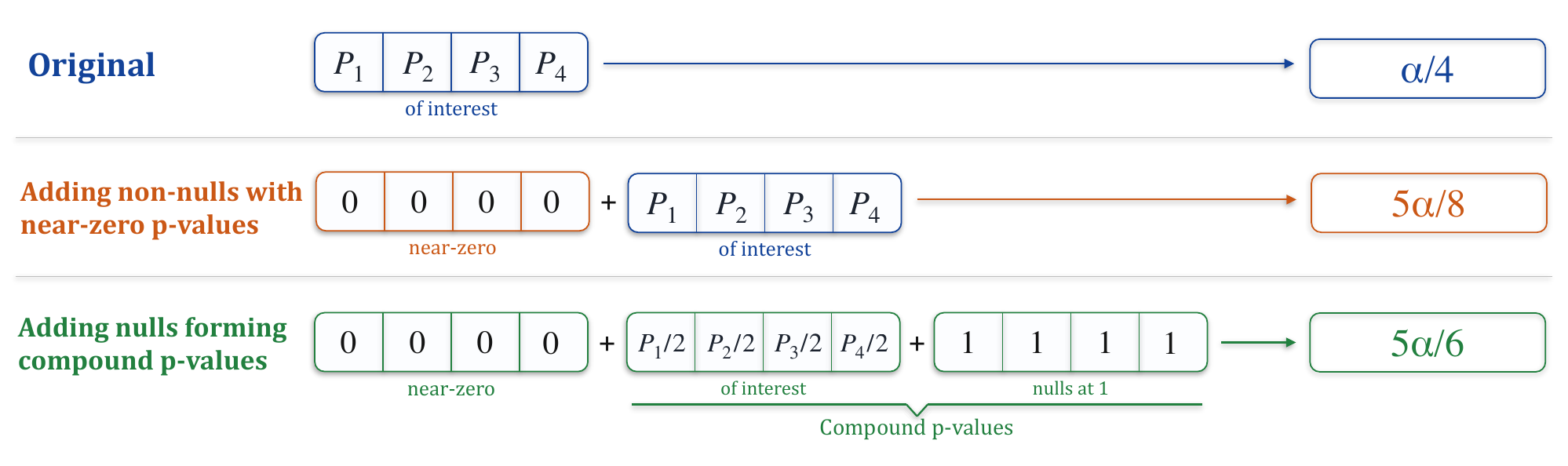}
    \caption{
    How FDR and compound validity can raise the effective rejection threshold
    for the same four null hypotheses of interest.
    Top: BH compares the smallest $P_i$ with $\alpha/4$.
    Middle: adding four near-zero non-null p-values  raises the first relevant
    BH critical value to $5\alpha/8$.
    Bottom: adding four conservative nulls and replacing each $P_i$ by $P_i/2$
    preserves compound validity. BH then compares $P_i/2$ with $5\alpha/12$,
    or equivalently compares $P_i$ with the displayed threshold $5\alpha/6$.
    }
    \label{fig:fdr-compound-schematic}
\end{figure}

\paragraph{Local FDR and boundary FDR.}
Even with individually valid p-values as in \eqref{eq:superuniform}, 
FDR control allows hypotheses supported by relatively weak evidence to be rejected
because they are accompanied by more confidently rejected non-nulls,
which enables ``cheating with FDR,'' as first explained by \citet{finner2001false}.
Compound validity further amplifies the concern that hypotheses with weak evidence may be rejected; see Figure~\ref{fig:fdr-compound-schematic} for an illustration.

Instead, a local perspective asks for the reliability of a particular discovery.
The local FDR provides such a hypothesis-specific measure in the classical
Bayesian two-groups model, but it depends on a probabilistic model for the
hypotheses and p-values \citep{Efron01122001}.

In a frequentist setting with fixed hypotheses, \citet{10.1093/biomet/asaf083} introduce the boundary false discovery rate (bFDR) as an operational analogue of the local FDR, defined as the probability that the last (boundary) rejection is a false discovery.
More formally,
let $R_\alpha$ denote the number of rejections made by a multiple testing procedure at level $\alpha$,
and $\sigma$ be the permutation such that
$P_{\sigma(1)} \le P_{\sigma(2)} \le \cdots \le P_{\sigma(m)}$.
Then the bFDR is defined as
\begin{equation}
\mathrm{bFDR}:= \PP\left(H_{0, \sigma(R_\alpha)} \text{ is true}, R_\alpha > 0\right).
\end{equation}
Compared to FDR, bFDR captures the quality of the rejection set boundary rather than the average quality of the rejection set.
For individually valid p-values, the difference can be substantial:
\cite{10.1214/24-AOS2359} point out the probability that the last rejection by the BH procedure is null at level $\alpha = 0.05$ could be as high as $30\%$.
Targeting bFDR guards against potential cheating with FDR by preventing strong discoveries from compensating for a false boundary rejection; see Remark~\ref{rmk:cheating-fdr} for further discussion.

\paragraph{Support-line (SL) procedure.} To control the bFDR,
\citet{10.1214/24-AOS2359} introduce the SL procedure.
Writing $P_{\sigma(0)} := 0$ for convenience, 
define
\begin{equation} \label{eq:sl-procedure}
R_\alpha^{\mathrm{SL}}
\equiv 
R_\alpha
:=\max\!\left\{\argmin_{k\in\{0,\ldots,m\}}
\left(P_{\sigma(k)}-\frac{\alpha k}{m}\right)\right\},
\qquad
\hat{\tau}_\alpha^{\mathrm{SL}}\equiv\hat{\tau}_\alpha
:=P_{\sigma(R_\alpha)}.
\end{equation}
The SL procedure rejects the $R_\alpha$ hypotheses with
$P_i\leq\hat{\tau}_\alpha$.
For independent p-values with exact uniform nulls, SL controls its bFDR at level $\alpha$ \citep{10.1093/biomet/asaf083}.

The goal of this work is to determine which bFDR guarantees of the SL procedure
persist under compound rather than individual p-value validity.
In Section~\ref{sec:density-condition}, we introduce
``density-compound p-values,'' which strengthen the average null CDF condition
in \eqref{eq:cp} to an average null density condition. Section~\ref{sec:bfdr} first
establishes exact bFDR control for weighted and compound-e-weighted p-values
(Theorem~\ref{thm:bfdr-weighted-pvals}). For general independent
density-compound p-values, Proposition~\ref{prop:cp-bfdr-counterexample} shows
that substantial finite-sample inflation can still occur, while
Theorems~\ref{thm:gamma-alpha-bfdr-control} and
\ref{thm:gamma-bfdr-control-posthoc} give upper bounds in terms of a tractable
null-distribution parameter and data-dependent refinements.
Theorem~\ref{thm:bfdr-dcperm} extends the bFDR theory to pooled permutation
p-values, with an extension for heterogeneous designs in
Appendix~\ref{subsec:sampled-permutation}.
In Section~\ref{sec:freq-lfdr},
Theorem~\ref{thm:lfdr-upper-bound-weak-dependence} establishes an asymptotic
frequentist local FDR guarantee for the SL threshold under weak dependence.
Sections~\ref{subsec:simulation-continuous} and~\ref{sec:applications}
illustrate the results through Gaussian mean simulations and applications
to regression discontinuity and gene expression.

\section{Density-Compound p-value Condition}\label{sec:density-condition}
Compound validity \eqref{eq:cp} controls the average null CDF,
whereas bFDR depends on the behavior of the null and non-null distributions
near the rejection boundary. This suggests strengthening average CDF control
to average density control. 
In Sections~\ref{subsec:continuous-dcp} and \ref{subsec:discrete-dcp},
we treat continuous and discrete cases in turn.

We write $[n]:=\{k\in\NN:1\leq k\leq n\}$. For each $i\in[m]$, let
$F_0^{(i)}$ denote the CDF of $P_i$ under its nominal null distribution. For
$i\notin\Nn_0$, this nominal distribution need not be uniquely defined and we can choose it as convenient without changing any of our claimed statistical guarantees.

\subsection{Continuous Density-Compound p-values}\label{subsec:continuous-dcp}
In this subsection, absolute continuity and densities are with
respect to Lebesgue measure.

\begin{definition}[Density-Compound p-values]
We say that
    $P_1,\ldots,P_m$ are density-compound p-values over an interval
    $I\subseteq[0,1]$ if, for every $i\in \Nn_0$,
    the null distribution of $P_i$ restricted to $I$ has a density
    $f_0^{(i)}$, and it holds that
    \begin{equation}\label{eq:density-condition}
        \frac{1}{m} \sum_{i \in \Nn_0} f^{(i)}_0(t) \leq 1
        \qquad\text{for Lebesgue-almost every }t\in I.
    \end{equation}
    We take $I=[0,1]$ unless otherwise specified.
\end{definition}

In fact, \citet{10.1214/009053606000001460} anticipated density-compound p-values
in his empirical Bayes analysis of FDR
(as also noted by \citealp{10.1214/24-AOS2431,10.1093/jrsssb/qkag116}):
the null densities may differ, provided their average behaves like
a reference null density $f_0$.
Here, \eqref{eq:density-condition} bounds the sum of the true-null
densities, divided by $m$, by the uniform density $f_0 = 1$.
Section~\ref{sec:freq-lfdr} makes this connection precise through a
frequentist local FDR interpretation of the SL threshold.

For multiple testing at level $\alpha$, we do not need \eqref{eq:density-condition} to hold for $I = [0,1]$. 
It is sufficient, for example, that $I = [0, \alpha]$. 
When it holds
over $[0,1]$, integration gives
\begin{equation}\label{eq:density-implies-compound}
    \frac{1}{m}\sum_{i\in\Nn_0}F_0^{(i)}(t)
    =\int_0^t\frac{1}{m}\sum_{i\in\Nn_0}f_0^{(i)}(s)\,ds
    \leq t,
\end{equation}
so density-compound p-values are compound p-values. The reverse implication
does not hold even for an absolutely continuous distribution. For example,
$P\sim\mathrm{Unif}[1/2,1]$ is superuniform, but its density equals $2$ on
$[1/2,1]$.

A useful partial reverse implication is the following.
Suppose the nominal null
CDFs $\{F_0^{(i)}\}_{i=1}^m$ are absolutely continuous. Differentiation gives the following implication:
\begin{equation}
\text{If }\, \;\frac{1}{m}\sum_{i=1}^m F_0^{(i)}(t)=t \;\mbox{~for all $t\in[0,1]$,~~then~}  \;\frac{1}{m}\sum_{i=1}^m f_0^{(i)}(t)=1,
\label{eq:comp_to_density}
\end{equation}
where the equality of densities holds Lebesgue-almost everywhere.
Hence \eqref{eq:density-condition} holds over $I=[0,1]$ since
$\mathcal N_0\subseteq[m]$.
Although elementary, this observation is useful for verifying the
density-compound condition: given $\{F_0^{(i)}\}_{i\in\mathcal N_0}$,
it suffices to exhibit \emph{any} collection of CDFs $\{F_0^{(i)}\}_{i\notin\mathcal N_0}$
satisfying the left-hand side of
\eqref{eq:comp_to_density}.

\begin{proposition}
\label{prop:density-condition-equivalent}
Suppose each $F_0^{(i)}$, $i\in\Nn_0$, is absolutely
continuous. Then $P_1,\ldots,P_m$ are density-compound p-values if and only if
there exist CDFs
$\{\widetilde F_0^{(i)}\}_{i\notin\Nn_0}$ such that
$$
   \frac{1}{m}\Bigg\{ \sum_{i\in\Nn_0}F_0^{(i)}(t)
    +\sum_{i\notin\Nn_0}\widetilde F_0^{(i)}(t) \Bigg\}
    =t,\qquad t\in[0,1].
$$
\end{proposition}

We next give two general constructions of density-compound p-values: one based on p-value weighting and the other on averaging nominal null CDFs. Both constructions are already known to yield compound p-values; we show that, under additional assumptions, they also satisfy the stronger density-compound condition.

\paragraph{Weighted p-values and Compound-e-weighted p-values.}
Let $U_1,\ldots,U_m$ be individually valid p-values. 
Given fixed weights $w_i\geq0$ with $\sum_{i=1}^m w_i=m$, 
the weighted p-values $P_i=(U_i/w_i)\wedge1$ \citep{10.1093/biomet/93.3.509,10.1214/18-AOS1765} 
are compound p-values as noted by \citet[Example~2.2]{ignatiadis2025asymptoticcompoundevaluesmultiple}. 
The compound p-values in Figure~\ref{fig:fdr-compound-schematic} are an example of this weighted construction.

More generally, we can replace fixed weights by compound e-values
\citep{ignatiadis2025asymptoticcompoundevaluesmultiple}.
Recall that e-values are $[0,\infty]$-valued random variables satisfying $\EE[E_i]\leq1$ for every $i\in\Nn_0$; see \citet{10.1561/3600000002} for a monograph treatment. Compound e-values relax these individual bounds to an aggregate bound, analogously to compound p-values.

\begin{definition}[Compound e-values]
\label{def:compound-e-values}
    Let $E_1,\ldots,E_m \in [0,\infty]$ be random variables. Then $E_1,\ldots,E_m$ are called compound e-values if
    $\sum_{i\in\Nn_0}\EE[E_i]\leq m$.
\end{definition}
The compound-e-weighted p-values $P_i=(U_i/E_i)\wedge1$ are compound p-values when $U_i$ is independent of $E_i$ for every $i\in\Nn_0$ \citep[Proposition~2.3]{ignatiadis2026tiny}. 
With exact uniform null p-values, they satisfy the stronger density-compound condition.

\begin{proposition}[Weighted constructions]
\label{prop:weighted-constructions}
\label{ex:weighted-pvalues}\label{ex:weighted-epvalues}
Suppose $U_i\sim\mathrm{Unif}[0,1]$ for every $i\in\Nn_0$. Let
$E_1,\ldots,E_m\in[0,\infty]$ be compound e-values such that $E_i$ is
independent of $U_i$ for every $i\in\Nn_0$. Then the compound-e-weighted
p-values
$
P_i=(U_i/E_i)\wedge1
$
are density-compound p-values over $[0,1)$. In particular, the conclusion
holds for weighted p-values with deterministic weights $w_i$ by taking $E_i=w_i$.
\end{proposition}

\paragraph{Averaging nominal null CDFs.}
Another important construction of compound p-values averages the nominal null
CDFs of the test statistics $X_1, \ldots, X_m$.
Let $F_{X_j,0}$ denote the nominal null CDF of test statistic $X_j$. From \citet[Proposition~7]{10.1093/jrsssb/qkag116}, it follows that
\begin{equation}\label{eq:average-cdf-construction}
        P_i=\frac{1}{m}\sum_{j=1}^m F_{X_j,0}(X_i)
\end{equation}
form compound p-values. If large values of $X_i$ provide evidence against the null, then we can apply the same construction with $-X_i$ in place of $X_i$.
Moreover, their universality result \citep[Proposition~S10]{10.1093/jrsssb/qkag116} 
shows this construction is essentially the only way to construct compound p-values that depend only on their respective test statistics.\footnote{More precisely, they show that for any compound p-values $P_1,\ldots,P_m$ such that $P_i$ depends only on $X_i$, there exist
functions $T_i$ such that
$P_i\geq m^{-1}\sum_{j\in\Nn_0}G_j(T_i(X_i))$ for every $i\in[m]$,
where $G_j$ is the null CDF of $T_j(X_j)$.
}

For the construction in \eqref{eq:average-cdf-construction}, absolute
continuity of the nominal null CDFs yields the stronger density-compound
condition.

\begin{proposition}[Average-CDF construction]
\label{prop:cp-ave-cdf-construction}
    For each $i\in[m]$, let test statistic $X_i$ have an absolutely continuous nominal null
    CDF $F_{X_i,0}$. Then the construction in \eqref{eq:average-cdf-construction}
    yields density-compound p-values.
\end{proposition}

We illustrate this with an example for Gaussian mean testing.
\begin{example}[Oracle Gaussian construction]
\label{ex:cp-eb-normal-known-var}
    For each $i\in[m]$, observe
    $Y_{i1},\ldots,Y_{in}\iid\Nn(\mu_i,\sigma_i^2)$ independent across units $i$ and test
    $H_{0,i}:\mu_i=0$.
    Write $\bar Y_i=n^{-1}\sum_{j=1}^nY_{ij}$.
    Assuming the empirical variance distribution
    $G=m^{-1}\sum_{j=1}^m\delta_{\sigma_j^2}$ is known,
    \citet{10.1093/jrsssb/qkag116} study the oracle compound p-values
    \begin{equation}\label{eq:oracle-gaussian-dcp}
        P_i^{\mathrm{oracle}}
        =\frac{1}{m}\sum_{j=1}^m
        2\bar\Phi\!\left(
            \frac{|\bar Y_i|}{\sigma_j/\sqrt n}
        \right),
    \end{equation}
    where $\Phi(\cdot)$ is the standard normal CDF and $\bar\Phi(\cdot)=1-\Phi(\cdot)$.
    Proposition~\ref{prop:cp-ave-cdf-construction}, applied to
    $X_i=-\sqrt n|\bar Y_i|$, shows that these are density-compound p-values.
\end{example}

In Appendix~\ref{subsec:additional-dcp}, we use Propositions~\ref{prop:density-condition-equivalent} and~\ref{prop:cp-ave-cdf-construction}
to show that two further constructions in the literature
on compound p-values \citep{ignatiadis2026empiricalbayesprepivotinggroup, 10.1214/24-AOS2431} in the setting of Example~\ref{ex:cp-eb-normal-known-var} are in fact density-compound p-values.
Section~\ref{subsec:simulation-continuous}
compares selected constructions in this setting of simultaneous one-sample t-tests.

\subsection{Discrete Density-Compound p-values} \label{subsec:discrete-dcp}
We now give the discrete analogue of \eqref{eq:density-condition}. Suppose
$P_1,\ldots,P_m$ are supported on the common grid
$
    \Gg_L:=\{k/L:k\in[L]\}
$
for some $L\in\NN$.

\begin{definition}[Discrete density-compound p-values]
    We say $P_1,\ldots,P_m$ are discrete density-compound p-values over
    $\Gg_L$ if
    \begin{equation}\label{eq:discrete-density-condition}
        \frac{1}{m}\sum_{i\in\Nn_0}
        \PP_{H_{0,i}}\left(P_i=\frac{k}{L}\right)
        \leq\frac{1}{L}
        \qquad\text{for every }k\in[L].
    \end{equation}
\end{definition}
Summing \eqref{eq:discrete-density-condition} over $k\leq\lfloor Lt\rfloor$
gives \eqref{eq:cp}, so these are also compound p-values in the sense of
Definition~\ref{def:compound-p-values}.
We next give
two important constructions, one based on Monte Carlo controls and one based
on permutation testing.
Both constructions pool test statistics across hypotheses to obtain finer resolution than individually valid p-values, 
in the sense of a smaller minimum attainable p-value.

\paragraph{Monte Carlo controls.}
In a setting where the null distribution of a test statistic $X_i$
is intractable, we can instead use Monte Carlo sampling to compute a p-value.
Formally, let
$\tilde X_i^{(1)},\ldots,\tilde X_i^{(K)}$
be $K$ i.i.d.\ draws from the nominal null distribution of $X_i$, generated
independently of $X_i$ and independently across hypotheses. Write
$Y_i=(X_i,\tilde X_i^{(1)},\ldots,\tilde X_i^{(K)})$ and define
\begin{equation}\label{eq:compound-mc-pval}
    P_i^{\mathrm{MC}}
    =\frac{1}{m(K+1)}\sum_{j=1}^m
    \left(
        \ind\left\{X_j\geq X_i\right\}
        +\sum_{k=1}^K\ind\left\{\tilde X_j^{(k)}\geq X_i\right\}
    \right).
\end{equation}
This construction is due to \citet{10.1093/jrsssb/qkag116}, who show that
these are compound p-values. The following proposition strengthens this
conclusion to discrete density-compound validity under the stated
no-ties assumption.

\begin{proposition}[Pooled Monte Carlo p-values]
\label{prop:compound-pvals-controls}
    Suppose the vectors $Y_1,\ldots,Y_m$ are independent and, for each
    $i\in\Nn_0$, the $K+1$ components of $Y_i$ are exchangeable. If the
    pooled vector $Y=(Y_1,\ldots,Y_m)$ has no ties almost surely, then
    $P_1^{\mathrm{MC}},\ldots,P_m^{\mathrm{MC}}$ in
    \eqref{eq:compound-mc-pval} are discrete density-compound p-values on
    $\Gg_{m(K+1)}$.
\end{proposition}

This pooled Monte Carlo construction has finer resolution than the
individually valid Monte Carlo p-value
$
    P_i^{\mathrm{indMC}}
    = (K+1)^{-1}\left(1+\sum_{k=1}^K\ind\left\{\tilde X_i^{(k)}\geq X_i\right\}\right)
$.

\paragraph{Two-sample permutation tests.}
In two-sample testing, permutation tests enjoy finite-sample validity 
without making parametric assumptions on the null distribution.
For each $i\in[m]$, suppose unit $i$ has $n$ outcomes $Z_i=(Z_{i1},\ldots,Z_{in})$, with the
first $n_1$ observations assigned to treatment and the remaining $n_0=n-n_1$ to
control. Let $T_i(Z_i)$ be a test statistic whose larger values indicate
stronger evidence against the null hypothesis $H_{0,i}$ that the components of $Z_i$ are exchangeable. Let $\mathfrak S(n)$ be the
symmetric group on $[n]$. For $\pi\in\mathfrak S(n)$, write
$(z^\pi)_\ell=z_{\pi^{-1}(\ell)}$ and set
$X_i^\pi=T_i(Z_i^\pi)$, with $X_i=X_i^{\mathrm{id}}$.
The usual
permutation p-value,
\begin{equation}\label{eq:perm-pval}
P_i^{\mathrm{perm}}
=
\frac{1}{n!}\sum_{\pi\in\mathfrak S(n)}
\ind\{X_i^\pi\geq X_i\},
\end{equation}
is individually valid under $H_{0,i}$.
To obtain finer resolution, we pool permutation statistics across
units\footnote{One formulation of Significance Analysis of
Microarrays (SAM) \citep{doi:10.1073/pnas.091062498,Storey2003} can be
viewed as applying BH to compound p-values $P_i^{\mathrm{dcperm}}$ of the form in
\eqref{eq:compound-perm-pval} \citep[Section~18.7.1]{hastie2009elements},
with recent work giving finite-sample FDR bounds \citep{10.1093/jrsssb/qkag116}.
We establish finite-sample bFDR bounds for SL on these pooled permutation
p-values in Theorem~\ref{thm:bfdr-dcperm} and illustrate the procedure in
Section~\ref{subsec:gene-expression}.}:
\begin{equation}\label{eq:compound-perm-pval}
    P_i^{\mathrm{dcperm}}
    =\frac{1}{mn!}\sum_{j=1}^m\sum_{\pi\in\mathfrak S(n)}
    \ind\{X_j^\pi\geq X_i\}.
\end{equation}
The pooled construction is discrete density-compound when, for every unit,
the test statistic attains the same number of distinct values over all
permutations, each value occurs equally often, and values from different
units do not coincide.
\begin{proposition}[Pooled permutation p-values]
\label{prop:dcperm-validity}
For a fixed $B\in\mathbb N$, assume that almost surely each
collection $(X_i^\pi)_{\pi\in\mathfrak S(n)}$ has $B$ distinct values, each
occurring $n!/B$ times, and that values from different units do not
coincide. 
Suppose $Z_1,\ldots,Z_m$ are
independent and the components of $Z_i$ are exchangeable for all
$i\in\Nn_0$. Then $P_1^{\mathrm{dcperm}},\ldots,P_m^{\mathrm{dcperm}}$ in
\eqref{eq:compound-perm-pval} are discrete density-compound p-values on
$\Gg_{mB}$.
\end{proposition}
The proof is in Appendix~\ref{subsec:proofs-main}. We refer to~\citet[Condition~1]{hemerik2018exact} for a discussion of ties in the context of a single permutation test.
The common multiplicity cancels out from~\eqref{eq:perm-pval} and~\eqref{eq:compound-perm-pval}, corresponding to
resolution $1/B$ instead of $1/n!$ for $P_i^{\mathrm{perm}}$, and resolution $1/(mB)$ instead of $1/(mn!)$ for $P_i^{\mathrm{dcperm}}$.
For statistics invariant under permutations within groups, 
the $n_1!n_0!$ permutations inducing the same treatment assignment yield the same statistic value.
If there are no
other ties, then $B=\binom n{n_1}$. For the absolute two-sample t-statistic in a
balanced design, exchanging the groups doubles this multiplicity, giving
$B=\binom n{n_1}/2$ in the absence of further ties.

The construction above uses a common two-sample design. To accommodate
different sample sizes across units, Appendix~\ref{subsec:sampled-permutation}
gives a randomized permutation construction and its bFDR guarantee, used in
the regression discontinuity application in Section~\ref{subsec:rd}.

\section{bFDR for Density-Compound p-values} \label{sec:bfdr}

We now turn to the primary objective of this paper: establishing bFDR
guarantees for the SL procedure applied to density-compound p-values.
Under independence and exactly uniform null p-values,
\citet{10.1093/biomet/asaf083} show that SL controls bFDR at level
$\bar\pi_0\alpha$, where $m_0=|\Nn_0|$ and $\bar\pi_0=m_0/m$. Here we seek 
bFDR guarantees when individual null uniformity is
replaced by the density-compound condition.

For BH, weighted and compound-e-weighted
p-values retain nominal FDR control under suitable independence
assumptions
\citep{10.1093/biomet/93.3.509, 10.1093/biomet/asad057,    ignatiadis2026tiny},
whereas general independent compound p-values need not
\citep{armstrong2025falsediscoveryrateadjustments, 10.1093/jrsssb/qkag116}.
We establish an analogous distinction for the bFDR of SL; that is,
weighted constructions retain nominal control under the assumptions
of Section~\ref{subsec:weighted-bfdr}, whereas the general continuous
and discrete settings require additional conditions, developed in
Sections~\ref{subsec:continuous-bfdr} and~\ref{subsec:discrete-bfdr}.

\subsection{bFDR for Weighted p-values and Compound-e-Weighted p-values}
\label{subsec:weighted-bfdr}
Proposition~\ref{prop:weighted-constructions} shows that weighted p-values and
compound-e-weighted p-values
form density-compound p-values.
We prove that the SL procedure on these p-values controls bFDR at level $\alpha$ under independence and exact uniform nulls.

\begin{theorem}[bFDR control for compound-e-weighted p-values] \label{thm:bfdr-weighted-pvals}
    Let $\alpha \in (0,1)$ and $U_1, \ldots, U_m$ be independent with $U_i \sim \mathrm{Unif}[0,1]$ for $i \in \Nn_0$.
    Suppose we have access to compound e-values $E_1, \ldots, E_m \in [0,\infty]$ as in Definition~\ref{def:compound-e-values}.
    Assuming $E = (E_1, \ldots, E_m)$ is independent of $U = (U_1, \ldots, U_m)$,
    the SL procedure $\Rr_\alpha$ on compound-e-weighted p-values $P_i = (U_i/E_i) \wedge 1$ 
    satisfies
    $$
    \mathrm{bFDR}(\Rr_\alpha) 
    \leq \frac{\alpha}{m}\sum_{i \in \Nn_0} \EE[E_i]
    \leq \alpha.
    $$
In particular, the same conclusion holds for weighted p-values with deterministic weights $w_i$ by
applying this result to $E_i = w_i$.
\end{theorem}

\begin{remark}[Compound e-weighted procedures]
Following the terminology of \citet{10.1093/biomet/asad057},
we use the prefix p- for a multiple testing procedure applied to p-values
$U_1,\ldots,U_m$, and the prefix ep- for the same procedure
applied to $P_i=(U_i/E_i)\wedge1,$ where $E_i$ are compound e-values.
For example, p-BH is ordinary BH, while ep-BH applies BH to
compound-e-weighted p-values.
In this terminology, Theorem~\ref{thm:bfdr-weighted-pvals}
establishes bFDR control for ep-SL under its stated assumptions.
\end{remark}

\subsection{bFDR for Continuous Density-Compound p-values}
\label{subsec:continuous-bfdr}

We return to independent density-compound p-values.
As a first attempt, we show in Appendix~\ref{subsec:proofs-main} the bound
\begin{equation}\label{prop:cp-bfdr-control}
    \mathrm{bFDR}(\Rr_\alpha)
    \leq m_0\alpha
    \wedge 2(\alpha + \alpha^2m)
    \wedge 1.
\end{equation}
While this upper bound may look uninformative at first glance as we typically expect $m_0$ and $m$ to be quite large, 
it turns out without further assumptions, this is actually sharp to the first order for small $\alpha$.

\begin{proposition}[bFDR lower bound] \label{prop:cp-bfdr-counterexample}
    Let $\alpha \in (0,1)$ satisfy $1/\alpha \in \NN$ and $m \geq 1/\alpha$.
    For any $m_0 \leq m + 1 - 1/\alpha$,
    there exist independent density-compound p-values $P_1, \ldots, P_m$
    with $m_0$ true nulls
    such that the SL procedure $\Rr_\alpha$ applied to them satisfies
    $$
    \mathrm{bFDR}(\Rr_\alpha) \geq 1 - (1-\alpha)^{m_0 \wedge \lfloor \frac{\alpha(m - m_0)}{1-\alpha} \rfloor}.
    $$
\end{proposition}
For small $\alpha$, the leading order of this lower bound is
$$
\min\left\{m_0 \alpha + O(m_0^2 \alpha^2),\; \frac{\alpha^2(m - m_0)}{1-\alpha} + O\left(\alpha+\frac{\alpha^4(m - m_0)^2}{(1-\alpha)^2}\right)\right\},
$$
which corroborates the upper bound in \eqref{prop:cp-bfdr-control} to the first order in the regime of $m_0 \leq m \alpha$, or when $m - m_0 \asymp m$ and $\alpha^2 m$ is bounded away from zero.

Proposition~\ref{prop:cp-bfdr-counterexample} shows that density-compound
p-values alone do not guarantee bFDR control at level $\alpha$, even under
independence. We need additional structure to rule out pathological constructions where individual null $P_i$ can concentrate near potential rejection boundaries
to drive up the bFDR.
To quantify how strongly an individual null distribution can concentrate near zero, 
define the null CDF envelope
\begin{equation} \label{eq:gamma}
    \gamma_S(t) := \max_{i \in S} F_0^{(i)}(t) \quad \mathrm{ for }\ t \in [0,1] \ \mathrm{ and }\ S \subseteq [m].
\end{equation}
Thus, $\gamma_{\Nn_0}(\alpha)$ is the largest probability that a true-null p-value falls in $[0,\alpha]$.
Requiring $\gamma_{\Nn_0}(t)\leq t$ for every $t\in[0,1]$ would recover individual validity,
but our results below allow $\gamma_{\Nn_0}(t)>t$ in general.
Moreover, Theorem~\ref{thm:gamma-alpha-bfdr-control} requires an upper bound only on
$\gamma_{\Nn_0}(\alpha)$, rather than on the entire function $\gamma_{\Nn_0}(\cdot)$.
Nonetheless, limiting $\gamma_{\Nn_0}(\alpha)$ effectively asks density-compound p-values
to behave more like individually valid p-values near $\alpha$.

The inflation factor $1/\big(1-\gamma_{\Nn_0}(\alpha)\big)$ below also appears in the FDR bound for BH
applied to independent compound p-values in \citet[Theorem S9]{10.1093/jrsssb/qkag116},
which also follows from the general FDR bounds of \citet[Theorem 1]{10.1214/18-EJS1441}.

\begin{theorem}[bFDR control under worst-case null CDF] \label{thm:gamma-alpha-bfdr-control}
    Suppose $P_1, \ldots, P_m$ are independent density-compound p-values.
    Then the SL procedure $\Rr_\alpha$ satisfies
    $$
    \mathrm{bFDR}(\Rr_\alpha) \leq \min_{\Bb \subseteq \Nn_0} \left\{|\Bb| \alpha + \frac{\alpha}{1 - \gamma_{\Nn_0 \setminus \Bb}(\alpha)}\right\} \leq \frac{\alpha}{1-\gamma_{\Nn_0}(\alpha)}.
    $$
\end{theorem}

In the construction of Proposition \ref{prop:cp-bfdr-counterexample}, 
$\gamma_{\mathcal N_0}(\alpha)$ can equal one, making the corresponding bound vacuous.
The intermediate bound provides more flexibility: in some cases, we can remove a small subset of nulls $\Bb$ with unusually large values of
$F_0^{(i)}(\alpha)$, to achieve a reduction in
$\gamma_{\Nn_0\setminus \Bb}(\alpha)$ and thereby sharpen the overall bound.

While $\gamma_{\Nn_0}(\alpha)$ is intractable since it relies on knowing $\Nn_0$, 
the monotonicity of $\gamma\mapsto\alpha/(1-\gamma)$ and the inequality $\gamma_{\Nn_0}(\cdot) \leq \gamma_{[m]}(\cdot)$ allow us to replace it by $\gamma_{[m]}(\alpha)$ in the bound.
Here, $\gamma_{[m]}(\alpha)$ is computable insofar as we can compute the null CDF of each $P_i$ under $H_{0,i}$.

The following result observes that since SL rejection thresholds are often much smaller than $\alpha$, 
evaluating the null CDF envelope $\gamma_{\Nn_0}(\cdot)$ 
at a suitable data-dependent threshold $\hat{\alpha}$ can sharpen the bound
 when $\hat{\alpha} < \alpha$.
The guarantee bounds by one the expected ratio of the boundary-error
indicator to the reported error level $\alpha/\big(1 - \gamma_{\Nn_0}(\hat \alpha)\big)$, in the spirit of post-hoc hypothesis
testing with e-values
\citep{doi:10.1073/pnas.2302098121, koning2025posthocalphahypothesistesting}.
Here, however, the SL procedure is run at a prespecified nominal
level $\alpha$; the data dependence enters only through the
subsequent error assessment.

For a vector $P=(P_1,\ldots,P_m)$, write $P_{-i}=(P_j)_{j\neq i}$;
expressions such as $(P_{-i},x,t)$ append the displayed coordinates.
For $\lambda\in[0,1]$, restricting SL to candidate thresholds at most $\lambda$
gives the rejection count and threshold
\begin{equation}\label{eq:modified-sl}
    R^{\hat{\tau}_{\alpha}^\lambda}
    = \max\!\left\{
    \argmin_{0 \leq k \leq m: P_{\sigma(k)} \leq \lambda}
    \left(P_{\sigma(k)} - \frac{\alpha k}{m}\right)
    \right\},
    \qquad
    \hat{\tau}_\alpha^\lambda(P) = P_{\sigma(R^{\hat{\tau}_\alpha^\lambda})}.
\end{equation}
For $\lambda=1$, this recovers the original SL threshold $\hat\tau_\alpha(P)$.
For an input vector of another length, $m$ is replaced by that length;
in particular, the denominator is $m+1$ for $(P_{-i},x,t)$.

Set $\tilde\alpha=\alpha(m+1)/m$ and define
\begin{equation}\label{eq:hat-alpha-minus}
    \hat{\alpha}_{-i} := \sup_{x,t\in[0,1]} \hat\tau_{\tilde\alpha}^{\alpha}(P_{-i},x,t),
    \qquad
    \hat{\alpha}_{-} := \max_{i\in\Nn_0} \hat{\alpha}_{-i}.
\end{equation}

\begin{theorem}[bFDR control under post-hoc worst-case null CDF] \label{thm:gamma-bfdr-control-posthoc}
    Suppose $P_1, \ldots, P_m$ are independent density-compound p-values.
    Then the SL procedure $\Rr_\alpha$ satisfies
    $$
    \EE\left[\big(1 -  \gamma_{\Nn_0}(\hat{\alpha}_{-})\big) \ind\{H_{0, \sigma(R_\alpha)} \mathrm{\ is \ true}\}\right] \leq \alpha.
    $$
\end{theorem}

Theorem \ref{thm:gamma-bfdr-control-posthoc} 
shows that a useful data-dependent correction is determined by a worst-case
leave-one-out rejection threshold.
To operationalize the bound, we may again replace the maximum over $\Nn_0$ 
with the maximum over $[m]$.

While computing $\hat{\alpha}_{-}$ results in the tightest guarantee, monotonicity implies the same result 
for any upper bound $\hat{\alpha}_{-} \leq \hat{u}$.
In particular, taking $\hat{u} = \alpha$ recovers the rightmost bound in
Theorem~\ref{thm:gamma-alpha-bfdr-control}.
As a computational shortcut, a middle ground is to leverage SL being more conservative than BH.
Specifically, we can take 
\begin{equation} \label{cor:gamma-bfdr-control-posthoc-bh}
\hat{\alpha}_{\mathrm{BH}}
=
\frac{\alpha}{m}
R_{\tilde\alpha}^{\mathrm{BH}}
\left(P_{-\sigma(m)},0,0\right),
\end{equation}
where $P_{-\sigma(m)}$ is the vector of p-values with the largest p-value removed
and $R_{\tilde\alpha}^{\mathrm{BH}}$ is the BH rejection count for the resulting
length-$(m+1)$ vector at level $\tilde\alpha$.
We verify the claim $\hat{\alpha}_{-} \leq \hat{\alpha}_{\mathrm{BH}}$ in Appendix~\ref{subsec:proofs-main}.

\begin{remark}[Comparison with results for BH] \label{rmk:comparison-bh}
It is instructive to compare the above results with those of
\citet{10.1093/jrsssb/qkag116} for BH applied to independent compound
p-values. They prove
$$
\mathrm{FDR}(\Rr_\alpha^\mathrm{BH})
\leq
\min\left\{1.93,\frac{1}{1-\gamma_{\Nn_0}(\alpha)}\right\}\alpha.
$$
Theorem~\ref{thm:gamma-alpha-bfdr-control} has the same
$\gamma_{\Nn_0}(\alpha)$-dependent form as this bound for the bFDR of SL under
the stronger density-compound condition. 
In contrast, their universal $1.93\alpha$ FDR bound remains
available when $\gamma_{\Nn_0}(\alpha)$ is close to one, whereas
Proposition~\ref{prop:cp-bfdr-counterexample} rules out an analogous universal
constant-factor bFDR guarantee. 
\end{remark}

\begin{remark}[Results under the global null] \label{rmk:global-null}
    Under the global null where $m_0 = m$, the worst-case bFDR inflation
    for independent density-compound p-values is relatively mild.
    Following ideas in \cite{10.1093/jrsssb/qkag116}, we show in Appendix~\ref{subsec:proofs-main}
    for every $\alpha\in[0,2/3]$, the $\mathrm{bFDR}$ is uniformly bounded above by
    $\alpha+2\alpha^2$ over all distributions of independent
    density-compound p-values under the global null.
    Moreover, for $m \geq 2$ and each such $\alpha$, there exists a distribution
    (possibly depending on $\alpha$) for which
    $
    \mathrm{bFDR}(\Rr_\alpha)
    \geq
    \alpha+\alpha^2/4.
    $
\end{remark}

\begin{remark}[Results for individually valid density-compound p-values] \label{rmk:individually-valid}
    If independent density-compound p-values are also individually valid as in~\eqref{eq:superuniform}, then
    $\gamma_{\Nn_0}(\alpha) \leq \alpha$ and Theorem~\ref{thm:gamma-alpha-bfdr-control}
    gives $\mathrm{bFDR}(\Rr_\alpha)\leq \alpha/(1-\alpha)
    =\alpha+\alpha^2+O(\alpha^3)$.
     We can show that an excess of order $\alpha^2$ is necessary.
    For every $\alpha\in(0,1)$ and $3\leq m_0\leq m/4+2$, we construct in
    Appendix~\ref{subsec:proofs-main} independent, individually valid
    density-compound p-values with $m_0$ true nulls and
    $\gamma_{\Nn_0}(\alpha)=\alpha$ that satisfy
    $$
    \mathrm{bFDR}(\Rr_\alpha)
    \geq \alpha+\frac{\alpha^2}{16}(1-3\alpha).
    $$
    Thus, the worst-case excess above $\alpha$ is of order $\alpha^2$, even as $\bar\pi_0\to0$ with a fixed $m_0\geq3$.
    In this construction, an oracle
    Storey-type adjustment \citep{10.1111/1467-9868.00346} that raises the SL
    level to $\alpha/\bar\pi_0$ can only exacerbate the excess.
\end{remark}

\begin{remark}[Cheating with FDR and Dirac-alternatives] \label{rmk:cheating-fdr}
    Cheating with FDR occurs when hypotheses with weak evidence are rejected due to the presence of many strong discoveries,
    as exemplified in Figure~\ref{fig:fdr-compound-schematic}.
    The Dirac-alternatives configuration of \cite{finner2009false} provides an extreme benchmark for this phenomenon,
    where $P_i = 0$ for all $i \notin \Nn_0$.
    We show in Appendix \ref{subsec:dirac-alternatives}
    that the SL procedure on independent compound p-values \eqref{eq:cp} under this configuration satisfies
    $$
    \mathrm{bFDR}(\Rr_\alpha) = \PP(\text{SL rejects any null}) \leq \alpha + 2\alpha^2.
    $$
    The key observation is that SL effectively discards the zeros when deciding which other hypotheses to reject.
    To see why, write $P=(0,\ldots,0,P_{\mathrm{rest}})$, where
    $P_{\mathrm{rest}}$ contains the positive p-values.
    Let $Q_k$ be their $k$th order statistic, with $Q_0=0$.
    SL rejects all zeros, and its objective for rejecting an additional $k$ hypotheses is
    $$
    Q_k-\frac{\alpha(m-m_0+k)}{m}
    =Q_k-\frac{\alpha\bar\pi_0 k}{m_0}-\alpha(1-\bar\pi_0),
    \qquad 0\leq k\leq m_0.
    $$
    In particular, $Q_k-(\alpha\bar\pi_0 k)/{m_0}$ is the SL objective for
    $P_{\mathrm{rest}}$ at level $\alpha\bar\pi_0$.
    Since the remaining term is constant in $k$, both procedures
    make the same rejections.
    See Example~\ref{ex:fdr-bfdr-compound}
    in Appendix~\ref{subsec:dirac-alternatives}
    for a comparison with BH in this setting.
\end{remark}

\subsection{bFDR for Discrete Density-Compound p-values} \label{subsec:discrete-bfdr}
For the discrete exact uniform case,
\cite{10.1093/biomet/asaf083} prove that if $P_i \iid \mathrm{Unif}(\Gg_L)$ for $i \in \Nn_0$,
then $\mathrm{bFDR}(\Rr_\alpha) \leq \bar{\pi}_0 \alpha + m^2/L$.
Their counterexamples and simulations show the inflation due to discreteness is unavoidable without further assumptions.
Using their results, we present the following general bound on the bFDR for discrete density-compound p-values.

\begin{proposition} \label{prop:discrete-cp-bfdr-bound}
    Suppose $P_1, \ldots, P_m$ are independent discrete density-compound
    p-values on $\Gg_L$.
    Then the SL procedure $\Rr_\alpha$ satisfies
     $$
     \mathrm{bFDR}(\Rr_\alpha) 
     \leq \frac{\alpha + m(m+1)/L}{1-\gamma_{\Nn_0}(\alpha)}
     $$
\end{proposition}

However, $O(m^2/L)$ is typically too large for the upper bound to be meaningful.
To refine this, we need to leverage the specific structure of particular constructions.
For instance, \citet{gao2026conformalnoveltydetectionfalse} refine the bFDR bound of conformal p-values
to $\bar{\pi}_0 \alpha + m/(K + 1)$ for testing $m$ hypotheses with $K$ null replicates total.

Instead, we can use the special structure of the pooled permutation p-values in
Proposition~\ref{prop:dcperm-validity}, yielding the following bound that avoids the
$O(m^2/L)$ discretization term.

Write $\widehat P_i:=n^{-1}\sum_{\ell=1}^n\delta_{Z_{i\ell}}$ for the
empirical distribution of unit $i$'s raw outcomes and
$\widehat P_{1:m}:=(\widehat P_1,\ldots,\widehat P_m)$.
Conditional on $\widehat P_{1:m}$, the full permutation collections are
fixed, but the observed statistic in each null unit remains random.
Define
$\gamma_S^{\mathrm{dcperm}}(t):=\max_{i\in S}
\PP_{H_{0,i}}(P_i^{\mathrm{dcperm}}\leq t\mid\widehat P_{1:m})$
for $S\subseteq[m]$ and $t\in[0,1]$.

\begin{theorem}[bFDR for pooled permutation p-values]
\label{thm:bfdr-dcperm}
Under the assumptions of Proposition~\ref{prop:dcperm-validity}, for
$\alpha\in(0,1)$ the SL procedure applied to
$P_1^{\mathrm{dcperm}},\ldots,P_m^{\mathrm{dcperm}}$ satisfies
\begin{equation}\label{eq:permutation-bfdr-bound}
\EE\left[
\left\{1-\gamma_{\Nn_0}^{\mathrm{dcperm}}(\hat{\alpha}_{-})\right\}
\ind\{H_{0,\sigma(R_\alpha)}\mathrm{\ is\ true}\}
\,\middle|\,\widehat P_{1:m}
\right]\leq\alpha,
\end{equation}
with $\hat{\alpha}_{-}$ defined by \eqref{eq:hat-alpha-minus}
using $P_i=P_i^{\mathrm{dcperm}}$.
\end{theorem}

The conditional null probabilities are computable from the permutation
statistics. Let $\mathcal V_i:=\{X_i^\pi:\pi\in\mathfrak S(n)\}$ and
write $V_{(1)}>\cdots>V_{(mB)}$ for the distinct values pooled across units.
For every $i\in\Nn_0$ and $t\in[0,1]$, equal multiplicity gives
\begin{equation}\label{eq:gamma-alpha-permutation}
\PP(P_i^{\mathrm{dcperm}}\leq t\mid\widehat P_{1:m})
=\frac{1}{B}\sum_{r=1}^{\lfloor tmB\rfloor}
\ind\{V_{(r)}\in\mathcal V_i\}.
\end{equation}
Taking the maximum of the right-hand side over all $i\in[m]$ gives
$\gamma_{[m]}^{\mathrm{dcperm}}(t)$, a computable upper bound on
$\gamma_{\Nn_0}^{\mathrm{dcperm}}(t)$.

Appendix~\ref{subsec:block-rank-verification} develops a general
compound-control construction and bFDR bound encompassing pooled Monte
Carlo \eqref{eq:compound-mc-pval}, fully enumerated permutation \eqref{eq:compound-perm-pval}, and randomized permutation p-values \eqref{eq:sampled-compound-perm-pval} used in
the regression discontinuity application of Section~\ref{subsec:rd}.

\section{Connection to Frequentist LFDR} \label{sec:freq-lfdr}
In this section, we interpret density-compound p-values through the lens of the local false discovery rate.
The key idea is that the SL procedure can be viewed as targeting a conservative estimate of local FDR at the rejection boundary.
For individually valid p-values with exact uniform nulls, the target is justified by standard upper bounds.
For density-compound p-values, the same target remains conservative because the sum of the true-null densities, divided by $m$, is bounded by the uniform density.
We use this insight to establish asymptotic frequentist local FDR
control at the SL rejection boundary under weak dependence
(Theorem~\ref{thm:lfdr-upper-bound-weak-dependence}).
Since we discuss several false discovery criteria in this section, 
we summarize them in Table \ref{tab:error-metrics}.

\begin{table}[t]
\centering
\small
\renewcommand{\arraystretch}{1.25}
\begin{tabularx}{\textwidth}
{
l
>{\raggedright\arraybackslash\hsize=0.72\hsize}X
>{\raggedright\arraybackslash\hsize=0.48\hsize}X
>{\raggedright\arraybackslash\hsize=1.8\hsize}X
}
\toprule
Quantity
&
Definition
&
Null status
&
Interpretation
\\
\midrule

FDR
&
$\EE\!\left[V_\alpha/(R_\alpha\vee 1)\right]$
&
Fixed
&
Expected proportion of false discoveries among the rejected hypotheses.
\\

LFDR
&
$\PP(H_i=0\mid P_i=t)$
&
Random
&
Posterior probability that a hypothesis is null given its observed p-value.
\\

$\mathrm{LFDR}^{\mathrm{freq}}_m$
&
$\displaystyle
\frac{\sum_{i\in\Nn_0} f^{(i)}(t)}
{\sum_{i=1}^m f^{(i)}(t)}$
&
Fixed
&
Relative frequency of nulls at p-value $t$, expressed as a ratio of summed densities.
\\
bFDR
&
$\PP\!\left(
H_{0,\sigma(R_\alpha)}\text{ is true}
\right)$
&
Fixed
&
Probability that the boundary rejection is null.
\\

\bottomrule
\end{tabularx}
\caption{
    Comparison of false discovery criteria. Here $V_\alpha$ denotes the number of false rejections.
    For brevity in defining the bFDR, we set $P_{\sigma(0)} := 0$ and declare $H_{0,\sigma(0)}$ false.
}
\label{tab:error-metrics}
\end{table}

\subsection{Background: Bayesian Local FDR and the SL procedure}
To start, we first recall local FDR within the Bayesian two-groups model \citep{Efron01122001}. Let $H_i \in \{0,1\}$ where $H_i = 0$ indicates the $i$-th null hypothesis is true
with prior probability
$
\mathbb{P}(H_i = 0) = \pi_0.
$
Suppose
$$
P_i \mid H_i = 0 \sim f_0, \qquad P_i \mid H_i = 1 \sim f_1,
$$
so that the marginal density of $P_i$ is
$
f(t) = \pi_0 f_0(t) + (1 - \pi_0) f_1(t).
$
Within the Bayesian model, the (Bayesian) local FDR is the posterior probability that a hypothesis is null given its p-value:
$$
\mathrm{LFDR}(t) = \mathbb{P}(H_i = 0 \mid P_i = t) = \frac{\pi_0 f_0(t)}{f(t)}.
$$
When the null density is uniform, $f_0(t) = 1$, and one takes the conservative bound $\pi_0 \le 1$, the local FDR is bounded by
\begin{equation} \label{eq:tg-lfdr-upper-bound}
\mathrm{LFDR}(t) \le \frac{1}{f(t)}.
\end{equation}
For rejection rules of the form $\{i \in [m] : P_i \le t\}$, 
the boundary rejection is the hypothesis with the largest rejected p-value. Focusing on the reliability of this boundary rejection
leads to the following population-level interpretation of the SL procedure. Let
$
\hat F_m(t) = m^{-1}\sum_{i=1}^m \ind\{P_i \le t\}
$
be the empirical CDF of the p-values.
Then, the SL threshold $\hat{\tau}_\alpha$ satisfies
$$
\hat{\tau}_\alpha = \max\left\{\argmax_{t \in [0,1]} \hat{G}_m(t)\right\}, \qquad \hat{G}_m(t) = \alpha \hat F_m(t) - t,
$$
over observed p-values.
Let $F$ denote the CDF corresponding to the mixture density $f$.
Replacing $\hat F_m$ by $F$ gives the population criterion maximized by SL,
$G(t)=\alpha F(t)-t$.
If $G$ has a unique interior maximizer $\tau_\alpha^*$ and $f$ is
continuous there, the first-order condition gives
$$
\frac{1}{f(\tau_\alpha^*)}=\alpha.
$$
Thus, under uniform nulls, the population SL boundary satisfies
$\mathrm{LFDR}(\tau_\alpha^*)\leq\alpha$.
The SL procedure can therefore be viewed as estimating this population
rejection boundary.

\subsection{Continuous Density-Compound p-values and Frequentist LFDR}
We return to the frequentist setting where hypotheses $\{H_{0,i}\}_{i=1}^m$ have a fixed set of null indices $\Nn_0$.
We now allow the p-values to have different marginal CDFs $F^{(i)}$
and densities $f^{(i)}$. Write
$\bar F_m(t)=m^{-1}\sum_{i=1}^m F^{(i)}(t)$ and
$\bar f_m(t)=m^{-1}\sum_{i=1}^m f^{(i)}(t)$ for their averages.
To assess discoveries locally without placing a prior on the
hypotheses, \citet{10.1093/biomet/asaf083} introduce frequentist local
FDR, a quantity that describes how the relative frequency of nulls varies
across the p-value scale. For $t\in[0,1]$ with
$\sum_{i=1}^m f^{(i)}(t)>0$, it is defined by
$$
\mathrm{LFDR}^{\mathrm{freq}}_m(t) := \frac{\sum_{i \in \Nn_0} f^{(i)}(t)}{\sum_{i=1}^m f^{(i)}(t)}.
$$
As in the Bayesian two-groups bound \eqref{eq:tg-lfdr-upper-bound},
it holds for density-compound p-values that
$\mathrm{LFDR}^{\mathrm{freq}}_m(t)\leq1/\bar f_m(t)$,
so $1/\bar f_m(t)$ remains a conservative local FDR target even when
the null distributions differ.

The population criterion maximized by SL is now $G_m(t)=\alpha\bar F_m(t)-t$.
If $G_m$ has a unique interior maximizer $\tau_{\alpha,m}^*$ and
$\bar f_m$ is continuous there, the same first-order argument gives
\begin{equation}\label{eq:sl-population-threshold}
\frac{1}{\bar f_m(\tau_{\alpha,m}^*)}=\alpha.
\end{equation}
Consequently, the density-compound condition yields
$\mathrm{LFDR}^{\mathrm{freq}}_m(\tau_{\alpha,m}^*)\leq\alpha$,
preserving the frequentist local FDR rationale for applying the SL procedure.

\subsection{Asymptotic Frequentist LFDR Control under Weak Dependence}
The preceding discussion gives a population-level interpretation of the
SL threshold. To transfer this interpretation to the empirical threshold,
we consider a triangular array of multiple testing problems as $m\to\infty$: for each $m$, we test hypotheses
$H_{0,1}^{(m)},\ldots,H_{0,m}^{(m)}$
using density-compound p-values $P_1^{(m)},\ldots,P_m^{(m)}$.
Thus, all quantities associated with each testing problem depend on $m$.
For this array, write
$
\hat F_m(t) = m^{-1}\sum_{i=1}^m \ind\{P_i^{(m)} \le t\}
$
and 
$\bar F_m(t)=m^{-1}\sum_{i=1}^m \PP(P_i^{(m)}\le t)$.
Let $\bar f_m$ denote the density corresponding to $\bar F_m$.

We impose uniform convergence of the empirical CDF to the average marginal CDF:
\begin{equation}\label{eq:weak-dependence}
    \Vert \hat F_m-\bar F_m\Vert_\infty=o_p(1),
\end{equation}
commonly used to formalize weak dependence in multiple testing
\citep[cf.][]{10.1111/j.1467-9868.2004.00439.x}.
For the SL procedure,
\citet[Proposition~8]{10.1214/24-AOS2359}
use the same uniform convergence condition, together with identification
of the population threshold, to establish consistency of the empirical
SL threshold.

Assume that, for each $m$, $G_m(\tau)=\alpha\bar F_m(\tau)-\tau$
has a unique maximizer $\tau_{\alpha,m}^*\in(0,1)$.
We further assume that these maximizers are uniformly well separated: for every $\eta>0$,
\begin{equation}\label{eq:separation-condition}
\liminf_{m\to\infty} \left\{G_m(\tau_{\alpha,m}^*) - \sup_{\tau\in[0,1]:|\tau-\tau_{\alpha,m}^*|\ge \eta} G_m(\tau) \right\} > 0,
\end{equation}
with the convention $\sup\varnothing=-\infty$.
Finally, we impose uniform local regularity of the average marginal
density: there exist constants $C_0,\eta_0>0$ such that, for all
sufficiently large $m$,
\begin{equation}\label{eq:local-density-regularity}
\left|
\bar f_m(\tau)-\bar f_m(\tau_{\alpha,m}^*)
\right|
\le
C_0|\tau-\tau_{\alpha,m}^*|
\quad
\text{whenever }
|\tau-\tau_{\alpha,m}^*|\le\eta_0.
\end{equation}

\begin{theorem}[Frequentist LFDR control under weak dependence]
\label{thm:lfdr-upper-bound-weak-dependence}
Consider a sequence of multiple testing problems as above, with $\alpha\in(0,1)$
fixed. Suppose each $G_m$ has a unique interior maximizer
$\tau_{\alpha,m}^*$ and \eqref{eq:weak-dependence}--\eqref{eq:local-density-regularity} hold.
If, for each $m$, $P_1^{(m)},\ldots,P_m^{(m)}$ are density-compound p-values, then, as $m\to\infty$,
$$
\mathrm{LFDR}^{\mathrm{freq}}_m(\hat\tau_\alpha)
\le \alpha+o_p(1),
$$
where $\hat\tau_\alpha$ is the SL rejection threshold at level $\alpha$
computed from $P_1^{(m)},\ldots,P_m^{(m)}$.
\end{theorem}

Unlike the finite-sample results in Section~\ref{sec:bfdr}, 
Theorem~\ref{thm:lfdr-upper-bound-weak-dependence} is not a bFDR guarantee. 
Rather, it shows that the SL rejection threshold $\hat{\tau}_\alpha$ retains the frequentist local FDR interpretation under a weak dependence condition. 
Establishing bFDR control under dependence conditions is left for future work.

\section{Simulation Study} \label{subsec:simulation-continuous}
We compare the bFDR, FDR, and power (defined as the expected proportion of non-nulls rejected) of SL and BH applied to four p-value constructions
in the Gaussian mean-testing setting of Example~\ref{ex:cp-eb-normal-known-var}.
For each of $m=2000$ independent units, we observe $n=5$ observations
$Y_{i1},\ldots,Y_{in}\iid\Nn(\mu_i,\sigma_i^2)$ and test $H_{0,i}:\mu_i=0$.
The null proportion is fixed at $\pi_0=0.99$, and the $20$ non-null means are
$\mu_i=\xi/\sqrt n$, with $\xi\in\{2,3.5,5,6.5,8\}$.
Standard deviations are drawn independently
from $\mathrm{Unif}(0.7,1.3)$.
We average over $10{,}000$ Monte Carlo repetitions per setting, applying both procedures
to the same p-values at nominal level $\alpha=0.1$.

The four constructions we consider are
\begin{enumerate*}[label=(\roman*)]
\item the usual two-sided t-test p-values $P_i^t$;
\item the density-compound p-values $P_i^{\mathrm{oracle}}$ in
\eqref{eq:oracle-gaussian-dcp}; \item
the weighted t-test p-values $P_i^{\mathrm{w}}$ of
\citet{westfall2004weighted}; 
\item and the oracle compound-e-weighted p-values
$P_i^{\mathrm{ew}}$ of \citet{10.1093/biomet/asad057}.
\end{enumerate*}
Appendix~\ref{subsec:simulation-details}
gives the formulas. We have bFDR control with SL for (i) by the results of~\citet{10.1093/biomet/asaf083}, for (iii)-(iv) by Theorem~\ref{thm:bfdr-weighted-pvals} and (ii) by Theorem~\ref{thm:gamma-alpha-bfdr-control} 
with the worst-case null envelope of approximately $\gamma_{[m]}(\alpha) = 0.2$, yielding a bFDR bound of $1.25\alpha$.

\begin{figure}[!ht]
    \centering
    \resizebox{\linewidth}{!}{\input{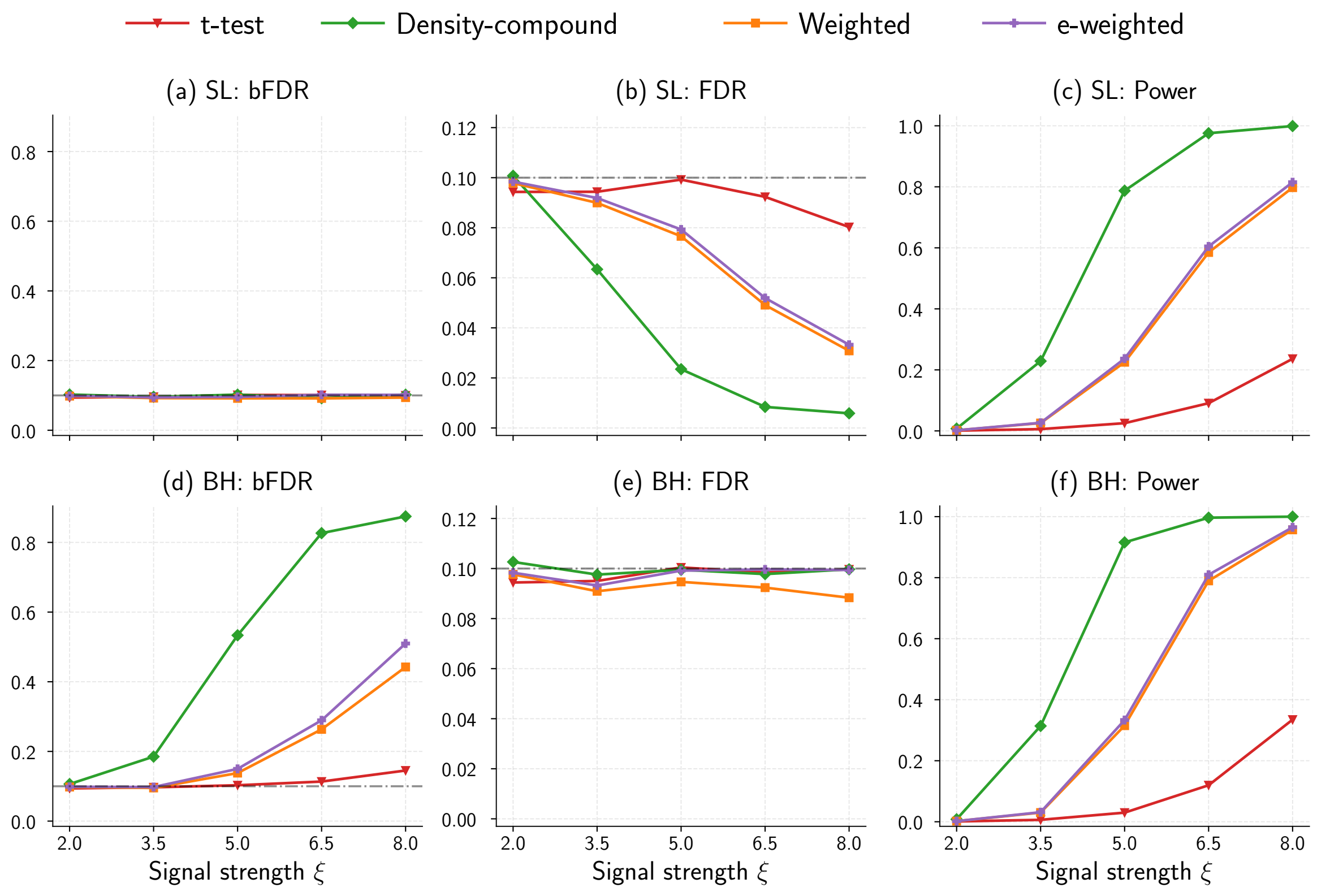}}
    \caption{Comparison of SL and BH with $m=2000$, $n=5$, $\pi_0=0.99$, and $\alpha=0.1$.
    The top row shows SL and the bottom row BH; columns show bFDR, FDR,
    and power.
    Gray lines mark the nominal level in the error panels.
    Each point averages $10{,}000$ repetitions.}
    \label{fig:t-test-0.3}
\end{figure}

Results are shown in Figure~\ref{fig:t-test-0.3}.
Panels (c) and (f) show substantial power gains over the t-test in this
low-sample, high-sparsity setting. At $\xi=5$, SL has power $0.79$ with the
density-compound p-values, $0.23$ and $0.24$ with the weighted and
compound-e-weighted p-values, and $0.03$ with the t-test p-values.
Panels (a)--(b) and (d)--(e) illustrate the distinction between the two
error criteria. SL's estimated bFDR stays near or below $0.1$, whereas
BH can have much larger bFDR despite FDR near the nominal level.
For the density-compound p-values at $\xi=5$, BH has FDR $0.1$ but
bFDR above $0.5$, compared with bFDR $0.1$ for SL.

\section{Applications} \label{sec:applications}
In this section, we apply the discrete permutation constructions of Section~\ref{subsec:discrete-dcp} to two applications: a regression discontinuity design in education economics (Section~\ref{subsec:rd}) and a differential gene expression analysis in high-throughput biology (Section~\ref{subsec:gene-expression}).
Both applications involve low-resolution permutation tests, where pooling
permutation statistics across hypotheses can provide finer p-value resolution
and thereby improve power.
We apply both SL and BH to each p-value construction, with SL targeting bFDR
control and BH serving as a comparison targeting FDR control.

\subsection{Regression Discontinuity Design} \label{subsec:rd}
We revisit the study of Romanian high-school admissions by
\citet{10.1257/aer.103.4.1289}, which uses a regression discontinuity design (RDD).
Students apply to tracks within schools, such as Mathematics,
and admission depends on a ``transition score,'' a weighted average of GPA and admission-test performance.
Each school/track has a year-specific cutoff, above which students are eligible for admission.
The outcome is performance on the Baccalaureate exam, the national examination taken at the end of high school.
The RDD seeks to assess causal effects by comparing outcomes for students just above and just below an admission cutoff; see \citet{IMBENS2008615} and~\citet{cattaneo2022regression} for background on RDD methods.

The original study adopts the continuity-based approach to analyzing RDDs~\citep{hahn2001identification}:
if mean potential outcomes vary continuously with the transition score at the cutoff,
a discontinuity in mean observed outcomes identifies a local causal effect of admission eligibility on later academic performance. \citet{10.1257/aer.103.4.1289} use local linear regression
to estimate average discontinuities pooled across cutoffs for different schools/tracks.

An alternative to local linear regression is the local randomization approach, which treats units
sufficiently close to the cutoff as if treatment assignment were randomized
\citep{CattaneoFrandsenTitiunik+2015+1+24,mattei2017regression}.
Under this assumption, each cutoff neighborhood forms a small
randomized experiment. We adopt this framework to test for
school/track-specific discontinuities using local permutation tests,
pooling observations across years.
The 2001--2003 data cover 135 towns and $6434$ town--school/track--year cutoffs.
After centering transition scores at their year-specific cutoffs and pooling across years,
there are $2575$ school/track combinations.\footnote{See \citet{10.1111/ectj.12092} for a school-level empirical Bayes analysis.}
We use a bandwidth of $h_1=0.05$.\footnote{Whereas \citet{10.1257/aer.103.4.1289} use a bandwidth of $h_0 = 1.0$, we use a smaller bandwidth to make the assumption of local randomization near the cutoff more plausible. This nonetheless remains a strong assumption~\citep{eckles2025noiseinduced}; see \citet{10.1093/restud/rdx062} for a formalization of approximate local randomization in RDDs.}

Before restricting the sample to $[-h_1,h_1]$, we require each combination to have at least one observation
at or below $-h_1$ and one at or above $h_1$.
Within $[-h_1,h_1]$, we exclude observations exactly at the cutoff and require at least $4$ observations
on each side and at most $20$ in total. These restrictions accentuate the low-sample regime
where individual permutation p-values have limited resolution, leaving $m=345$ quasi-experiments.

For each quasi-experiment, we enumerate all treatment assignments to
compute $P_i^{\mathrm{perm}}$ in \eqref{eq:perm-pval} and construct $P_i^{\mathrm{dcperm},K}$
as described in
Appendix~\ref{subsec:sampled-permutation}
by sampling $K=\binom{8}{4}-1=69$ distinct assignments uniformly
without replacement, excluding the observed assignment.
Together with the observed statistic, this gives 70 statistics per
quasi-experiment. Both constructions use the (negative) one-sided
Welch t-test p-value as the test statistic \citep{10.1093/biomet/29.3-4.350}, 
with higher Baccalaureate
scores to the right of the cutoff as the alternative.

\begin{figure}[!ht]
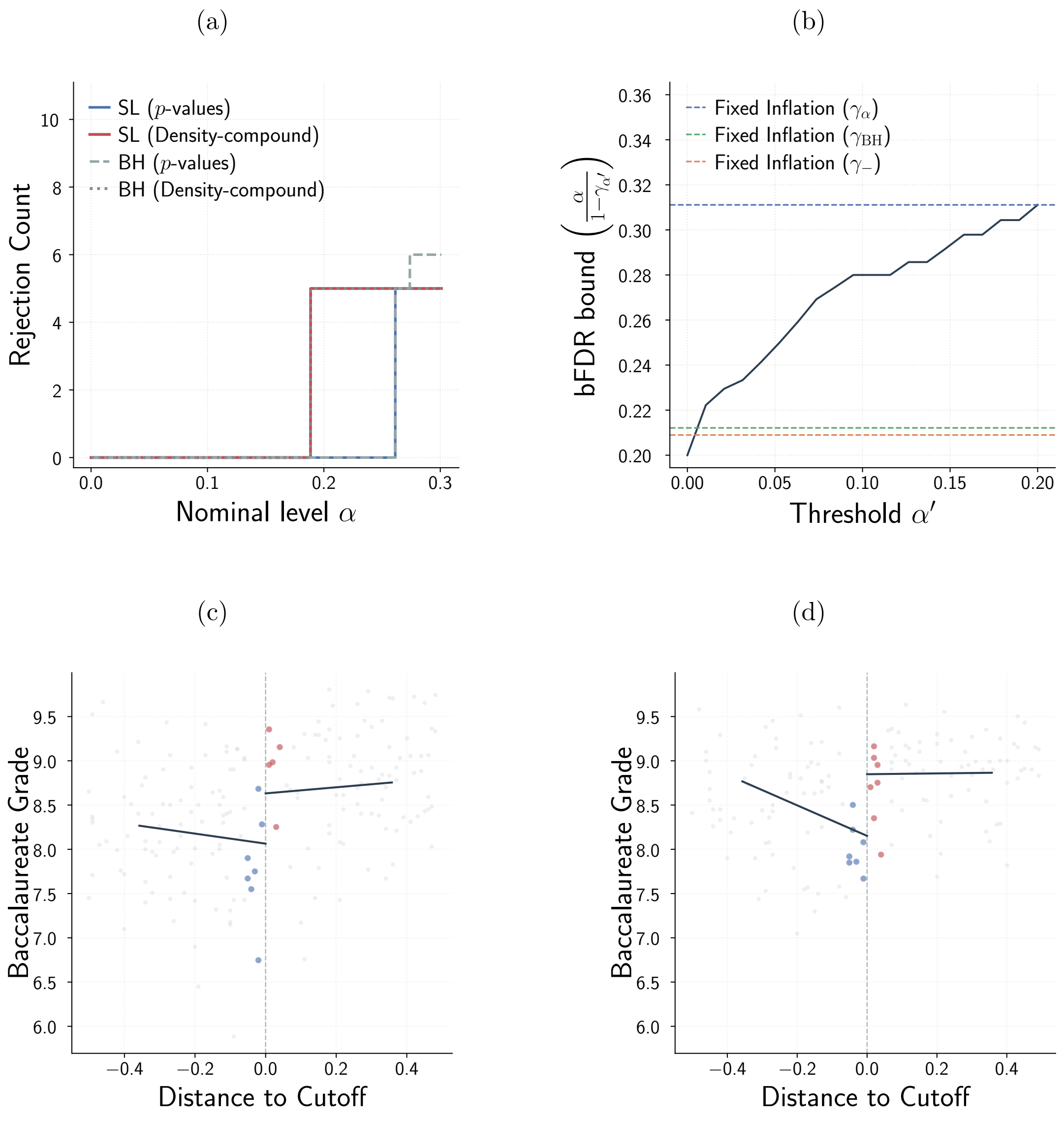

    \centering
    \begin{subfigure}[b]{0.48\textwidth}
        \centering
        \caption{}\label{fig:rejection-curves_rd}
        \resizebox{\linewidth}{!}{\input{images/rd_rejection_curves.tex}}
    \end{subfigure}
    \hfill
    \begin{subfigure}[b]{0.48\textwidth}
        \centering
        \caption{}\label{fig:rejection-inflation_rd}
        \resizebox{\linewidth}{!}{\input{images/rd_inflation_curve.tex}}
    \end{subfigure}

    \vspace{2ex}

    \begin{subfigure}[b]{0.48\textwidth}
        \centering
        \caption{}\label{fig:rd-rejection-strongest}
        \resizebox{\linewidth}{!}{\clipbox{0 0 426.5bp 0}{\input{images/rd_local_fits.tex}}}
    \end{subfigure}
    \hfill
    \begin{subfigure}[b]{0.48\textwidth}
        \centering
        \caption{}\label{fig:rd-rejection-boundary}
        \resizebox{\linewidth}{!}{\clipbox{426.5bp 0 0 0}{\input{images/rd_local_fits.tex}}}
    \end{subfigure}
    \caption{Comparison of permutation p-values and density-compound p-values
    in the RDD application. (a) Rejection counts versus
    nominal level $\alpha$ for SL and BH. (b) Theoretical bFDR bound versus threshold
    $\alpha'$ at $\alpha=0.2$. 
    (c)--(d) Local RDD fits for the density-compound discoveries with the
    smallest and largest rejected p-values, respectively.
    In (c)--(d), colored points lie in $[-h_1, h_1]$, while gray points show other
    points within distance $0.5$ of the cutoff. Solid lines are local linear fits with bandwidths selected by \texttt{rdrobust} \citep{RJ-2015-004}.}
    \label{fig:rd-rejections}
\end{figure}

Figure~\ref{fig:rejection-curves_rd} compares the SL and BH rejection curves.
At the illustrative level $\alpha=0.2$, $P_i^{\mathrm{perm}}$ yield no rejections while $P_i^{\mathrm{dcperm},K}$ reject $5$ quasi-experiments under SL.
Figure \ref{fig:rejection-inflation_rd} shows the theoretical bFDR bound vs. the threshold $\alpha^\prime$ at $\alpha = 0.2$.
Using $\gamma_\alpha := \gamma_{[m]}(\alpha)$ gives a theoretical bFDR bound of about $0.31$ (see \eqref{eq:gamma-alpha-block} in Appendix~\ref{subsec:block-rank-verification} for the formula).
The post-hoc choices of Corollary~\ref{cor:bfdr-sampled-dcperm} $(\gamma_{-} := \gamma_{[m]}(\hat{\alpha}_{-}))$ in Appendix~\ref{subsec:sampled-permutation}
and Equation \eqref{cor:gamma-bfdr-control-posthoc-bh} ($\gamma_{\mathrm{BH}} := \gamma_{[m]}(\hat{\alpha}_{\mathrm{BH}})$) 
yield a bound of about $0.21$.
Figures~\ref{fig:rd-rejection-strongest} and~\ref{fig:rd-rejection-boundary} respectively
show the density-compound discoveries with the smallest and largest
rejected p-values at $\alpha=0.2$,
both visually consistent with upward discontinuities.

Appendix~\ref{subsec:additional-plots-rd} provides additional robustness checks.
Comparison with deterministic compound p-values constructed via full enumeration of group assignments
suggests that randomly sampling group assignments has little effect
on our conclusions (Appendix Figure~\ref{fig:rd-pvalue-distributions_cp}).
A falsification test placing artificial cutoffs between adjacent real
cutoffs yields no rejections at $\alpha=0.2$ for the
baseline bandwidth $h_1=0.05$ or the smaller bandwidths considered,
but produces rejections at artificial cutoffs at some larger bandwidths
(Appendix Figure~\ref{fig:rd-placebo-rejection-curves}).
When testing 
multiple cutoffs,
increasing the bandwidth can improve power, but render the local
randomization assumption less plausible. 
Density-compound p-values
offer a way to attain higher power relative to individual permutation p-values while retaining a smaller bandwidth.

\subsection{Differential Gene Expression Analysis} \label{subsec:gene-expression}
We analyze the macrophage gene-expression dataset of \cite{Rangel-Salazar2011}, 
which compares gene expression in 3 unstimulated control samples and 9 lipoprotein-stimulated samples measured on microarrays. 
The biological goal of the original study was to identify genes whose expression is repressed in response to lipoprotein stimulation. 
After preprocessing, following Supplement C of \cite{ling2025empiricalpartiallybayessample}, 
$m = 20{,}989$ genes remain for analysis.

\begin{figure}[!ht]
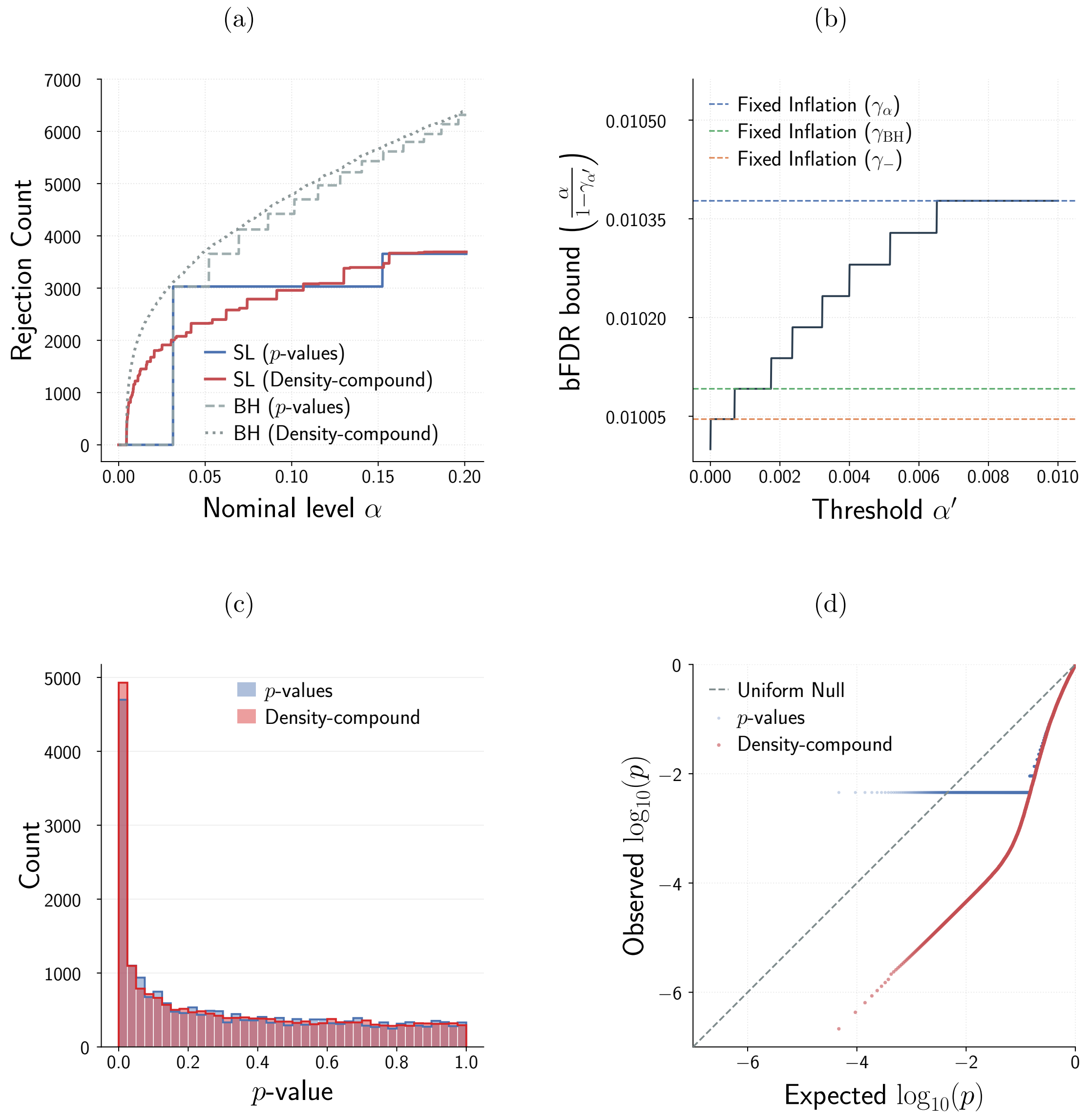

    \centering
    \begin{subfigure}[b]{0.48\textwidth}
        \centering
        \caption{}\label{fig:rejection-curves}
        \resizebox{\linewidth}{!}{\input{images/rejection_curves.tex}}
    \end{subfigure}
    \hfill
    \begin{subfigure}[b]{0.48\textwidth}
        \centering
        \caption{}\label{fig:rejection-inflation}
        \resizebox{\linewidth}{!}{\input{images/inflation_alpha_prime.tex}}
    \end{subfigure}

    \vspace{2ex}

    \begin{subfigure}[b]{0.48\textwidth}
        \centering
        \caption{}\label{fig:pvalue-histogram}
        \resizebox{\linewidth}{!}{\input{images/pvalue_histogram.tex}}
    \end{subfigure}
    \hfill
    \begin{subfigure}[b]{0.48\textwidth}
        \centering
        \caption{}\label{fig:pvalue-qq}
        \resizebox{\linewidth}{!}{\input{images/qq_tail_plot.tex}}
    \end{subfigure}
    \caption{Comparison of permutation p-values and density-compound p-values
    in the gene-expression application. (a) Rejection counts versus nominal
    level $\alpha$. (b) Theoretical bFDR bound versus threshold $\alpha'$ at
    $\alpha=0.01$. (c) Histogram of p-values. (d) $\log_{10}$ Q-Q plot of
    p-values.}
    \label{fig:master-simulation}
\end{figure}

For each gene, we enumerate all $\binom{12}{3}=220$ treatment assignments
and use the absolute equal-variance two-sample t-statistic. Each
assignment is induced by $3!9!$ permutations, whose common multiplicity
cancels from the full permutation sums.
We then construct both the individually valid permutation p-values $P_i^\mathrm{perm}$ defined in \eqref{eq:perm-pval} 
and the discrete density-compound permutation p-values $P_i^\mathrm{dcperm}$ defined in \eqref{eq:compound-perm-pval}.
Figure~\ref{fig:rejection-curves} compares the SL and BH rejection curves.
Under SL at $\alpha = 0.01$, $P_i^\mathrm{perm}$ yield no rejections due to low resolution.
In contrast, $P_i^\mathrm{dcperm}$ reject $1{,}222$ genes.
Figure \ref{fig:rejection-inflation} shows the theoretical bFDR bound vs. the threshold $\alpha^\prime$ at $\alpha = 0.01$.
Using $\gamma_\alpha := \gamma_{[m]}(\alpha)$ gives a theoretical bFDR bound of about $0.0104$ (see \eqref{eq:gamma-alpha-permutation} for the formula).
The post-hoc choices of Theorem~\ref{thm:bfdr-dcperm} $(\gamma_{-} := \gamma_{[m]}(\hat{\alpha}_{-}))$
and Equation \eqref{cor:gamma-bfdr-control-posthoc-bh} ($\gamma_{\mathrm{BH}} := \gamma_{[m]}(\hat{\alpha}_{\mathrm{BH}})$) 
yield a bound even closer to $\alpha = 0.01$.

Beyond small $\alpha$, the rejection curve for compound p-values is smoother as a function of $\alpha$ 
while the rejection curve for permutation p-values is stepwise constant.
However, the number of rejections for density-compound p-values does not uniformly dominate that of individually valid permutation p-values. 
The histogram and Q-Q plot of p-values and compound p-values are shown in Figures 
\ref{fig:pvalue-histogram} and \ref{fig:pvalue-qq}.

\section{Discussion and Future Work} \label{sec:discussion}
In this work, we introduced density-compound p-values, a density-level strengthening of compound p-values 
motivated by bFDR control and its connection to local FDR.
Under independence, we established nominal-level bFDR control for
SL applied to weighted and compound-e-weighted p-values, and
bounds involving the null CDF envelope $\gamma$ for general
density-compound p-values.
Our lower bound shows why the density-compound condition alone
does not ensure nominal-level bFDR control.
We also developed guarantees for pooled permutation constructions
and a frequentist local FDR interpretation under weak dependence.
The simulation and applications illustrate the power gains these
constructions can offer in low-sample or low-resolution settings.
Several questions remain for future work.

\paragraph{General error metrics.}
FDR can be written as the probability that a uniformly selected rejection is null
while bFDR is the probability that the boundary rejection is null.
This suggests a continuum of error criteria interpolating between FDR and bFDR.
For weights $w_j=w_j(R_{\alpha})\geq0$ satisfying
$\sum_{j=1}^{R_{\alpha}}w_j=1$ when $R_\alpha>0$, one could define
$$
\mathrm{FDR}_w
=
\PP\left(H_{0, \sigma(J)} \text{ is true}, R_{\alpha} > 0 \right)
=
\EE\left[
    \sum_{j=1}^{R_{\alpha}}
    w_j(R_{\alpha})
    \ind\{H_{0, \sigma(j)} \text{ is true}\}
\right],
$$
where, conditional on the p-values and $R_\alpha>0$, we draw $J$ with
probabilities $w_j(R_\alpha)$; set $J=0$ when $R_\alpha=0$.
Uniform weights recover FDR, while placing all mass on $j=R_\alpha$ recovers bFDR.
Other choices, such as weights increasing in $j$, would emphasize discoveries near the rejection boundary.
This differs from classical weighted FDR \citep{a5f3b3b3-ba89-31a3-b019-58d85d361941}, where weights are assigned to hypotheses before observing the data,
whereas weights here are assigned to positions in the rejected set and depend on the random variable $R_{\alpha}$. 
Developing powerful procedures to control such criteria in finite samples is an interesting direction.

\paragraph{Extending discrete density-compound p-values to heterogeneous grids.}
Our discrete theory focuses on settings where the pooled ranks lie on a
common grid, obtained either by full enumeration with a common number of
distinct statistic values or by retaining a common number of statistics
for each hypothesis.
This covers common-design experiments as in Section~\ref{subsec:gene-expression},
and the construction in Section~\ref{subsec:rd}, which samples the same number
of permutation statistics for each hypothesis. Extending bFDR control directly
to heterogeneous grids remains open.

\paragraph{bFDR control under dependence.}
The finite-sample bFDR results in this paper rely on independence.
However, in applications like the gene expression analysis in Section \ref{subsec:gene-expression}, the test statistics are often correlated.
The BH procedure on p-values is known to control FDR under a type of positive dependence, namely PRDS \citep{10.1214/aos/1013699998}.
\cite{10.1093/jrsssb/qkag116} show that BH on compound p-values under PRDS could have an $O(\log m)$ inflation in the finite-sample FDR.
In Section~\ref{sec:freq-lfdr}, we showed that under a weak dependence condition expressed through uniform convergence of the empirical CDF, the SL rejection threshold retains its intended frequentist LFDR interpretation asymptotically, which suggests that some robustness to dependence is possible.
Identifying the right assumptions for turning this frequentist local FDR result into bFDR control under dependence is left for future work.

\section*{Reproducibility}
Numerical results in this paper are fully reproducible. 
The first author acknowledges the use of GPT-5.6 Sol in the development of the code,
provided on GitHub under the following link: \url{https://github.com/jerryzhu509/cp-lfdr-paper}.

\section*{Acknowledgements}
We thank Daniel Xiang for helpful discussions.
R.F.B.\ was supported by the Office of Naval Research via grant N00014-24-1-2544. N.I.\ gratefully acknowledges support from NSF via grant DMS 2443410.

\bibliography{cp.bib}

@article{armstrong2025falsediscoveryrateadjustments,
      title={False Discovery Rate Adjustments for Average Significance Level Controlling Tests}, 
      author={Timothy B. Armstrong},
      year={2022},
      journal={arXiv preprint},
      volume={arXiv:2209.13686}
}

@article{10.1093/jrsssb/qkag116,
    author = {Barber, Rina Foygel and Samworth, Richard J},
    title = {False discovery rate control with compound p-values},
    journal = {Journal of the Royal Statistical Society Series B: Statistical Methodology},
    pages = {qkag116},
    year = {2026},
    month = {07},
    issn = {1369-7412},
    doi = {10.1093/jrsssb/qkag116},
    url = {https://doi.org/10.1093/jrsssb/qkag116},
    eprint = {https://academic.oup.com/jrsssb/advance-article-pdf/doi/10.1093/jrsssb/qkag116/70454239/qkag116.pdf},
}

@article{10.1093/biomet/asaf083,
    author = {Xiang, Daniel and Soloff, Jake A and Fithian, William},
    title = {A frequentist local false discovery rate},
    journal = {Biometrika},
    volume = {113},
    number = {1},
    pages = {asaf083},
    year = {2026},
    month = {02},
    issn = {1464-3510},
    doi = {10.1093/biomet/asaf083},
    url = {https://doi.org/10.1093/biomet/asaf083},
    eprint = {https://academic.oup.com/biomet/article-pdf/113/1/asaf083/65551958/asaf083.pdf},
}

@article{gao2026conformalnoveltydetectionfalse,
      title={Reliable conformal novelty detection at the decision boundary}, 
      author={Zijun Gao and Etienne Roquain and Daniel Xiang},
      year={2026},
      journal={arXiv preprint},
      volume={arXiv:2601.02610}
}

@article{10.1214/24-AOS2359,
author = {Jake A. Soloff and Daniel Xiang and William Fithian},
title = {{The edge of discovery: Controlling the local false discovery rate at the margin}},
volume = {52},
journal = {The Annals of Statistics},
number = {2},
publisher = {Institute of Mathematical Statistics},
pages = {580 -- 601},
year = {2024},
doi = {10.1214/24-AOS2359},
}

@article{10.1093/biomet/asad057,
    author = {Ignatiadis, Nikolaos and Wang, Ruodu and Ramdas, Aaditya},
    title = {E-values as unnormalized weights in multiple testing},
    journal = {Biometrika},
    volume = {111},
    number = {2},
    pages = {417-439},
    year = {2024},
    month = {09},
    issn = {1464-3510},
    doi = {10.1093/biomet/asad057},
    eprint = {https://academic.oup.com/biomet/article-pdf/111/2/417/57467467/asad057.pdf},
}

@article{finner2009false,
  title = {On the False Discovery Rate and an Asymptotically Optimal Rejection Curve},
  author = {Finner, Helmut and Dickhaus, Thorsten and Roters, Markus},
  year = {2009},
  journal = {The Annals of Statistics},
  volume = {37},
  number = {2},
  pages = {596--618},
}

@article{finner2001false,
author = {Finner, Helmut and Roters, M.},
title = {On the False Discovery Rate and Expected Type {{I}} Errors},
journal = {Biometrical Journal},
volume = {43},
number = {8},
pages = {985-1005},
doi = {https://doi.org/10.1002/1521-4036(200112)43:8<985::AID-BIMJ985>3.0.CO;2-4},
url = {https://onlinelibrary.wiley.com/doi/abs/10.1002/1521-4036%28200112%2943%3A8%3C985%3A%3AAID-BIMJ985%3E3.0.CO%3B2-4},
eprint = {https://onlinelibrary.wiley.com/doi/pdf/10.1002/1521-4036%28200112%2943%3A8%3C985%3A%3AAID-BIMJ985%3E3.0.CO%3B2-4},
year = {2001}
}

@article{10.1214/18-AOS1765,
author = {Aaditya K. Ramdas and Rina F. Barber and Martin J. Wainwright and Michael I. Jordan},
title = {{A unified treatment of multiple testing with prior knowledge using the p-filter}},
volume = {47},
journal = {The Annals of Statistics},
number = {5},
publisher = {Institute of Mathematical Statistics},
pages = {2790 -- 2821},
year = {2019},
doi = {10.1214/18-AOS1765},
}

@article{ignatiadis2025asymptoticcompoundevaluesmultiple,
      title={Asymptotic and compound e-values: multiple testing and empirical {Bayes}}, 
      author={Nikolaos Ignatiadis and Ruodu Wang and Aaditya Ramdas},
      year={2025},
      journal={arXiv preprint},
      volume={arXiv:2409.19812}
}

@Article{Rangel-Salazar2011,
author={Rangel-Salazar, Rub{\'e}n
and Wickstr{\"o}m-Lindholm, Marie
and Aguilar-Salinas, Carlos A.
and Alvarado-Caudillo, Yolanda
and D{\o}ssing, Kristina BV
and Esteller, Manel
and Labourier, Emmanuel
and Lund, Gertrud
and Nielsen, Finn C.
and Rodr{\'i}guez-R{\'i}os, Dalia
and Sol{\'i}s-Mart{\'i}nez, Martha O.
and Wrobel, Katarzyna
and Wrobel, Kazimierz
and Zaina, Silvio},
title={Human native lipoprotein-induced de novo {DNA} methylation is associated with repression of inflammatory genes in {THP-1} macrophages},
journal={BMC Genomics},
year={2011},
month={Nov},
day={25},
volume={12},
number={1},
pages={582},
issn={1471-2164},
doi={10.1186/1471-2164-12-582},
}

@article{ling2025empiricalpartiallybayessample,
      title={Empirical partially {Bayes} two sample testing}, 
      author={Wanyi Ling and Wufang Hong and Nikolaos Ignatiadis},
      year={2025},
      journal={arXiv preprint},
      volume={arXiv:2510.00432}
}

@article{10.1214/24-AOS2431,
author = {Nikolaos Ignatiadis and Bodhisattva Sen},
title = {Empirical partially {Bayes} multiple testing and compound $\chi^2$ decisions},
volume = {53},
journal = {The Annals of Statistics},
number = {1},
publisher = {Institute of Mathematical Statistics},
pages = {1 -- 36},
year = {2025},
doi = {10.1214/24-AOS2431},
}

@article{888cd474-50a6-33fd-a789-415b80e67e78,
 ISSN = {00359246},
 author = {Yoav Benjamini and Yosef Hochberg},
 journal = {Journal of the Royal Statistical Society Series B: Statistical Methodology},
 number = {1},
 pages = {289--300},
 publisher = {[Royal Statistical Society, Oxford University Press]},
 title = {Controlling the False Discovery Rate: A Practical and Powerful Approach to Multiple Testing},
 volume = {57},
 year = {1995}
}

@article{Efron01122001,
author = {Bradley Efron and Robert Tibshirani and John D Storey and Virginia Tusher},
title = {Empirical {Bayes} Analysis of a Microarray Experiment},
journal = {Journal of the American Statistical Association},
volume = {96},
number = {456},
pages = {1151--1160},
year = {2001},
publisher = {Taylor \& Francis},
doi = {10.1198/016214501753382129},
eprint = { 
    
        https://doi.org/10.1198/016214501753382129
    
    

}

}

@article{10.1214/009053606000001460,
author = {Bradley Efron},
title = {{Size, power and false discovery rates}},
volume = {35},
journal = {The Annals of Statistics},
number = {4},
publisher = {Institute of Mathematical Statistics},
pages = {1351 -- 1377},
year = {2007},
doi = {10.1214/009053606000001460},
}

@article{10.1093/biomet/93.3.509,
    author = {Genovese, Christopher R. and Roeder, Kathryn and Wasserman, Larry},
    title = {False discovery control with p-value weighting},
    journal = {Biometrika},
    volume = {93},
    number = {3},
    pages = {509-524},
    year = {2006},
    month = {09},
    issn = {0006-3444},
    doi = {10.1093/biomet/93.3.509},
    eprint = {https://academic.oup.com/biomet/article-pdf/93/3/509/1080969/933509.pdf},
}

@Inbook{Storey2003,
author={Storey, John D.
and Tibshirani, Robert},
title={SAM Thresholding and False Discovery Rates for Detecting Differential Gene Expression in DNA Microarrays},
bookTitle={The Analysis of Gene Expression Data: Methods and Software},
year={2003},
publisher={Springer New York},
address={New York, NY},
pages={272--290},
isbn={978-0-387-21679-9},
doi={10.1007/0-387-21679-0_12},
}

@book{hastie2009elements,
  title={The Elements of Statistical Learning: Data Mining, Inference, and Prediction},
  author={Hastie, T. and Tibshirani, R. and Friedman, J.H.},
  isbn={9780387848846},
  edition = {2nd},
  lccn={2008941148},
  series={Springer series in statistics},
  year={2009},
  publisher={Springer}
}

@article{10.1257/aer.103.4.1289,
Author = {Pop-Eleches, Cristian and Urquiola, Miguel},
Title = {Going to a Better School: Effects and Behavioral Responses},
Journal = {American Economic Review},
Volume = {103},
Number = {4},
Year = {2013},
Month = {June},
Pages = {1289-1324},
DOI = {10.1257/aer.103.4.1289},}

@article{10.1111/ectj.12092,
    author = {Gu, Jiaying and Shen, Shu},
    title = {Oracle and adaptive false discovery rate controlling methods for one‐sided testing: theory and application in treatment effect evaluation},
    journal = {The Econometrics Journal},
    volume = {21},
    number = {1},
    pages = {11-35},
    year = {2018},
    month = {02},
    issn = {1368-4221},
    doi = {10.1111/ectj.12092},
    eprint = {https://academic.oup.com/ectj/article-pdf/21/1/11/27684863/ectj0011.pdf},
}

@article{CattaneoFrandsenTitiunik+2015+1+24,
title = {Randomization Inference in the Regression Discontinuity Design: An Application to Party Advantages in the {U.S. Senate}},
author = {Matias D. Cattaneo and Brigham R. Frandsen and Rocío Titiunik},
pages = {1--24},
volume = {3},
number = {1},
journal = {Journal of Causal Inference},
doi = {doi:10.1515/jci-2013-0010},
year = {2015},
lastchecked = {2026-06-07}
}

@article{mattei2017regression,
  title={Regression discontinuity designs as local randomized experiments},
  author={Mattei, Alessandra and Mealli, Fabrizia},
  journal={Observational Studies},
  volume={3},
  number={2},
  pages={156--173},
  year={2017},
  publisher={University of Pennsylvania Press}
}

@article{RJ-2015-004,
  author = {Calonico, Sebastian and Cattaneo, Matias D. and Titiunik, Rocío},
  title = {rdrobust: An {R} Package for Robust Nonparametric Inference in Regression-Discontinuity Designs},
  journal = {The R Journal},
  year = {2015},
  doi = {10.32614/RJ-2015-004},
  volume = {7},
  issue = {1},
  issn = {2073-4859},
  pages = {38-51}
}

@article{10.1561/3600000002,
    author = {Ramdas, Aaditya and Wang, Ruodu},
    title = {Hypothesis Testing with E-values},
    journal = {Foundations and Trends in Statistics},
    volume = {1},
    number = {1-2},
    pages = {1-390},
    year = {2025},
    month = {07},
    issn = {2978-4212},
    doi = {10.1561/3600000002},
    eprint = {https://www.emerald.com/ftstat/article-pdf/1/1-2/1/11146063/3600000002en.pdf},
}

@article{IMBENS2008615,
title = {Regression discontinuity designs: A guide to practice},
journal = {Journal of Econometrics},
volume = {142},
number = {2},
pages = {615-635},
year = {2008},
issn = {0304-4076},
doi = {https://doi.org/10.1016/j.jeconom.2007.05.001},
author = {Guido W. Imbens and Thomas Lemieux}
}

@article{westfall2004weighted,
  title={Weighted {FWE}-controlling methods in high-dimensional situations},
  author={Westfall, Peter H and Kropf, Siegfried and Finos, Livio},
  journal={Lecture Notes-Monograph Series},
  pages={143--154},
  year={2004},
  publisher={JSTOR}
}

@article{10.1093/biomet/29.3-4.350,
    author = {Welch, B. L.},
    title = {The Significance of the Difference Between Two Means When the Population Variances Are Unequal},
    journal = {Biometrika},
    volume = {29},
    number = {3-4},
    pages = {350-362},
    year = {1938},
    month = {02},
    issn = {0006-3444},
    doi = {10.1093/biomet/29.3-4.350},
    eprint = {https://academic.oup.com/biomet/article-pdf/29/3-4/350/712866/29-3-4-350.pdf},
}

@article{10.1093/restud/rdx062,
    author = {Canay, Ivan A and Kamat, Vishal},
    title = {Approximate Permutation Tests and Induced Order Statistics in the Regression Discontinuity Design},
    journal = {The Review of Economic Studies},
    volume = {85},
    number = {3},
    pages = {1577-1608},
    year = {2018},
    month = {07},
    issn = {0034-6527},
    doi = {10.1093/restud/rdx062},
    eprint = {https://academic.oup.com/restud/article-pdf/85/3/1577/25077990/rdx062.pdf},
}

@article{a5f3b3b3-ba89-31a3-b019-58d85d361941,
 ISSN = {03036898, 14679469},
 author = {Yoav Benjamini and Yosef Hochberg},
 journal = {Scandinavian Journal of Statistics},
 number = {3},
 pages = {407--418},
 publisher = {[Board of the Foundation of the Scandinavian Journal of Statistics, Wiley]},
 title = {Multiple Hypotheses Testing with Weights},
 urldate = {2026-06-26},
 volume = {24},
 year = {1997}
}

@article{10.1214/aos/1013699998,
author = {Yoav Benjamini and Daniel Yekutieli},
title = {{The control of the false discovery rate in multiple testing under dependency}},
volume = {29},
journal = {The Annals of Statistics},
number = {4},
publisher = {Institute of Mathematical Statistics},
pages = {1165 -- 1188},
year = {2001},
doi = {10.1214/aos/1013699998},
}

@article{10.1214/18-EJS1441,
author = {Sebastian D{\"o}hler and Guillermo Durand and Etienne Roquain},
title = {{New FDR bounds for discrete and heterogeneous tests}},
volume = {12},
journal = {Electronic Journal of Statistics},
number = {1},
publisher = {Institute of Mathematical Statistics and Bernoulli Society},
pages = {1867 -- 1900},
year = {2018},
doi = {10.1214/18-EJS1441},
URL = {https://doi.org/10.1214/18-EJS1441}
}

@article{doi:10.1073/pnas.2302098121,
author = {Peter D. Grünwald},
title = {Beyond {N}eyman–{P}earson: E-values enable hypothesis testing with a data-driven alpha},
journal = {Proceedings of the National Academy of Sciences},
volume = {121},
number = {39},
pages = {e2302098121},
year = {2024},
doi = {10.1073/pnas.2302098121},
URL = {https://www.pnas.org/doi/abs/10.1073/pnas.2302098121},
eprint = {https://www.pnas.org/doi/pdf/10.1073/pnas.2302098121},}

@article{koning2025posthocalphahypothesistesting,
      title={Post-hoc $\alpha$ Hypothesis Testing and the Post-hoc $p$-value}, 
      author={Nick W. Koning},
      year={2025},
      journal={arXiv preprint},
      volume={arXiv:2312.08040},
}

@article{10.1111/j.1467-9868.2004.00439.x,
    author = {Storey, John D. and Taylor, Jonathan E. and Siegmund, David},
    title = {Strong Control, Conservative Point Estimation and Simultaneous Conservative Consistency of False Discovery Rates: A Unified Approach},
    journal = {Journal of the Royal Statistical Society Series B: Statistical Methodology},
    volume = {66},
    number = {1},
    pages = {187-205},
    year = {2004},
    month = {02},
    issn = {1369-7412},
    doi = {10.1111/j.1467-9868.2004.00439.x},
    url = {https://doi.org/10.1111/j.1467-9868.2004.00439.x},
    eprint = {https://academic.oup.com/jrsssb/article-pdf/66/1/187/49727331/jrsssb_66_1_187.pdf},
}

@article{hemerik2018exact,
    author = {Hemerik, Jesse and Goeman, Jelle},
    title = {Exact Testing with Random Permutations},
    journal = {TEST},
    volume = {27},
    number = {4},
    pages = {811--825},
    year = {2018},
    doi = {10.1007/s11749-017-0571-1},
    url = {https://doi.org/10.1007/s11749-017-0571-1}
}

@article{ignatiadis2026tiny,
  title = {Tiny but Uniform Improvements of Adaptive {{BH}} Procedures via Compound E-Values},
  author = {Ignatiadis, Nikolaos and Wang, Ruodu and Ramdas, Aaditya},
  year = {2026},
  journal = {arXiv preprint},
  volume = {arXiv:2603.21424}
}

@article{ignatiadis2026empiricalbayesprepivotinggroup,
      title={Empirical {Bayes} prepivoting under group invariance: false discovery rate control and moderated t-tests}, 
      author={Nikolaos Ignatiadis and Etienne Roquain},
      year={2026},
      journal = {arXiv preprint},
      volume={arXiv:2609.26764}, 
}

@article{doi:10.1073/pnas.091062498,
author = {Virginia Goss Tusher  and Robert Tibshirani  and Gilbert Chu },
title = {Significance analysis of microarrays applied to the ionizing
 radiation response},
journal = {Proceedings of the National Academy of Sciences},
volume = {98},
number = {9},
pages = {5116-5121},
year = {2001},
doi = {10.1073/pnas.091062498},
URL = {https://www.pnas.org/doi/abs/10.1073/pnas.091062498},
eprint = {https://www.pnas.org/doi/pdf/10.1073/pnas.091062498}}

@article{eckles2025noiseinduced,
  title = {Noise-Induced Randomization in Regression Discontinuity Designs},
  author = {Eckles, Dean and Ignatiadis, Nikolaos and Wager, Stefan and Wu, Han},
  year = {2025},
  journal = {Biometrika},
  volume = {112},
  number = {2},
  pages = {asaf003},
}

@article{hahn2001identification,
author = {Hahn, Jinyong and Todd, Petra and Van der Klaauw, Wilbert},
title = {Identification and Estimation of Treatment Effects with a Regression-Discontinuity Design},
journal = {Econometrica},
volume = {69},
number = {1},
pages = {201-209},
doi = {https://doi.org/10.1111/1468-0262.00183},
url = {https://onlinelibrary.wiley.com/doi/abs/10.1111/1468-0262.00183},
eprint = {https://onlinelibrary.wiley.com/doi/pdf/10.1111/1468-0262.00183},
year = {2001}
}

@article{cattaneo2022regression,
  title = {Regression Discontinuity Designs},
  author = {Cattaneo, Matias D. and Titiunik, Roc{\'i}o},
  year = {2022},
  journal = {Annual Review of Economics},
  volume = {14},
  number = {1},
  pages = {821--851},
}

@article{10.1111/1467-9868.00346,
    author = {Storey, John D.},
    title = {A Direct Approach to False Discovery Rates},
    journal = {Journal of the Royal Statistical Society Series B: Statistical Methodology},
    volume = {64},
    number = {3},
    pages = {479-498},
    year = {2002},
    month = {08},
    issn = {1369-7412},
    doi = {10.1111/1467-9868.00346},
    url = {https://doi.org/10.1111/1467-9868.00346},
    eprint = {https://academic.oup.com/jrsssb/article-pdf/64/3/479/49722049/jrsssb_64_3_479.pdf},
}

\appendix

\section{Auxiliary Results}\label{sec:auxiliary}

\subsection{Additional Density-Compound p-value Constructions} \label{subsec:additional-dcp}
We verify that two constructions from the compound p-value literature
also satisfy the stronger density-compound condition.

\paragraph{Orthogonal rotation compound p-values.}
Let $Z_1,\ldots,Z_m\in\mathbb R^n$ be independent
random vectors. Assume that, for each $i\in\mathcal N_0$,
$$
QZ_i\overset d=Z_i
\qquad\text{for every }Q\in O(n),
$$
where $O(n)=\{Q\in\mathbb R^{n\times n}:Q^\top Q=I_n\}$
is the orthogonal group. Following
\citet{ignatiadis2026empiricalbayesprepivotinggroup}, let
$\widehat S:\RR^n\to\RR$ be a score function learned using only the norms
$(\|Z_1\|_2,\ldots,\|Z_m\|_2)$. Let $H$ be uniformly distributed on $O(n)$, 
independently of the data, and define
\begin{equation}\label{eq:rotation-dcp}
    \xi_j(t):=\PP_H\big(\widehat S(HZ_j)\geq t\big),
    \qquad
    P_i^{\mathrm{rot}}:=\frac1m\sum_{j=1}^m\xi_j(\widehat S(Z_i)),
\end{equation}
where $\PP_H$ denotes probability over $H$ alone.
Assume the survival functions $\xi_j$ are almost surely absolutely continuous.
Conditional on $(\|Z_1\|_2,\ldots,\|Z_m\|_2)$, 
the score function $\widehat S$ and the survival functions $\xi_j$ are deterministic. 
For each $i \in \Nn_0$, rotational invariance gives
$$\xi_i(t)=\mathbb P\big(\widehat S(Z_i)\ge t\mid \Vert Z_1\Vert_2,\ldots,\Vert Z_m\Vert_2\big).$$
Proposition~\ref{prop:cp-ave-cdf-construction}, applied conditionally with
$X_i=-\widehat S(Z_i)$, therefore shows that
$P_1^{\mathrm{rot}},\ldots,P_m^{\mathrm{rot}}$ are density-compound p-values.

We present a particular instance of $P_i^{\mathrm{rot}}$ in the Gaussian setting of Example~\ref{ex:cp-eb-normal-known-var}.
Take
$Z_i=(Y_{i1},\ldots,Y_{in})$, $n\geq2$, and
$\widehat S(z)=(\sum_{k=1}^n z_k/n)^2$.
Writing $U_j^2=\|Z_j\|_2^2/n$, the identity
$\widehat S(HZ_j)/U_j^2 \mid Z_j \sim\mathrm{Beta}(1/2,(n-1)/2)$ gives the
data-driven construction
\begin{equation}\label{eq:data-driven-gaussian-dcp}
    P_i^{\mathrm{rot}}
    =\frac1m\sum_{j=1}^m
    \bar F_{\mathrm{Beta}(1/2,(n-1)/2)}
    \!\left(\bar Y_i^2/U_j^2\right),
\end{equation}
where $\bar F$ denotes the survival function.

\paragraph{Empirically partially Bayes p-values.}
In the Gaussian setting of Example~\ref{ex:cp-eb-normal-known-var}, let
$S_i^2$ be the sample variance and suppose the empirical variance
distribution $G=m^{-1}\sum_{j=1}^m\delta_{\sigma_j^2}$ is known. Define
\begin{equation}\label{eq:epb-dcp}
    P_i^{\mathrm{PB}}
    =
    \frac{
        \sum_{j=1}^m
        2\bar\Phi\!\left(|\bar Y_i|/(\sigma_j/\sqrt n)\right)
        f_{S^2\mid\sigma^2}(S_i^2\mid\sigma_j^2)
    }{
        \sum_{j=1}^m f_{S^2\mid\sigma^2}(S_i^2\mid\sigma_j^2)
    },
\end{equation}
where $f_{S^2\mid\sigma^2}(\cdot\mid\sigma^2)$ is the density of the
sample variance conditional on $\sigma^2$. 
Let $F_{0,\mathrm{PB}}^{(i)}$ denote the CDF of $P_i^{\mathrm{PB}}$ under $H_{0,i}:\mu_i=0$.
Proposition 4 of
\citet{10.1214/24-AOS2431} shows that the corresponding nominal null CDFs
satisfy
$
    \sum_{i=1}^m F_{0,\mathrm{PB}}^{(i)}(t)=mt
$
for every $t\in[0,1]$. Since these CDFs are absolutely continuous,
Proposition~\ref{prop:density-condition-equivalent} implies that
$P_1^{\mathrm{PB}},\ldots,P_m^{\mathrm{PB}}$ are density-compound p-values.

\subsection{Compound Controls and bFDR}
\label{subsec:block-rank-verification}

We give a common representation of pooled Monte Carlo and 
permutation p-values and identify conditions for density-compound validity and bFDR bounds.
For each $i\in[m]$, let $Y_i=(X_i^{(0)},\ldots,X_i^{(K)})$, where
$X_i^{(0)}$ is observed and the remaining $K$ statistics serve as controls.
Write $Y_{1:m}:=(Y_1,\ldots,Y_m)$ and $L:=m(K+1)$. 
Define the compound control p-value for unit $i$ as
\begin{equation}\label{eq:pooled-block-rank-pval}
P_i^{\mathrm{CC}}
=\frac{1}{m(K+1)}\sum_{j=1}^m\sum_{k=0}^K
\ind\{X_j^{(k)}\geq X_i^{(0)}\}.
\end{equation}
Write $\widehat{\mathcal P}_i:=(K+1)^{-1}
\sum_{k=0}^K\delta_{X_i^{(k)}}$ for the empirical distribution of unit
$i$'s statistics, and
$\widehat{\mathcal P}_{1:m}:=(\widehat{\mathcal P}_1,\ldots,
\widehat{\mathcal P}_m)$.
Define the normalized upper rank
$\rho(x):=L^{-1}\sum_{j=1}^m\sum_{k=0}^K
\ind\{X_j^{(k)}\geq x\}$ and the unit-specific grid
$\mathcal G_{i,K+1}:=\{\rho(X_i^{(k)}):0\leq k\leq K\}$.
Note $P_i^{\mathrm{CC}}=\rho(X_i^{(0)})$
and that each $\mathcal G_{i,K+1}$ is a subset of the common grid $\Gg_L = \left\{k/L: k \in [L]\right\}$ defined
in Section~\ref{subsec:discrete-dcp}.

We assume that almost surely
\begin{equation}\label{eq:grid-partition-condition}
|\mathcal G_{i,K+1}|=K+1\quad \mathrm{for\ all\ }i\in[m],
\qquad
\bigsqcup_{i=1}^m\mathcal G_{i,K+1}=\mathcal G_L,
\end{equation}
and, writing $X_{-i}^{(0)}=(X_j^{(0)})_{j\neq i}$, that
\begin{equation}\label{eq:block-null-uniformity}
P_i^{\mathrm{CC}}\mid
\widehat{\mathcal P}_{1:m},X_{-i}^{(0)}
\sim\mathrm{Unif}(\mathcal G_{i,K+1}),
\qquad \mathrm{for\ all\ }i\in\Nn_0.
\end{equation}
Condition \eqref{eq:grid-partition-condition} rules out ties among the pooled statistics.
Condition \eqref{eq:block-null-uniformity} requires
that, given the unordered statistics from every unit and the observed statistics from the other units, 
a null unit's observed rank is uniform over its $K+1$ possible ranks.

These conditions imply discrete density-compound validity and the following conditional bFDR bound.

\begin{lemma}
\label{lem:block-rank-implies-discrete-dcp}
If
\eqref{eq:grid-partition-condition}--\eqref{eq:block-null-uniformity} hold, then
$P_1^{\mathrm{CC}},\ldots,P_m^{\mathrm{CC}}$ are discrete
density-compound p-values on $\mathcal G_L$.
\end{lemma}

For the bFDR bound, define
\begin{equation} \label{eq:gamma-alpha-block}
\gamma_S^{\mathrm{CC}}(t):=\max_{i\in S}
\PP_{H_{0,i}}(P_i^{\mathrm{CC}}\leq t\mid
\widehat{\mathcal P}_{1:m})
\quad \mathrm{ for }\ t \in [0,1] \ \mathrm{ and }\ S \subseteq [m].
\end{equation}
Define $\hat{\alpha}_{-}$ by \eqref{eq:hat-alpha-minus}
using $P_i=P_i^{\mathrm{CC}}$.

\begin{theorem}[bFDR for compound controls]
\label{thm:bfdr-cp-controls}
Fix $\alpha\in(0,1)$ and suppose
\eqref{eq:grid-partition-condition}--\eqref{eq:block-null-uniformity} hold.
Then the SL procedure applied to the p-values in
\eqref{eq:pooled-block-rank-pval} satisfies
\begin{equation}\label{eq:block-rank-bfdr-bound}
\EE\left[
\left\{1-\gamma_{\Nn_0}^{\mathrm{CC}}(\hat{\alpha}_{-})\right\}
\ind\{H_{0,\sigma(R_\alpha)}\text{ is true}\}
\,\middle|\,\widehat{\mathcal P}_{1:m}
\right]\leq\alpha.
\end{equation}
\end{theorem}

For $i\in\Nn_0$, \eqref{eq:block-null-uniformity} implies
$$
\PP(P_i^{\mathrm{CC}}\leq t\mid\widehat{\mathcal P}_{1:m})
=\frac{|\mathcal G_{i,K+1}\cap[0,t]|}{K+1}.
$$
Then, a computable upper bound on $\gamma_{\Nn_0}^{\mathrm{CC}}(t)$ is
$\gamma_{[m]}^{\mathrm{CC}}(t)$.

\paragraph{Monte Carlo controls.}
The following lemma covers pooled Monte Carlo p-values \eqref{eq:compound-mc-pval} and will also be used
for randomized permutations in Appendix~\ref{subsec:sampled-permutation}.

\begin{lemma}
\label{lem:exchangeable-block-verification}
Suppose the vectors
$Y_i=(X_i^{(0)},\ldots,X_i^{(K)})$, $i\in[m]$, are independent and that
$Y_i$ is exchangeable for every $i\in\Nn_0$. If the $m(K+1)$ pooled
statistics are distinct almost surely, then
\eqref{eq:grid-partition-condition} and
\eqref{eq:block-null-uniformity} hold.
\end{lemma}

For the Monte Carlo construction, take $X_i^{(0)}=X_i$ and
$X_i^{(k)}=\widetilde X_i^{(k)}$ for $k\in[K]$, so that
$P_i^{\mathrm{CC}}=P_i^{\mathrm{MC}}$. Under the assumptions of
Proposition~\ref{prop:compound-pvals-controls},
Lemmas~\ref{lem:exchangeable-block-verification} and
\ref{lem:block-rank-implies-discrete-dcp} give density-compound validity,
and Theorem~\ref{thm:bfdr-cp-controls} gives the corresponding bFDR bound.

\paragraph{Full permutation enumeration.}
For each $i$, let $\mathcal V_i$ be the set of distinct values in
$\{X_i^\pi:\pi\in\mathfrak S(n)\}$. Under the assumptions of
Proposition~\ref{prop:dcperm-validity}, $|\mathcal V_i|=B$. Take $K=B-1$
and let $Y_i$ list each value in $\mathcal V_i$ once, with $X_i^{(0)}=X_i$.

\begin{lemma}
\label{lem:permutation-block-verification}
Under the hypotheses of Proposition~\ref{prop:dcperm-validity}, these
vectors satisfy \eqref{eq:grid-partition-condition} and
\eqref{eq:block-null-uniformity}, and
$P_i^{\mathrm{CC}}=P_i^{\mathrm{dcperm}}$.
\end{lemma}

Theorem~\ref{thm:bfdr-cp-controls} 
underlies the bFDR guarantee for $P_i^{\mathrm{dcperm}}$ in Theorem~\ref{thm:bfdr-dcperm}.

\subsection{Randomized Permutation p-values for Heterogeneous Designs}
\label{subsec:sampled-permutation}

We now allow the sample size and the number of distinct permutation
statistics to vary across units. Sampling the same number of statistics
from each unit gives a common pooled grid, allowing us to apply the
compound control result of Appendix~\ref{subsec:block-rank-verification}.

For unit $i$, let $Z_i\in\mathbb R^{n_i}$ be the outcome vector and let $T_i$ be its test statistic.
Write $X_i^\pi=T_i(Z_i^\pi)$ for $\pi\in\mathfrak S(n_i)$, with
$(z^\pi)_\ell=z_{\pi^{-1}(\ell)}$, and set $X_i=X_i^{\mathrm{id}}$.
To avoid sampling permutations that always give the same statistic, 
define the equivalence relation
$$
\pi\sim_i\pi'
\quad\Longleftrightarrow\quad
T_i(z^\pi)=T_i(z^{\pi'})
\quad\text{for every }z\in\mathbb R^{n_i},
\qquad
\Ss_i:=\{[\pi]_i:\pi\in\mathfrak S(n_i)\},
$$
where $[\pi]_i$ is the equivalence class containing $\pi$, and $B_i:=|\Ss_i|$.

Fix one representative from each class, taking $\mathrm{id}$ for
$[\mathrm{id}]_i$. Choose $K+1\leq\min_i B_i$, set
$\pi_{i0}=\mathrm{id}$, and draw $\pi_{i1},\ldots,\pi_{iK}$ uniformly
without replacement from the remaining $B_i-1$ representatives,
independently of the data and across units. Set
$X_i^{(k)}=T_i(Z_i^{\pi_{ik}})$; its value does not depend on which
representative was chosen within the class.
Define
\begin{equation}\label{eq:sampled-compound-perm-pval}
P_i^{\mathrm{dcperm},K}
=\frac{1}{m(K+1)}\sum_{j=1}^m\sum_{k=0}^K
\ind\{X_j^{(k)}\geq X_i^{(0)}\}.
\end{equation}
With $Y_i=(X_i^{(0)},\ldots,X_i^{(K)})$, these are the compound-control
p-values $P_i^{\mathrm{CC}}$ in \eqref{eq:pooled-block-rank-pval}.

\begin{proposition}[Randomized permutation p-values]
\label{prop:sampled-dcperm-validity}
Suppose $Z_1,\ldots,Z_m$ are independent and the components of $Z_i$
are exchangeable for $i\in\Nn_0$.
If the $m(K+1)$ pooled statistics are distinct almost surely,
then \eqref{eq:grid-partition-condition} and
\eqref{eq:block-null-uniformity} hold, and the p-values in
\eqref{eq:sampled-compound-perm-pval} are discrete density-compound p-values
on
$\mathcal G_{m(K+1)}$.
\end{proposition}

Set $\gamma_S^{\mathrm{dcperm},K}:=\gamma_S^{\mathrm{CC}}$ as defined in \eqref{eq:gamma-alpha-block}.
Theorem~\ref{thm:bfdr-cp-controls} then gives the following bFDR
bound.

\begin{corollary}[bFDR for randomized permutations]
\label{cor:bfdr-sampled-dcperm}
Under the assumptions of Proposition~\ref{prop:sampled-dcperm-validity},
for $\alpha\in(0,1)$ the SL procedure applied to
$P_i^{\mathrm{dcperm},K}$ satisfies
$$
\EE\left[
\left\{1-\gamma_{\Nn_0}^{\mathrm{dcperm},K}(\hat{\alpha}_{-})\right\}
\ind\{H_{0,\sigma(R_\alpha)}\mathrm{\ is\ true}\}
\,\middle|\,\widehat{\mathcal P}_{1:m}
\right]\leq\alpha,
$$
with $\hat{\alpha}_{-}$ defined by \eqref{eq:hat-alpha-minus}
using $P_i=P_i^{\mathrm{dcperm},K}$.
\end{corollary}

In the RDD application in Section~\ref{subsec:rd}, $T_i$ is the negative
one-sided Welch t-test p-value. Its classes correspond to the
$B_i=\binom{n_i}{n_{i1}}$ treatment assignments, where $n_{i1}$ is the
number of observations to the right of the cutoff. Drawing $K=69$
distinct assignments other than the observed assignment implements the
construction above, giving $70$ statistics per unit.

\subsection{Dirac-alternatives Configuration} \label{subsec:dirac-alternatives}
We first prove the bound in Remark~\ref{rmk:cheating-fdr}, then illustrate the contrast between BH and SL
for this configuration.

\begin{proof}[Proof for Remark \ref{rmk:cheating-fdr}]
Let $m_0=|\Nn_0|$ and $\bar\pi_0=m_0/m$. The claim is trivial if
$m_0=0$ or $\alpha=1$, so suppose $m_0>0$ and $\alpha<1$.
When $P_i=0$ for all $i\notin\Nn_0$, the minimum of $P_{\sigma(k)}-\alpha k/m$
over $0\leq k\leq m-m_0$ is $-\alpha(m-m_0)/m$. Thus
\begin{align*}
    \mathrm{bFDR}(\Rr_\alpha)
    &= \PP(\text{SL rejects any null}) \\
    &= \PP\left(\exists k \geq m - m_0 + 1: P_{\sigma(k)} - \frac{\alpha k}{m} \leq \alpha(\bar{\pi}_0 - 1)\right) \\
    &= \PP\left(\exists k \geq m - m_0 + 1: P_{\sigma(k)} \leq \frac{\alpha (m_0 - m + k)}{m}\right) \\
    &= \PP\left(\exists k \geq 1: P_{(k)}^{(0)} \leq \frac{\alpha k}{m_0} \cdot \frac{m_0}{m}\right) \\
    &= \PP\left(\text{BH at level } \alpha \bar{\pi}_0 \text{ on nulls makes at least one discovery}\right),
\end{align*}
where $P_{(k)}^{(0)}$ denotes the $k$th smallest null p-value.
When $P_1,\ldots, P_m$ are independent and exactly uniform under the null, it is well known that the above probability equals $\bar{\pi}_0 \alpha$.

For the compound case, set $P_i^\star=(P_i/\bar\pi_0)\wedge 1$ for $i\in\Nn_0$.
For $t\in[0,1)$, \eqref{eq:cp} gives
$$
\frac{1}{m_0}\sum_{i\in\Nn_0}\PP(P_i^\star\leq t)
=\frac{1}{m_0}\sum_{i\in\Nn_0}\PP(P_i\leq\bar\pi_0 t)\leq t.
$$
The bound also holds at $t=1$, so the $P_i^\star$ are independent compound p-values.
BH at level $\alpha < 1$ on the $P_i^\star$ makes the same rejections as BH at level
$\alpha\bar\pi_0$ on the null $P_i$, since clipping $P_i/\bar\pi_0$ at $1$ leaves comparisons
to the rescaled cutoffs $\alpha k/m_0<1$ unchanged for all $k \in [m_0]$. The global null bound of
\cite{10.1093/jrsssb/qkag116} thus gives $\mathrm{bFDR}(\Rr_\alpha)\leq \alpha+2\alpha^2$.
\end{proof}

The following example shows that replacing uniform null p-values by a 
compound construction can increase BH's bFDR while preserving its FDR.

\begin{example}[BH and SL under the Dirac-alternatives configuration]
\label{ex:fdr-bfdr-compound}
Consider $m=100$ hypotheses, with $20$ non-null p-values equal to zero and
$m_0=80$ null p-values
$P_1^{(0)},\ldots,P_{80}^{(0)}\overset{\mathrm{iid}}{\sim}\mathrm{Unif}[0,1]$.
Fix $\alpha=0.05$ and write
$P=(P_1^{(0)},\ldots,P_{80}^{(0)},0,\ldots,0)$.
The standard BH identity gives
$\mathrm{FDR}(\Rr_\alpha^{\mathrm{BH}}(P))=\bar\pi_0\alpha=0.04$.
Writing $P_{(1)}^{(0)}\leq\cdots\leq P_{(80)}^{(0)}$ for the ordered
null p-values, we have
$$
\mathrm{bFDR}(\Rr_\alpha^{\mathrm{BH}}(P))
=
\PP\left(
\exists k\in[80]:
P_{(k)}^{(0)}\leq\frac{\alpha(20+k)}{100}
\right)
\approx 0.571,
$$
where the probability is evaluated numerically.

Now replace $P$ by
$\widetilde P=(\widetilde P_1^{(0)},\ldots,\widetilde P_{80}^{(0)},0,\ldots,0)$,
where
$$
\widetilde P_1^{(0)}=\frac{P_1^{(0)}}{80},
\qquad
\widetilde P_i^{(0)}=1,\quad i=2,\ldots,80
$$
so that $\widetilde P_1,\ldots,\widetilde P_{100}$ form compound p-values.
The only null that BH can reject is the first one, which is rejected
exactly when $P_1^{(0)}/80\leq21\alpha/100$. Thus,
$$
\mathrm{bFDR}(\Rr_\alpha^{\mathrm{BH}}(\widetilde P))
=
\PP\left(P_1^{(0)}\leq\frac{80\cdot21\alpha}{100}\right)
=0.84.
$$
Whenever this null is rejected, there are $21$ discoveries, exactly one
of which is false. Hence
$$
\mathrm{FDR}(\Rr_\alpha^{\mathrm{BH}}(\widetilde P))
=\frac{0.84}{21}=0.04.
$$
By the proof of Remark~\ref{rmk:cheating-fdr}, SL has bFDR $0.04$
for the uniform null p-values.
For $\widetilde P$, SL rejects the first null exactly when
$P_1^{(0)}/80\leq\alpha/100$, or equivalently $P_1^{(0)}\leq0.04$.
Its bFDR is therefore also $0.04$.
The results are summarized below:
$$
\begin{array}{c|cc}
& \text{Individually valid p-values $(P)$} & \text{Compound p-values ($\widetilde P$)} \\
\hline
\mathrm{BH\ FDR}  & 0.04 & 0.04 \\
\mathrm{BH\ bFDR} & \approx 0.571 & 0.84 \\
\mathrm{SL\ bFDR} & 0.04 & 0.04
\end{array}
$$
\end{example}
The contrast shows that compound validity \eqref{eq:cp} can raise BH's bFDR while
leaving SL's unchanged.

\subsection{bFDR Control under a Constant \texorpdfstring{$\gamma$}{gamma}}
The following extension of Theorem~\ref{thm:gamma-alpha-bfdr-control}
allows some null CDFs at $\alpha$ to exceed a fixed $\gamma\in(0,1)$,
at a cost depending on the number of such nulls.

\begin{theorem}[bFDR control under a constant $\gamma$] \label{thm:gamma-bfdr-control}
    Fix $\alpha, \gamma \in (0,1)$ and define $\Nn_\gamma := \{i \in \Nn_0: \PP(P_i \leq \alpha) > \gamma\}$ with $m_\gamma := |\Nn_\gamma|$.
    Suppose $P_1, \ldots, P_m$ are independent density-compound p-values.
    Then the SL procedure $\Rr_\alpha$ satisfies
    $$
    \mathrm{bFDR}(\Rr_\alpha) \leq m_\gamma \alpha + \frac{\alpha}{1-\gamma}.
    $$
\end{theorem}
Suitable choices of $\gamma$ recover
Theorem~\ref{thm:gamma-alpha-bfdr-control} and yield the term
$2(\alpha+\alpha^2m)$ in the bound \eqref{prop:cp-bfdr-control}.

\section{Proofs of Technical Results} \label{sec:proofs}

\subsection{Notation}
\label{subsec:proof-notation}
We collect notation used throughout the proofs. For $P=(P_1,\ldots,P_m)$,
recall that $P_{-i}=(P_j)_{j\neq i}$, and expressions such as
$(P_{-i},x,t)$ append the displayed coordinates.
We use the ordering $P_{\sigma(1)}\leq\cdots\leq P_{\sigma(m)}$ and
the convention $P_{\sigma(0)}=0$ from the main text.
We write $\hat\tau_\beta^\lambda(q)$ for the modified SL threshold
\eqref{eq:modified-sl} applied to a vector $q$ at level $\beta$, with candidate
thresholds at most $\lambda$. For $\lambda=1$, this gives the unrestricted
SL threshold $\hat\tau_\beta(q)=\hat\tau_\beta^1(q)$ in \eqref{eq:sl-procedure}.

For the auxiliary sets and thresholds below, fix $\alpha\in(0,1)$ and
$P\in[0,1]^m$. Let $\lambda \in [0,1]$. Write
$$
A_{-i}^\lambda
:=\{t\in[0,1]:\hat\tau_\alpha^\lambda(P_{-i},t)=t\},
\qquad A_{-i}:=A_{-i}^1.
$$
For the length-$(m+1)$ vector $(P,t)$, set $\tilde\alpha:=\alpha(m+1)/m$ and
$$
\tilde A_P^\lambda
:=\{t\in[0,1]:\hat\tau_{\tilde\alpha}^\lambda(P,t)=t\}.
$$
These sets contain the values of the inserted coordinate for which the
SL threshold equals that coordinate.
Recall $\hat\alpha_{-i}$ from \eqref{eq:hat-alpha-minus}:
$$
\hat{\alpha}_{-i}
:=
\sup_{x,t\in[0,1]}
\hat\tau_{\tilde\alpha}^{\alpha}(P_{-i},x,t)
$$
and also define
$\bar\alpha_{-i}$ and $\tilde\alpha_P$ as follows:
$$
\bar{\alpha}_{-i}
:=
\sup_{x\in[0,1]}
\hat\tau_\alpha(P_{-i},x),
\qquad
\tilde\alpha_P
:=
\sup_{t\in[0,1]}
\hat\tau_{\tilde\alpha}^{\alpha}(P,t).
$$
Unless another measure is specified, $\mu$ denotes Lebesgue measure on $[0,1]$.

\subsection{Technical Lemmas}\label{subsec:technical-lemmas}
We restate the following technical lemmas from \cite{10.1214/24-AOS2359} and \cite{10.1093/biomet/asaf083}.
\begin{lemma}[Lemma 2 of \cite{10.1214/24-AOS2359}] \label{lem:lem2}
    Let $\alpha > 0$.
    Fix any $P_{-i} \in [0, 1]^{m-1}$. If $P_i \mid P_{-i} \sim \mathrm{Unif}[0,1]$, then
    $$
    \PP\left(P_{\sigma(R^{\hat{\tau}_\alpha})} = P_i \mid P_{-i}\right) \leq \frac{\alpha}{m}
    $$
    with equality if $\alpha \leq 1$.
\end{lemma}

The proof of Lemma A.2 in \cite{10.1093/biomet/asaf083} also gives the following result.
\begin{lemma}[Lemma A.2 of \cite{10.1093/biomet/asaf083}] \label{lem:lem2-discrete}
    Fix any $P_{-i} \in [0, 1]^{m-1}$. 
    If $P_i \mid P_{-i} \sim \mathrm{Unif}(\Gg_L)$, then
    $$
    \PP\left(P_{\sigma(R^{\hat{\tau}_\alpha})} = P_i \mid P_{-i}\right) \leq \frac{\alpha}{m} + \frac{m}{L},
    $$
    for any $L \in \mathbb{N}$.
\end{lemma}

For the modified SL procedure \eqref{eq:modified-sl}, we prove a strengthened version
of Lemma~\ref{lem:lem2} in terms of weighted p-values.

\begin{lemma}[Modified Lemma 2, weighted] \label{lem:modified-lem2}
    Let $\alpha>0$. Fix $U_{-i}\in [0,1]^{m-1}$ and weights $w_1, \ldots, w_m \in[0,\infty]$.
    Let $\lambda \in (0,1)$.
    If $U_i \mid U_{-i} \sim \mathrm{Unif}[0,1]$, then
    the SL threshold $\hat{\tau}_{\alpha}^\lambda$ defined in \eqref{eq:modified-sl}, applied to $P_j = \left(U_j/w_j\right) \wedge 1$, satisfies
    $$
    \PP\left(P_{\sigma\big(R^{\hat{\tau}_{\alpha}^\lambda}\big)} = P_i \mid U_{-i}\right)
    \leq \frac{\alpha}{m} \cdot w_i.
    $$
    Here, we use the conventions $0/0:=0$, $a/0:=\infty$ for $a>0$, and
    $a/\infty:=0$ for $a\in[0,1]$. 
\end{lemma}

\begin{proof}[Proof of Lemma \ref{lem:modified-lem2}]
    If $w_i=0$, then $P_i=1$ almost surely, since
    $\PP(U_i=0\mid U_{-i})=0$. As $\lambda<1$, the event on the
    left has probability zero. If $w_i=\infty$, the bound is trivial. 
    Hence suppose $0<w_i<\infty$.
    Let $\lambda_i = 1/w_i \wedge \lambda$.
    Since $P_{\sigma(R^{\hat{\tau}_{\alpha}^\lambda})} \leq\lambda$ and $P_i\leq 1/w_i$,
    it holds that
    \begin{equation} \label{eq:imp-2}
    P_{\sigma\big(R^{\hat{\tau}_{\alpha}^\lambda}\big)} = P_i \implies P_i\leq\lambda_i.
    \end{equation}
    Define $m_i := \sum_{j=1}^m \ind\{P_j \leq \lambda_i\}$ and $P_{\sigma(0)} := 0 \leq P_{\sigma(1)} \leq \ldots \leq P_{\sigma(m_i)} \leq \lambda_i$.
    Conditional on $U_{-i}$ and $\{P_i\leq\lambda_i\}$, note that
    $m_i=1+\sum_{j\neq i}\ind\{P_j\leq\lambda_i\}$ is fixed, and it holds that
    \begin{equation} \label{eq:implication}
    P_{\sigma\big(R^{\hat{\tau}_{\alpha}^\lambda}\big)} = P_i
    \implies
    P_{\sigma\big(R^{\hat{\tau}_{\alpha}^{\lambda_i}}\big)} = P_i
    \end{equation}
    since, on the left event, the rejection threshold $P_i$ is unaffected by discarding candidates larger than $\lambda_i$.
    Moreover, we have
    $$
     R^{\hat{\tau}_\alpha^{\lambda_i}}
     = \max\left\{\argmin_{k=0,\ldots,m_i} \left(\frac{P_{\sigma(k)}}{\lambda_i} - \frac{\alpha k}{\lambda_i m_i}\cdot \frac{m_i}{m}\right)\right\}
     = \max\left\{\argmin_{k=0,\ldots,m_i} \left(\frac{P_{\sigma(k)}}{\lambda_i} - \frac{\alpha_i k}{m_i}\right)\right\},
    $$
     where $\alpha_i = {\alpha m_i}/(\lambda_i m)$.
     By the hypothesis that $U_i \mid U_{-i} \sim \mathrm{Unif}[0, 1]$, we have
     $U_i \mid \{U_{-i}, P_i \leq \lambda_i\} \sim \mathrm{Unif}[0, w_i\lambda_i]$.
     It then holds that
     $P_i/\lambda_i \mid \{U_{-i}, P_i \leq \lambda_i\} \sim \mathrm{Unif}[0, 1]$,
     so we can apply Lemma \ref{lem:lem2} to the ordinary SL procedure at level $\alpha_i$
     on $P_{\sigma(1)}/\lambda_i, \ldots, P_{\sigma(m_i)}/\lambda_i$.
    Therefore,
     \begin{align*}
     \PP\left(P_{\sigma\big(R^{\hat{\tau}_{\alpha}^\lambda}\big)} = P_i \mid U_{-i}\right)
     &= \PP\left(P_i \leq \lambda_i\right) \PP\left(P_{\sigma\big(R^{\hat{\tau}_{\alpha}^\lambda}\big)} = P_i \mid U_{-i}, P_i \leq \lambda_i\right) \\
     &\leq \PP\left(P_i \leq \lambda_i\right) \PP\left(P_{\sigma\big(R^{\hat{\tau}_{\alpha}^{\lambda_i}}\big)}/\lambda_i = P_i/\lambda_i \mid U_{-i}, P_i \leq \lambda_i\right) \\
     &\leq \PP\left(P_i \leq \lambda_i\right) \frac{\alpha_i}{m_i}\\
     &\leq \frac{\alpha}{m} w_i,
     \end{align*}
     where the equality is from \eqref{eq:imp-2} along with $U_i \indeps U_{-i}$,
     the first inequality follows from the implication \eqref{eq:implication},
    the second inequality follows from Lemma \ref{lem:lem2},
     and the last inequality uses $\PP(P_i \leq \lambda_i) \leq w_i \lambda_i$.
\end{proof}

\begin{lemma}[Leave-one-out stability] \label{lem:universal-A}
Fix $\alpha\in(0,1)$ and let
$P=(P_1,\ldots,P_m)\in[0,1]^m$.
If $P_i>\hat\alpha_{-i}$, then:
\begin{enumerate}[label=(\roman*)]
\item $\tilde A_P^u=A_{-i}$ for every
$u\in[\hat\alpha_{-i},\alpha]$;
\item
$\tilde\alpha_P=\bar\alpha_{-i}$ and
$\tilde A_P^{\tilde\alpha_P}=A_{-i}$.
\end{enumerate}
\end{lemma}

\begin{proof}[Proof of Lemma \ref{lem:universal-A}]
For a vector $q=(q_1,\ldots,q_d)$, let
$$
\hat F_q(s)
:=
\frac1d\sum_{\ell=1}^d\ind\{q_\ell\leq s\},
\qquad
\hat G_{\beta,q}(s)
:=
\beta\hat F_q(s)-s,
$$
and let
$
\mathcal C_q^\lambda
:=
\{0\}\cup\{q_\ell:q_\ell\leq\lambda\}.
$
In this notation,
the modified SL procedure can be expressed as
$$
\hat\tau_\beta^\lambda(q)
=
\max\!\left\{
\argmax_{s\in\mathcal C_q^\lambda}
\hat G_{\beta,q}(s)
\right\}.
$$
The largest-maximizer convention gives the following identity:
\begin{equation}\label{eq:threshold-truncation}
\hat\tau_\beta^b(q)\leq a\leq b
\quad\Longrightarrow\quad
\hat\tau_\beta^a(q)=\hat\tau_\beta^b(q).
\end{equation}
Indeed, the selected maximizer remains eligible after truncation at $a$.
Any removed candidate is larger and therefore has a strictly smaller
$\hat G_{\beta,q}$ value than the selected maximizer.

We first show that $\bar\alpha_{-i}\leq\hat\alpha_{-i}$. For any $q$,
each candidate $s>\alpha$ satisfies
$$
\hat G_{\alpha,q}(s)
\leq
\alpha-s
<0
\leq
\hat G_{\alpha,q}(0),
$$
so $\hat\tau_\alpha(q)\leq\alpha$. Since
$\tilde\alpha/(m+1)=\alpha/m$, augmenting $(P_{-i},t)$ by the value
$1$ and changing the nominal level from $\alpha$ to $\tilde\alpha$
leave both the candidate set and the maximization criterion unchanged after truncation
at $\alpha$. Hence, for every $t\in[0,1]$,
$$
\hat\tau_\alpha(P_{-i},t)
=
\hat\tau_\alpha^\alpha(P_{-i},t)
=
\hat\tau_{\tilde\alpha}^\alpha(P_{-i},1,t)
\leq\hat\alpha_{-i}.
$$
Taking the supremum over $t$ yields
$
\bar\alpha_{-i}\leq\hat\alpha_{-i}.
$

Now assume $P_i>\hat\alpha_{-i}$ and fix $t\in[0,1]$. Write
$$
a_t:=\hat\tau_\alpha(P_{-i},t),
\qquad
b_t:=\hat\tau_{\tilde\alpha}^\alpha(P_{-i},P_i,t).
$$
By definition,
$$
a_t\leq\bar\alpha_{-i}\leq\hat\alpha_{-i}<P_i,
\qquad
b_t\leq\hat\alpha_{-i}<P_i.
$$
Identity \eqref{eq:threshold-truncation} and equality of the maximization criteria below
$\hat\alpha_{-i}$ give
\begin{align*}
b_t
=
\hat\tau_{\tilde\alpha}^{\hat\alpha_{-i}}
    (P_{-i},P_i,t)
=
\hat\tau_\alpha^{\hat\alpha_{-i}}(P_{-i},t)
=
a_t.
\end{align*}
The middle equality holds because, for
$s\leq\hat\alpha_{-i}<P_i$, the eligible candidates coincide and
$
\hat G_{\tilde\alpha,(P_{-i},P_i,t)}(s)
=
\hat G_{\alpha,(P_{-i},t)}(s).
$

If $u\in[\hat\alpha_{-i},\alpha]$, then
$b_t\leq\hat\alpha_{-i}\leq u$, so another application of
\eqref{eq:threshold-truncation} gives
$$
\hat\tau_{\tilde\alpha}^u(P,t)
=
b_t
=
a_t
=
\hat\tau_\alpha(P_{-i},t).
$$
Consequently, it holds that
$$
t\in\tilde A_P^u
\iff
t\in A_{-i},
$$
which proves (i).
Finally,
$$
\tilde\alpha_P
=
\sup_{t\in[0,1]}b_t
=
\sup_{t\in[0,1]}a_t
=
\bar\alpha_{-i}.
$$
Since $b_t\leq\tilde\alpha_P$ for every $t$, identity \eqref{eq:threshold-truncation}
also gives
$$
\hat\tau_{\tilde\alpha}^{\tilde\alpha_P}(P,t)
=
b_t
=
a_t.
$$
Thus
$
\tilde A_P^{\tilde\alpha_P}=A_{-i}
$
holds, proving (ii).
\end{proof}

\begin{lemma}[Geometry of the SL rejection region]
\label{lem:sl-geometry}
Let $\alpha \in (0,1)$.
Fix $P_{-i}\in (0,1]^{m-1}$ with distinct coordinates.
Then there exist intervals $I_0$ and $I_j$, $j\neq i$, such that
$$
    A_{-i}
    =
    I_0
    \sqcup
    \bigsqcup_{j \in [m] \setminus \{i\}} I_j,
$$
where $I_0$ is either empty, $[0,\delta_0)$, or $[0,\delta_0]$, and each
$I_j$ is either empty, $[P_j,P_j+\delta_j)$, or $[P_j,P_j+\delta_j]$ for some
$\delta_j\geq 0$.
\end{lemma}

\begin{proof}[Proof of Lemma \ref{lem:sl-geometry}]
    Let $Q_{\sigma(1)} < Q_{\sigma(2)} < \ldots < Q_{\sigma(m-1)}$ be the order statistics of $P_{-i}$. 
    Define $\Delta_j := {\alpha j}/{m} - Q_{\sigma(j)}$ for $j=0,\ldots,m-1$ with $Q_{\sigma(0)} := 0$.
Fix $k\in\{0,\ldots,m-1\}$ and $t\in[Q_{\sigma(k)},Q_{\sigma(k+1)})$, where $Q_{\sigma(m)} := 1$ for notational convenience.
Then the largest rank corresponding to the value $t$ is $k+1$. Therefore,
by definition of the SL procedure, we have
    $$
    t \in A_{-i}\cap [Q_{\sigma(k)}, Q_{\sigma(k+1)})
    \iff
    \begin{cases}
        Q_{\sigma(k)} \leq t < Q_{\sigma(k+1)},\\[2pt]
        \displaystyle\frac{\alpha(k+1)}{m} - t \geq \max_{0 \leq j \leq k} \Delta_j,\\[4pt]
        \displaystyle\frac{\alpha(k+1)}{m} - t >
        \max_{k+1 \leq j \leq m-1}
        \left\{\frac{\alpha(j+1)}{m}-Q_{\sigma(j)}\right\},
    \end{cases}
    $$
where the last inequality is strict because if the rank $k+1$ score ties a larger
rank score, the largest-maximizer convention selects the larger rank.
Equivalently,
$$
    A_{-i}\cap[Q_{\sigma(k)},Q_{\sigma(k+1)})
    =
    [Q_{\sigma(k)},Q_{\sigma(k+1)})
    \cap(-\infty,a_k]
    \cap(-\infty,b_k),
$$
where
$$
    a_k
    :=
    \frac{\alpha(k+1)}{m}
    -
    \max_{0\leq j\leq k}\Delta_j
$$
and
$$
    b_k
    :=
    \frac{\alpha k}{m}
    -
    \max_{k+1\leq j\leq m-1}\Delta_j, \quad b_{m-1} := \infty.
$$
Define $\delta_k := \left[(a_k \wedge b_k) - Q_{\sigma(k)}\right]_+$.
If $Q_{\sigma(k)} \leq a_k < b_k$, set $I_k := [Q_{\sigma(k)}, Q_{\sigma(k)} + \delta_k]$.
Otherwise, set $I_k := [Q_{\sigma(k)}, Q_{\sigma(k)} + \delta_k)$.
By construction, we have
$$
    A_{-i}\cap[Q_{\sigma(k)},Q_{\sigma(k+1)})
    = I_k.
$$
As $[Q_{\sigma(k)},Q_{\sigma(k+1)})$ are disjoint for $k=0,\ldots,m-1$, the intervals $I_k$ are also disjoint.
Taking the disjoint union over $k=0,\ldots,m-1$ gives the desired
decomposition. Since $Q_{\sigma(1)},\ldots,Q_{\sigma(m-1)}$ are exactly the coordinates of
$P_{-i}$, this proves the claim.
\end{proof}

\subsection{Proofs of Main Results}\label{subsec:proofs-main}
\begin{proof}[Proof of Proposition~\ref{prop:density-condition-equivalent}]
Let $m_0=|\Nn_0|$. 
Suppose $P_1,\ldots,P_m$ are density-compound p-values.
For $m_0=m$, the density bound and
$\int_0^1\sum_{i\in\Nn_0}f_0^{(i)}(t)\,dt=m$ imply
$\sum_{i\in\Nn_0}F_0^{(i)}(t)=mt$, so taking the empty collection of CDFs suffices.
If $m_0<m$, define
$$
H(t):= \frac{1}{m-m_0}\left(mt-\sum_{i\in\Nn_0}F_0^{(i)}(t)\right).
$$
Then, $H$ is absolutely continuous since $\{F_0^{(i)}\}_{i\in\Nn_0}$ are absolutely continuous.
Moreover,
$$
H'(t)=\frac{1}{m-m_0}\left(m-\sum_{i\in\Nn_0}f_0^{(i)}(t)\right)\geq0
\quad\text{for almost every }t\in[0,1]
$$
by \eqref{eq:density-condition}. Thus $H$ is nondecreasing, and its
definition gives $H(0)=0$ and $H(1)=1$. Taking
$\widetilde F_0^{(i)}(t)=H(t)$ for every $i\notin\Nn_0$
gives CDFs satisfying the required identity.

Conversely, the existence of $\{\widetilde F_0^{(i)}\}_{i\notin\Nn_0}$ such that the identity holds gives
$$
\sum_{i\in\Nn_0}
\bigl(F_0^{(i)}(t)-F_0^{(i)}(s)\bigr)
\leq m(t-s), \qquad 0\leq s<t\leq1.
$$
Dividing by $t-s$ and letting $t\downarrow s$ yields
\eqref{eq:density-condition} on $[0,1]$ almost everywhere.
\end{proof}

\begin{proof}[Proof of Proposition \ref{prop:weighted-constructions}]
    Fix $i\in\Nn_0$. For $t\in[0,1)$, independence of $E_i$ and $U_i$ gives 
    $$ 
    F_{P_i}(t) = \EE\left[\PP(U_i\le tE_i\mid E_i)\right] = \EE[(tE_i)\wedge1]. 
    $$ 
    Since 
    $
    (tE_i)\wedge1 = \int_0^t E_i\ind\{sE_i<1\}\,ds, 
    $ 
    Tonelli's theorem yields 
    $$ 
    F_{P_i}(t) = \int_0^t \EE\left[E_i\ind\{sE_i<1\}\right]ds. 
    $$ 
    Thus $F_{P_i}$ is absolutely continuous on $[0,1)$ with its Lebesgue density satisfying 
    $$ 
    f_{P_i}(t) = \EE\left[E_i\ind\{tE_i<1\}\right] \le \EE[E_i]
    $$ 
    for Lebesgue-almost every $t \in [0,1)$.
    Consequently,
    $$
    \frac1m\sum_{i\in\Nn_0}f_{P_i}(t)
    \leq\frac1m\sum_{i\in\Nn_0}\EE[E_i]
    \leq1,
    $$
    which proves the claim.
\end{proof}

\begin{proof}[Proof of Proposition \ref{prop:cp-ave-cdf-construction}]
    Let
    $p(x)=m^{-1}\sum_{j=1}^mF_{X_j,0}(x)$ and let $G_i$ be the nominal null
    CDF of $p(X_i)$. For $t\in(0,1)$, continuity of $p$ gives a generalized
    inverse $x_t$ satisfying $p(x_t)=t$. If $p$ is constant on an interval,
    every $F_{X_i,0}$ is constant there as well, so the choice of $x_t$ does
    not affect the following calculation:
    \begin{equation*}
        \sum_{i=1}^mG_i(t)
        =\sum_{i=1}^mF_{X_i,0}(x_t)
        =mp(x_t)=mt.
    \end{equation*}
    This calculation implies that every $G_i$ is Lipschitz and hence absolutely continuous.
    Applying
    Proposition~\ref{prop:density-condition-equivalent} completes the proof.
\end{proof}

\begin{proof}[Proof of Proposition \ref{prop:compound-pvals-controls}]
    Lemma~\ref{lem:exchangeable-block-verification} verifies
    \eqref{eq:grid-partition-condition} and
    \eqref{eq:block-null-uniformity}. The conclusion then follows from
    Lemma~\ref{lem:block-rank-implies-discrete-dcp}.
\end{proof}

\begin{proof}[Proof of Proposition \ref{prop:dcperm-validity}]
Lemma~\ref{lem:permutation-block-verification} verifies
\eqref{eq:grid-partition-condition} and \eqref{eq:block-null-uniformity}
with $K=B-1$ and identifies $P_i^{\mathrm{dcperm}}=P_i^{\mathrm{CC}}$.
Lemma~\ref{lem:block-rank-implies-discrete-dcp} then proves the claim.
\end{proof}

\begin{proof}[Proof of Theorem \ref{thm:bfdr-weighted-pvals}]
    By Lemma \ref{lem:modified-lem2} with any $\lambda \in (\alpha, 1)$ and $w_i = E_i$, we have
    \begin{align*}
        \mathrm{bFDR}(\Rr_\alpha)
        = \sum_{i \in \Nn_0} \PP\left(P_{\sigma(R^{\hat{\tau}_\alpha})} = P_i\right)
        &= \sum_{i \in \Nn_0} \EE\left[\PP\left(P_{\sigma(R^{\hat{\tau}_\alpha}\big)} = P_i \mid U_{-i}, E_{1:m}\right)\right] \\
        &\leq \sum_{i \in \Nn_0} \EE\left[\frac{\alpha E_i}{m}\right] \\
        &= \frac{\alpha}{m}\sum_{i \in \Nn_0} \EE[E_i] \\
        &\leq \alpha,
    \end{align*}
    where the last inequality is by definition of compound e-values.
\end{proof}

\begin{proof}[Proof of Claim \eqref{prop:cp-bfdr-control}]
    We first show $\mathrm{bFDR}(\Rr_\alpha) \leq m_0 \alpha \wedge 1$.
    We write
    \begin{align*}
        \mathrm{bFDR}(\Rr_\alpha)
        = \sum_{i \in \Nn_0} \PP\left(P_{\sigma(R^{\hat{\tau}_\alpha})} = P_i\right)
        &= \sum_{i \in \Nn_0} \EE\left[\PP\left(P_{\sigma(R^{\hat{\tau}_\alpha})} = P_i \mid P_{-i}\right)\right] \\
        &= \EE\left[\sum_{i \in \Nn_0} \int_0^1 \ind\{t \in A_{-i}\} f^{(i)}_0(t) dt\right].
    \end{align*}
    Under condition \eqref{eq:density-condition}, it holds that $\sup_{t \in [0,1]} f^{(i)}_0(t) \leq m$ for all $i \in \Nn_0$.
    Therefore, we have the bound
    $$
    \mathrm{bFDR}(\Rr_\alpha)
    \leq \EE\left[\sum_{i \in \Nn_0} m \mu(A_{-i})\right] \wedge 1
    = m \EE\left[\sum_{i \in \Nn_0} \frac{\alpha}{m}\right] \wedge 1
    = m_0 \alpha \wedge 1,
    $$
    where the first equality is because $\mu(A_{-i}) = {\alpha}/{m}$ by Lemma \ref{lem:lem2}.

    It remains to show $\mathrm{bFDR}(\Rr_\alpha) \leq 2(\alpha + \alpha^2 m)$.
    By the compound p-value property \eqref{eq:cp}, it holds that for any $\gamma \in (0, 1)$,
    $$
    \gamma m_\gamma \leq \sum_{i \in \Nn_\gamma} \PP(P_i \leq \alpha) \leq m \alpha,
    $$
    where $\Nn_\gamma$ and $m_\gamma$ are as in Theorem~\ref{thm:gamma-bfdr-control}.
    Theorem \ref{thm:gamma-bfdr-control}
    implies the upper bound $\mathrm{bFDR}(\Rr_\alpha) \leq {m \alpha^2}/{\gamma} + {\alpha}/{(1-\gamma)}$.
    The optimal choice $\gamma^* = {\sqrt{\alpha m}}/(\sqrt{\alpha m} + 1)$
    to minimize the right-hand side gives
    $$
    \mathrm{bFDR}(\Rr_\alpha) \leq \alpha (1+\sqrt{\alpha m})^2
    \leq 2(\alpha + \alpha^2m),
    $$
    as claimed.
\end{proof}

\begin{proof}[Proof of Proposition \ref{prop:cp-bfdr-counterexample}]
    Let $k = 1/\alpha$ be an integer. Fix $m \geq 1/\alpha$ and $m_0 \leq m + 1 - k$, which guarantees at least $k-1$ non-nulls.
    The case $m_0=0$ is immediate, so suppose $m_0\geq1$. 
    To begin, take a null p-value defined as
    $$
    P_1 \sim \mathrm{Unif}\left[\frac{\alpha (k - 1)}{m}, \frac{\alpha (2k - 1)}{m}\right]
    $$
    with non-nulls defined as
    $$
    P_2 = P_3 = \ldots = P_{k} = \frac{\alpha (k-1)}{m}.
    $$
    Generalizing this construction to $m_0 \geq 2$ nulls, at each stage $j=1,\ldots, j^* - 1$ when possible, define
    \begin{align*}
        P_{jk + 1} \sim \mathrm{Unif}\left[\frac{\alpha ((j+1)k - 1)}{m}, \frac{\alpha ((j+2)k - 1)}{m}\right]
    \end{align*}
    and $k-1$ non-nulls
    \begin{align*}
        P_{jk + 2} = \ldots = P_{(j+1)k} = \frac{\alpha ((j+1)k - 1)}{m},
    \end{align*}
    where all null p-values are drawn independently.
    These $j^*$ null p-values are uniform over disjoint intervals 
    $$
    \left[\frac{\alpha ((j+1)k - 1)}{m}, \frac{\alpha ((j+2)k - 1)}{m}\right] \quad \mathrm{for} \ 0 \leq j \leq j^* - 1.
    $$
    They have density $f^{(i)}_0(t) = m/(\alpha k) = m$ on their supports, so condition \eqref{eq:density-condition} holds
    over the union of these intervals.

    By construction, the non-null p-values are never below the support line $y = \alpha j/{m}$ for $j=1,\ldots, m$,
    so the last rejected p-value must be one of the nulls in the event that at least one null p-value falls below the support line.
    Then,
    \begin{align*}
        \mathrm{bFDR}(\Rr_\alpha)
        \geq \PP\left(\bigcup_{j=0}^{j^* - 1} \left\{P_{jk + 1} \in \left(\frac{\alpha ((j+1)k - 1)}{m}, \frac{\alpha (j+1)k}{m}\right]\right\}\right) 
        = 1 - (1-\alpha)^{j^*}.
    \end{align*}
    The hard constraints on $j^*$ are that $j^* \leq m_0$ since we only have $m_0$ nulls,
    and the total number of p-values is at most $m$,
    i.e. 
    $$
    kj^* \leq m \iff j^* \leq m \alpha.
    $$
    When $m_0 \leq m \alpha$, we can do the construction for all $m_0$ nulls.
    Otherwise, if $m_0 > m \alpha$, then we can only do the construction for at most 
    $$
    j^* = \left\lfloor \frac{m - m_0}{k-1} \right\rfloor = \left\lfloor \frac{\alpha(m - m_0)}{1-\alpha} \right\rfloor
    $$
    nulls since each null requires $k-1$ non-nulls and we only have $m - m_0$ total non-nulls.
    We set the remaining $m_0 - j^*$ nulls to be $\mathrm{Unif}\left[{\alpha ((j^* + 1)k - 1)}/{m}, 1\right]$
    and non-nulls to be $1$.
    For $t \in \left({\alpha ((j^*+1)k-1)}/{m}, 1\right)$, we verify condition \eqref{eq:density-condition} holds:
    \begin{align*}
        \frac{1}{m} \sum_{i \in \Nn_0} f^{(i)}_0(t)
        = \frac{m_0 - j^*}{m} \cdot \frac{m}{m - \alpha ((j^* + 1)k - 1)}
        = \frac{m_0 - j^*}{m - j^* - 1 + \alpha}
        \leq 1,
    \end{align*}
    which holds because $m - m_0 \geq 1$.
    Together, we have
    $$
    \mathrm{bFDR}(\Rr_\alpha)
    \geq 1 - (1-\alpha)^{m_0 \wedge \lfloor \frac{\alpha(m - m_0)}{1-\alpha} \rfloor},
    $$
    as claimed.
\end{proof}

\begin{proof}[Proof of Theorem~\ref{thm:gamma-alpha-bfdr-control}]
    Fix any $\Bb \subseteq\Nn_0$ and take
    $
    \gamma
    =
    \gamma_{\Nn_0\setminus \Bb}(\alpha).
    $
    Then
    $$
    \Nn_\gamma
    =
    \left\{
        i\in\Nn_0:
        \PP(P_i\le\alpha)>
        \gamma_{\Nn_0\setminus \Bb}(\alpha)
    \right\}
    \subseteq \Bb,
    $$
    so $m_\gamma\le |\Bb|$.
    Applying Theorem~\ref{thm:gamma-bfdr-control} yields
    $$
    \mathrm{bFDR}(\Rr_\alpha)
    \le
    |\Bb|\alpha
    +
    \frac{\alpha}
    {1-\gamma_{\Nn_0\setminus \Bb}(\alpha)}.
    $$
    Since this holds for every $\Bb\subseteq\Nn_0$, taking the minimum over
    $\Bb$ gives the intermediate bound.
    Taking $\Bb=\varnothing$ gives
    $$
    \mathrm{bFDR}(\Rr_\alpha)
    \le
    \frac{\alpha}
    {1-\gamma_{\Nn_0}(\alpha)},
    $$
    which proves the final bound.
\end{proof}

\begin{proof}[Proof of Theorem \ref{thm:gamma-bfdr-control-posthoc}]
    Recall the auxiliary thresholds from Section~\ref{subsec:proof-notation}.
    For each $i\in\Nn_0$, $\hat{\alpha}_{-i}$ is $P_{-i}$-measurable, so
    $P_i \indeps P_{-i}$ and the definition of $\gamma_{\Nn_0}$ give
    $\PP(P_i>\hat{\alpha}_{-i}\mid P_{-i})
    \geq 1-\gamma_{\Nn_0}(\hat{\alpha}_{-i})$.
    Thus, we have
    $$
    \begin{aligned}
    \EE\left[\big(1-\gamma_{\Nn_0}(\hat{\alpha}_{-i})\big)
        \PP(P_i\in A_{-i}\mid P_{-i})\right]
    &\leq \EE\left[
        \PP(P_i>\hat{\alpha}_{-i}\mid P_{-i})
        \PP(P_i\in A_{-i}\mid P_{-i})\right] \\
    &= \EE\left[
        \ind\{P_i>\hat{\alpha}_{-i}\}
        \PP(P_i\in A_{-i}\mid P_{-i})\right],
    \end{aligned}
    $$
    where the equality follows from the tower property.

    On the event $\{P_i > \hat{\alpha}_{-i}\}$, Lemma \ref{lem:universal-A} $(ii)$
    gives $A_{-i} = \tilde{A}_P^{\tilde{\alpha}_P}$ and
    $\tilde\alpha_P=\bar\alpha_{-i}\leq \hat{\alpha}_{-i}$. By monotonicity of
    $t \mapsto \gamma_{\Nn_0}(t)$,
    \begin{align*}
        \EE\left[\bigg(1 - \gamma_{\Nn_0}(\hat{\alpha}_{-})\bigg) \ind\{\sigma(R_\alpha) \in \Nn_0\}\right]  
        &= \EE\left[\bigg(1 - \gamma_{\Nn_0}(\hat{\alpha}_{-})\bigg) \sum_{i \in \Nn_0} \ind\left\{P_{\sigma(R_\alpha)} = P_i\right\}\right] \\
        &= \sum_{i\in\Nn_0}\EE\left[\EE\left[\big(1-\gamma_{\Nn_0}(\hat{\alpha}_{-})\big)
        \ind\{P_i\in A_{-i}\}\,\middle|\,P_{-i}\right]\right] \\
        &\leq \sum_{i \in \Nn_0} \EE\left[\big(1 - \gamma_{\Nn_0}(\hat{\alpha}_{-i})\big) \PP\left(P_i \in A_{-i} \mid P_{-i}\right)\right] \\
        &\leq \sum_{i \in \Nn_0} \EE\left[\ind\{P_i > \hat{\alpha}_{-i}\} \PP\left(P_i \in A_{-i} \mid P_{-i}\right)\right] \\
        &\leq \sum_{i \in \Nn_0} \EE\left[\int_0^1 \ind\{t \in \tilde{A}_P^{\tilde{\alpha}_P}\} f^{(i)}_0(t) dt\right] \\
        &\leq m \cdot \EE\left[\mu(\tilde{A}_P^{\tilde{\alpha}_P})\right] \\
        &\leq \alpha.
    \end{align*}
    To see the last inequality, note $\mu(\tilde A_P^{\tilde\alpha_P})=0$ if $\tilde\alpha_P=0$.
    Otherwise, conditional on $P$, $\mu(\tilde A_P^{\tilde\alpha_P})$ is the probability that an
    independent $T\sim\mathrm{Unif}[0,1]$ is the SL threshold on $(P,T)$ at
    level $\tilde\alpha$, with the cap $\lambda = \tilde\alpha_P$. 
    Lemma~\ref{lem:modified-lem2}
    with unit weights implies 
    $\mu(\tilde A_P^{\tilde\alpha_P}) \leq \tilde\alpha/(m+1)=\alpha/m$.
\end{proof}

\begin{proof}[Proof of Claim \eqref{cor:gamma-bfdr-control-posthoc-bh}]
    Recall $\tilde\alpha=\alpha(m+1)/m$. For every $i\in\Nn_0$ and
    $x,t\in[0,1]$, the SL threshold is bounded by the corresponding BH
    critical value:
    $$
    \hat\tau_{\tilde\alpha}^{\alpha}(P_{-i},x,t)
    \leq
    \frac{\alpha}{m}
    R_{\tilde\alpha}^{\mathrm{BH}}(P_{-i},x,t).
    $$
    Since BH is coordinatewise monotone, its rejection count is maximized by
    setting both free coordinates to zero and removing the largest p-value:
    $$
    R_{\tilde\alpha}^{\mathrm{BH}}(P_{-i},x,t)
    \leq
    R_{\tilde\alpha}^{\mathrm{BH}}(P_{-i},0,0)
    \leq
    R_{\tilde\alpha}^{\mathrm{BH}}(P_{-\sigma(m)},0,0),
    $$
    where $P_{-\sigma(m)}$ is the vector of p-values with the largest p-value removed.
    Taking the supremum over $x,t$ and then the maximum over $i$ proves
    $\hat{\alpha}_{-}\leq \hat{\alpha}_{\mathrm{BH}}$.
\end{proof}

\begin{proof}[Proof for Remark \ref{rmk:global-null}]
    Under the global null,
    it holds
    \begin{align*}
    \mathrm{bFDR}(\Rr_\alpha)
    = \PP\left(R^{\hat{\tau}_\alpha} > 0\right)
    = \PP\left(\exists k \in [m]: P_{(k)} \leq \frac{\alpha k}{m}\right)
    = \PP\left(\text{BH makes a rejection}\right).
    \end{align*}
    The results of \cite{10.1093/jrsssb/qkag116} under the global null yield an upper bound of $\alpha + 2\alpha^2$
    for independent compound p-values.
    For the lower bound, take independently drawn
    $$
    P_1 \sim \begin{cases} \mathrm{Unif}[0,\alpha/m] & \text{w.p. } \alpha, \\ 
    \mathrm{Unif}[\alpha/m, 2\alpha/m] & \text{w.p. } \alpha/2, \\
    \mathrm{Unif}[2\alpha/m, 1] & \text{otherwise},
    \end{cases} \quad
    P_2 \sim \begin{cases} \mathrm{Unif}[\alpha/m, 2\alpha/m] & \text{w.p. } \alpha/2, \\
    \mathrm{Unif}[2\alpha/m, 1] & \text{otherwise},
    \end{cases}
    $$
    and 
    $P_3,\ldots,P_m \sim \mathrm{Unif}[2\alpha/m, 1].$
    These p-values satisfy \eqref{eq:density-condition} since the average density is exactly $1$.
    Moreover, the bFDR is at least $\alpha + \alpha^2/4$
    from the union of events $\{P_1 \leq \alpha/m\}$ and $\{P_1, P_2 \in [\alpha/m, 2\alpha/m]\}$.
\end{proof}

\begin{proof}[Proof for Remark \ref{rmk:individually-valid}]
    Let $m_1=m-m_0$ and label the first $m_0$ p-values as nulls.
    Set the non-null p-values to
    $$
    P_{m_0+1}=\cdots=P_m=\frac{\alpha m_1}{m}.
    $$
    Write $I_1=(\alpha m_1/m,\alpha(m_1+1)/m]$,
    $I_2=(\alpha(m_1+1)/m,\alpha(m_1+2)/m]$, and
    $I_3=(\alpha(m_1+2)/m,\alpha]$.
    Draw the $m_0$ null p-values independently with
    $$
        P_1,P_2 \sim
        \begin{cases}
            \mathrm{Unif}(I_1) & \text{w.p. } \alpha/2, \\
            \mathrm{Unif}(I_2) & \text{w.p. } \alpha/4, \\
            \mathrm{Unif}(I_3) & \text{w.p. } \alpha/4, \\
            \mathrm{Unif}[\alpha,1] & \text{otherwise},
        \end{cases} \quad
        P_3 \sim
        \begin{cases}
            \mathrm{Unif}(I_2) & \text{w.p. } \alpha/2, \\
            \mathrm{Unif}(I_3) & \text{w.p. } \alpha/2, \\
            \mathrm{Unif}[\alpha,1] & \text{otherwise},
        \end{cases}
    $$
    and, for $4\leq i\leq m_0$,
    $$
        P_i \sim
        \begin{cases}
            \mathrm{Unif}(I_3) & \text{w.p. } \alpha, \\
            \mathrm{Unif}[\alpha,1] & \text{otherwise}.
        \end{cases}
    $$
    To verify \eqref{eq:density-condition}, note that the null densities
    are zero below $\alpha m_1/m$. On the remaining intervals, we have
    $$
        \sum_{i\in\Nn_0} f_0^{(i)}(t)=
        \begin{cases}
            \left({\alpha}/{2}+{\alpha}/{2}\right){m}/{\alpha}
                =m, & t\in I_1, \\[4pt]
            \left({\alpha}/{4}+{\alpha}/{4}+{\alpha}/{2}\right)
                {m}/{\alpha}=m, & t\in I_2, \\[4pt]
            \left({\alpha}/{4}+{\alpha}/{4}+{\alpha}/{2}
                +(m_0-3)\alpha\right)\dfrac{m}{(m_0-2)\alpha}=m, & t\in I_3, \\[4pt]
            \dfrac{m_0(1-\alpha)}{1-\alpha}=m_0, & t\in(\alpha,1],
        \end{cases}
    $$
    so these p-values are density-compound.

    To check individual validity \eqref{eq:superuniform}, the null CDFs are zero up to
    $\alpha m_1/m$ and continuously piecewise linear on each interval. 
    It thus suffices to check the validity at the interval endpoints.
    At the right endpoints
    of $I_1$ and $I_2$, the first two null CDFs satisfy
    \begin{align*}
        F_0^{(1)}\left(\frac{\alpha(m_1+1)}{m}\right)
        &= F_0^{(2)}\left(\frac{\alpha(m_1+1)}{m}\right)
        =\frac{\alpha}{2}\leq\frac{\alpha(m_1+1)}{m}, \\
        F_0^{(1)}\left(\frac{\alpha(m_1+2)}{m}\right)
        &= F_0^{(2)}\left(\frac{\alpha(m_1+2)}{m}\right)
        =\frac{3\alpha}{4}\leq\frac{\alpha(m_1+2)}{m},
    \end{align*}
    where the inequalities follow
    from $m_0\leq m/4+2$ and $m\geq4$. 
    $F_0^{(3)}$ is zero at
    the right endpoint of $I_1$ and equals $\alpha/2$ at the right endpoint
    of $I_2$. The remaining null CDFs are zero at both endpoints.
    Every null CDF satisfies $F_0^{(i)}(\alpha)=\alpha$ and has slope one
    above $\alpha$. Hence, the null p-values are individually valid and $\gamma_{\Nn_0}(\alpha)=\alpha$.

    The non-nulls $P_{m_0+1},\ldots,P_m$ occupy the first $m_1$ ranks, for which
    $P_{\sigma(k)}-\alpha k/m$ is nonnegative over $k \in [m_1]$. A null in $I_1$ or two nulls
    in $I_2$ give a negative value at rank $m_1+1$ or $m_1+2$, respectively, almost surely.
    By independence, the bFDR admits the lower bound
    \begin{align*}
        \mathrm{bFDR}(\Rr_\alpha)
        \geq 1-(1-\alpha/2)^2+(\alpha/4)^2
        + 2\cdot {\alpha}/{2} \cdot {\alpha}/{4}\left(1-{3\alpha}/{4}\right)
        = \alpha+\frac{\alpha^2}{16}(1-3\alpha)
    \end{align*}
    from the disjoint union of $\{P_1\in I_1\text{ or }P_2\in I_1\}$,
    $\{P_1,P_2\in I_2\}$, and
    $$
        \{P_3,P_1\in I_2,\ P_2>\alpha(m_1+2)/m\}
        \sqcup
        \{P_3,P_2\in I_2,\ P_1>\alpha(m_1+2)/m\}.
    $$
    This concludes the proof.
\end{proof}

\begin{proof}[Proof of Proposition \ref{prop:discrete-cp-bfdr-bound}]
    On the grid $\{k/L : k \in [L]\}$, let $\mu$ denote the normalized counting measure, i.e. $\mu(\{k/L\}) = 1/L$ for $k \in [L]$.
    Condition \eqref{eq:discrete-density-condition} states that $P_i$ has a density $f^{(i)}$ with respect to $\mu$ satisfying
    $$
    \sum_{i \in \Nn_0} f^{(i)}(t) \leq m, \quad \forall t \in \Gg_L.
    $$
    By Lemma \ref{lem:lem2-discrete}, it holds that
    $$
    \mu(A_{-i}\cap\Gg_L) \leq \frac{\alpha}{m} + \frac{m}{L}.
    $$
    Noting $\Nn_{\gamma_{\Nn_0}(\alpha)} = \emptyset$, the rest of the proof proceeds exactly as the proof of Theorem~\ref{thm:gamma-bfdr-control}
    replacing integrals with respect to the Lebesgue measure by integrals with respect to $\mu$ and applying the above bound on $\mu(A_{-i}\cap\Gg_L)$.
\end{proof}

\begin{proof}[Proof of Theorem~\ref{thm:bfdr-dcperm}]
Condition on $\widehat P_{1:m}$. In the notation of
Lemma~\ref{lem:permutation-block-verification}, the sets $\mathcal V_i$
and empirical distributions $\widehat{\mathcal P}_i$ are then fixed.
For every $i\in\Nn_0$, exchangeability and equal multiplicity give
$\PP(X_i=x\mid\widehat P_i)=1/B$ for $x\in\mathcal V_i$.
Independence across units therefore implies
$$
\PP(X_i=x\mid\widehat P_{1:m},X_{-i}^{(0)})
=\widehat{\mathcal P}_i(\{x\})
=\frac1B,
\qquad x\in\mathcal V_i.
$$
The rank-map argument in the proof of
Lemma~\ref{lem:permutation-block-verification} now verifies
\eqref{eq:grid-partition-condition} and
\eqref{eq:block-null-uniformity} under this conditioning, with
$K=B-1$ and $P_i^{\mathrm{CC}}=P_i^{\mathrm{dcperm}}$.
Applying Theorem~\ref{thm:bfdr-cp-controls} conditionally on
$\widehat P_{1:m}$ completes the proof.
\end{proof}

\begin{proof}[Proof of Theorem \ref{thm:lfdr-upper-bound-weak-dependence}]
    Observe that $\hat\tau_\alpha$ is the largest maximizer of $\hat G_m(\tau) = \alpha \hat F_m(\tau) - \tau$ over $[0,1]$.
    We first establish $|\hat\tau_\alpha-\tau_{\alpha,m}^*|=o_p(1)$.

    Fix $\eta > 0$ and define $\Delta_m(\eta) = G_m(\tau_{\alpha,m}^*) - \sup_{\tau\in[0,1]:|\tau-\tau_{\alpha,m}^*|\ge \eta} G_m(\tau)$.
    By the uniform separation condition \eqref{eq:separation-condition}, it holds that $\liminf_{m\to\infty} \Delta_m(\eta) > 0$.
    Since
    $$
    \Vert \hat G_m - G_m\Vert_\infty = \alpha \Vert \hat F_m - \bar F_m\Vert_\infty = o_p(1),
    $$
    we have $\PP\left(\Vert \hat G_m - G_m\Vert_\infty < \frac{1}{2}\Delta_m(\eta)\right) \to 1$.
    On the event $\Vert \hat G_m - G_m\Vert_\infty < \frac{1}{2}\Delta_m(\eta)$, for every $\tau\in[0,1]$ satisfying $|\tau-\tau_{\alpha,m}^*|\ge \eta$, it holds that
    $$ 
    \hat G_m(\tau) \le G_m(\tau)+\|\hat G_m-G_m\|_\infty < G_m(\tau_{\alpha,m}^*)-\frac{1}{2}\Delta_m(\eta). 
    $$
    Moreover,
    $$
    \hat G_m(\tau_{\alpha,m}^*) \ge G_m(\tau_{\alpha,m}^*) - \|\hat G_m - G_m\|_\infty > G_m(\tau_{\alpha,m}^*) - \frac{1}{2}\Delta_m(\eta).
    $$
    Therefore, no maximizer of $\hat G_m$ can lie outside $\{\tau:|\tau-\tau_{\alpha,m}^*|<\eta\}$, which implies
    $$
    \PP\left(|\hat\tau_\alpha-\tau_{\alpha,m}^*|\ge \eta\right)
    \leq \PP\left(\Vert \hat G_m - G_m\Vert_\infty \ge \frac{1}{2}\Delta_m(\eta)\right) \to 0,
    $$
    showing $|\hat\tau_\alpha-\tau_{\alpha,m}^*|=o_p(1)$ as $\eta>0$ was arbitrary.
    
    Next, the local Lipschitz condition on $\bar f_m$ gives
    $$ 
    \bar f_m(\hat\tau_\alpha) = \bar f_m(\tau_{\alpha,m}^*)+o_p(1) = \frac{1}{\alpha}+o_p(1),
    $$ 
    where we used $\bar f_m(\tau_{\alpha,m}^*) = 1/\alpha$ by the first-order condition for the unique maximizer of $G_m$.
    By condition \eqref{eq:density-condition}, it holds that
    \begin{align*}
    \mathrm{LFDR}^{\rm freq}_m(\hat\tau_\alpha) 
    \le \frac{1}{\bar f_m(\hat\tau_\alpha)}
    = \frac{1}{\bar f_m(\tau_{\alpha,m}^*)+o_p(1)}
    = \alpha + o_p(1),
    \end{align*}
    which establishes the desired result.
\end{proof}

\subsection{Proofs of Auxiliary Results}
\label{subsec:proofs-auxiliary}
We use the notation introduced with each result in Appendix~\ref{sec:auxiliary}.

\begin{proof}[Proof of Lemma~\ref{lem:block-rank-implies-discrete-dcp}]
Fix $q\in\mathcal G_L$. By
\eqref{eq:block-null-uniformity}, the tower property, and
\eqref{eq:grid-partition-condition},
\begin{align*}
\frac{1}{m}\sum_{i\in\Nn_0}
\PP(P_i^{\mathrm{CC}}=q)
&=
\frac{1}{m}\sum_{i\in\Nn_0}
\EE\left[
\PP\left(
P_i^{\mathrm{CC}}=q
\,\middle|\,
\widehat{\mathcal P}_{1:m},X_{-i}^{(0)}
\right)
\right] \\
&=
\frac{1}{m(K+1)}
\EE\left[
\sum_{i\in\Nn_0}
\ind\{q\in\mathcal G_{i,K+1}\}
\right] \\
&\leq \frac{1}{m(K+1)}
=\frac{1}{L},
\end{align*}
which proves the claim.
\end{proof}

\begin{proof}[Proof of Theorem \ref{thm:bfdr-cp-controls}]
    Condition throughout on $\widehat{\mathcal P}_{1:m}$.
    Write $P_i=P_i^{\mathrm{CC}}$ and
    $\gamma_{\Nn_0}=\gamma_{\Nn_0}^{\mathrm{CC}}$.
    Use $A_{-i}$, $\hat\alpha_{-i}$, and $\tilde\alpha_P$ from
    Section~\ref{subsec:proof-notation} for this vector $P$.

    By \eqref{eq:block-null-uniformity},
    for each $i\in\Nn_0$,
    $P_i\mid \widehat{\mathcal P}_{1:m},X_{-i}^{(0)}$ is uniform on $\mathcal G_{i,K+1}$.
    The unit-specific grids partition $\mathcal G_L$ for $L=m(K+1)$
    by \eqref{eq:grid-partition-condition}. Since
    $P_i\in\mathcal G_{i,K+1}$ for every $i$ and these grids are disjoint,
    the observed p-values $P_1,\ldots,P_m$ are distinct.
    Write
    $$
        Q_i:=\PP\left(
            P_i\in A_{-i}\cap\mathcal G_{i,K+1}
            \,\middle|\,\widehat{\mathcal P}_{1:m},X_{-i}^{(0)}
        \right)
        =\frac{|A_{-i}\cap\mathcal G_{i,K+1}|}{K+1}.
    $$
    Conditional uniformity gives
    $\PP(P_i\leq\hat{\alpha}_{-i}\mid \widehat{\mathcal P}_{1:m},X_{-i}^{(0)})
    \leq\gamma_{\Nn_0}(\hat{\alpha}_{-i})$.
    Since $Q_i$ and $\hat{\alpha}_{-i}$ are measurable with respect to
    $\widehat{\mathcal P}_{1:m}$ and $X_{-i}^{(0)}$, the tower property yields
    \begin{align*}
    \EE\left[
        \{1-\gamma_{\Nn_0}(\hat{\alpha}_{-i})\}Q_i\,\middle|\,\widehat{\mathcal P}_{1:m}
    \right]
    &\leq \EE\left[
        Q_i\PP\big(P_i>\hat{\alpha}_{-i}\mid \widehat{\mathcal P}_{1:m},X_{-i}^{(0)}\big)
        \,\middle|\,\widehat{\mathcal P}_{1:m}
    \right] \\
    &=\EE\left[\ind\{P_i>\hat{\alpha}_{-i}\}Q_i\,\middle|\,\widehat{\mathcal P}_{1:m}\right].
    \end{align*}
    On the event $\{P_i > \hat{\alpha}_{-i}\}$, Lemma
    \ref{lem:universal-A} $(ii)$ gives
    $A_{-i}=\tilde A_P^{\tilde\alpha_P}$ and
    $\tilde\alpha_P=\bar\alpha_{-i}\leq \hat{\alpha}_{-i}$. By monotonicity of
    $t \mapsto \gamma_{\Nn_0}(t)$,
    \begin{align*}
        &\EE\left[\bigg(1 - \gamma_{\Nn_0}(\hat{\alpha}_{-})\bigg) \ind\{\sigma(R_\alpha) \in \Nn_0\}
        \,\middle|\,\widehat{\mathcal P}_{1:m}\right] \\
        &\quad= \EE\left[\bigg(1 - \gamma_{\Nn_0}(\hat{\alpha}_{-})\bigg) \sum_{i \in \Nn_0} \ind\left\{P_{\sigma(R_\alpha)} = P_i\right\}
        \,\middle|\,\widehat{\mathcal P}_{1:m}\right] \\
        &\quad= \sum_{i\in\Nn_0}\EE\left[\EE\left[\big(1-\gamma_{\Nn_0}(\hat{\alpha}_{-})\big)
        \ind\{P_i\in A_{-i}\}\,\middle|\,\widehat{\mathcal P}_{1:m},X_{-i}^{(0)}\right]
        \,\middle|\,\widehat{\mathcal P}_{1:m}\right] \\
        &\quad\leq \sum_{i \in \Nn_0} \EE\left[\big(1 - \gamma_{\Nn_0}(\hat{\alpha}_{-i})\big) Q_i
        \,\middle|\,\widehat{\mathcal P}_{1:m}\right] \\
        &\quad\leq \sum_{i \in \Nn_0} \EE\left[\ind\{P_i > \hat{\alpha}_{-i}\} Q_i
        \,\middle|\,\widehat{\mathcal P}_{1:m}\right] \\
        &\quad= \sum_{i \in \Nn_0} \EE\left[\ind\{P_i > \hat{\alpha}_{-i}\} \frac{1}{K+1} \left|\tilde{A}_P^{\tilde{\alpha}_P} \cap \mathcal G_{i, K+1}\right|
        \,\middle|\,\widehat{\mathcal P}_{1:m}\right],
    \end{align*}
    where the last equality uses the pointwise identity
    $A_{-i}=\tilde A_P^{\tilde\alpha_P}$ on
    $\{P_i>\hat\alpha_{-i}\}$.
    Define
    $$
        E:=\bigcup_{i\in\Nn_0}\{P_i>\hat\alpha_{-i}\},
        \qquad
        \tilde B_P^{\tilde\alpha_P}:=
        \begin{cases}
            \tilde A_P^{\tilde\alpha_P}\setminus\{P_1,\ldots,P_m\},
            &\text{on }E,\\
            \emptyset,&\text{on }E^c.
        \end{cases}
    $$
    On $E$, choose any $i_0\in\Nn_0$ with $P_{i_0}>\hat\alpha_{-i_0}$.
    By Lemma~\ref{lem:sl-geometry},
    there exist an interval $I_0$ of the form
    $\emptyset$, $[0,\delta_0)$, or $[0,\delta_0]$, and disjoint intervals
    $I_j$, $j\in[m]$, $j\neq i_0$, of the form $\emptyset$, $[P_j,P_j+\delta_j)$, or
    $[P_j,P_j+\delta_j]$, such that
    $$
        \tilde{A}_P^{\tilde{\alpha}_P}=A_{-i_0}
        =
        I_0
        \sqcup
        \bigsqcup_{j=1}^m I_j,
    $$
    where we define $I_{i_0} := \emptyset$.
    For each $i\in\Nn_0$, on $\{P_i>\hat\alpha_{-i}\}$ we have
    $P_i>\tilde\alpha_P$, so $P_i\notin\tilde A_P^{\tilde\alpha_P}$.
    Since $P_j\in\mathcal G_{j,K+1}$ and
    $\Gg_{j, K+1} \cap \Gg_{i, K+1} = \emptyset$ for every $j\neq i$,
    it follows that
    $$
    \tilde{A}_P^{\tilde{\alpha}_P}\cap\mathcal G_{i,K+1}
    =\tilde{B}_P^{\tilde{\alpha}_P}\cap\mathcal G_{i,K+1}.
    $$
    Therefore,
    \begin{align*}
    &\EE\left[\bigg(1 - \gamma_{\Nn_0}(\hat{\alpha}_{-})\bigg) \ind\{\sigma(R_\alpha) \in \Nn_0\}
    \,\middle|\,\widehat{\mathcal P}_{1:m}\right] \\
    &\quad\leq \sum_{i \in \Nn_0} \EE\left[\ind\{P_i > \hat\alpha_{-i}\} \frac{1}{K+1} \left|\tilde{B}_P^{\tilde{\alpha}_P} \cap \mathcal G_{i, K+1}\right|
    \,\middle|\,\widehat{\mathcal P}_{1:m}\right] \\
    &\quad\leq \frac{1}{K+1} \EE\left[\sum_{i=1}^m \left|\tilde{B}_P^{\tilde{\alpha}_P} \cap \mathcal G_{i, K+1}\right|
    \,\middle|\,\widehat{\mathcal P}_{1:m}\right] \\
    &\quad= \frac{1}{K+1} \EE\left[\left|\tilde{B}_P^{\tilde{\alpha}_P} \cap \mathcal G_L\right|
    \,\middle|\,\widehat{\mathcal P}_{1:m}\right] \\
    &\quad\leq \frac{L}{K+1} \EE\left[\mu(\tilde{B}_P^{\tilde{\alpha}_P})
    \,\middle|\,\widehat{\mathcal P}_{1:m}\right] \\
    &\quad\leq \alpha,
    \end{align*}
    where the second inequality used $\tilde B_P^{\tilde\alpha_P}=\emptyset$ on $E^c$,
    the equality used the unit-specific grids forming a partition of
    $\mathcal G_L$, and the third inequality used the elementary inequality
    $$
    \left| (I_j \setminus \{P_j\}) \cap \mathcal G_L \right| \leq L\delta_j
    $$
    for $j\geq1$ and all $\delta_j \geq 0$, together with
    $|I_0\cap\mathcal G_L|\leq L\delta_0$ since $0\notin\mathcal G_L$.
    The last inequality follows from
    $\tilde B_P^{\tilde\alpha_P}\subseteq\tilde A_P^{\tilde\alpha_P}$
    and $\mu(\tilde A_P^{\tilde\alpha_P}) \leq \alpha/m$, as shown in the last step
    of the proof of Theorem~\ref{thm:gamma-bfdr-control-posthoc}.
\end{proof}

\begin{proof}[Proof of Lemma~\ref{lem:exchangeable-block-verification}]
Distinctness gives pooled ranks $1/L,\ldots,1$, with $K+1$ from each
unit, proving \eqref{eq:grid-partition-condition}.
For $i\in\Nn_0$, exchangeability makes $X_i^{(0)}$ uniform over the
entries of $Y_i$ conditional on $\widehat{\mathcal P}_i$. Independence
across units therefore gives, for each value $x$ in $Y_i$,
$$
\PP\left(
X_i^{(0)}=x
\,\middle|\,
\widehat{\mathcal P}_{1:m},X_{-i}^{(0)}
\right)
=
\PP\left(X_i^{(0)}=x\mid\widehat{\mathcal P}_i\right)
=
\frac1{K+1}.
$$
The rank map $\rho$ is fixed given
$\widehat{\mathcal P}_{1:m}$ and sends the distinct entries of $Y_i$ bijectively to
$\mathcal G_{i,K+1}$, proving \eqref{eq:block-null-uniformity}.
\end{proof}

\begin{proof}[Proof of Lemma~\ref{lem:permutation-block-verification}]
Fix $i\in\Nn_0$, and let $\Pi_i$ be a uniform permutation independent
of the data. Conditional on $Z_i$, the statistic $T_i(Z_i^{\Pi_i})$
is uniform over $\mathcal V_i$, since each value occurs $n!/B$ times.
The set $\mathcal V_i$ is unchanged by permutations, so exchangeability
of $Z_i$ gives, for each $x\in\mathcal V_i$,
$$
\PP(X_i=x\mid\mathcal V_i)
=\PP(T_i(Z_i^{\Pi_i})=x\mid\mathcal V_i)
=\frac1B.
$$
Independence across units then implies
$$
\PP\left(
X_i=x\mid\widehat{\mathcal P}_{1:m},X_{-i}^{(0)}
\right)
=\frac1B,
\qquad x\in\mathcal V_i.
$$
For $K=B-1$, the map $\rho$ defined above sends the $mB$ distinct retained
values bijectively to $\mathcal G_{mB}$, proving
\eqref{eq:grid-partition-condition}. It is fixed conditional on
$\widehat{\mathcal P}_{1:m}$, so applying $\rho$ to the last display gives
\eqref{eq:block-null-uniformity}. Finally, each distinct value
occurs $n!/B$ times in the full permutation sum, so
$$
P_i^{\mathrm{dcperm}}
=\frac{1}{mB}\sum_{j=1}^m\sum_{v\in\mathcal V_j}
\ind\{v\geq X_i\}
=\rho(X_i)=P_i^{\mathrm{CC}},
$$
as needed.
\end{proof}

\begin{proof}[Proof of Proposition~\ref{prop:sampled-dcperm-validity}]
First, the identity $(z^\sigma)^\pi=z^{\pi\sigma}$ gives
$$
\pi\sim_i\pi'
\quad\Longleftrightarrow\quad
\pi\sigma\sim_i\pi'\sigma.
$$
Right multiplication therefore permutes the classes and can map any
class bijectively to any other. Hence all $B_i$ classes have size
$n_i!/B_i$.

Fix $i\in\Nn_0$ and condition on the outcomes arranged in increasing
order, denoted by $z_i$. By exchangeability, we may represent
$Z_i=z_i^{\Pi_i}$, where $\Pi_i$ is a uniform permutation independent
of the sampled permutations.

Set $D_{ik}:=[\pi_{ik}\Pi_i]_i$. Equal class sizes make
$D_{i0}=[\Pi_i]_i$ uniform over $\Ss_i$.
For fixed $\Pi_i$, 
the map $[\pi]_i\mapsto[\pi\Pi_i]_i$ sends the identity class to 
$D_{i0}$. Thus, $D_{i1},\ldots,D_{iK}$ form a uniform sample without replacement from $\Ss_i\setminus\{D_{i0}\}$.
Then, for distinct
$d_0,\ldots,d_K\in\Ss_i$,
$$
\PP\left(
(D_{i0},\ldots,D_{iK})=(d_0,\ldots,d_K)
\,\middle|\,z_i
\right)
=\frac{1}{B_i(B_i-1)\cdots(B_i-K)}.
$$
This probability is symmetric in $d_0,\ldots,d_K$. Since
$X_i^{(k)}=T_i(z_i^{\pi_{ik}\Pi_i})$ depends only on $z_i$ and $D_{ik}$,
it holds that $Y_i$ is exchangeable conditional on $z_i$, and hence unconditionally.
Since units are independent, Lemmas~\ref{lem:exchangeable-block-verification}
and~\ref{lem:block-rank-implies-discrete-dcp} complete the proof.
\end{proof}

\begin{proof}[Proof of Theorem \ref{thm:gamma-bfdr-control}]
    By Lemma~\ref{lem:lem2}, it holds that $\mu(A_{-i})=\alpha/m$.
    Write $\tilde P=(P,t)$ for the augmented vector appearing in
    the definition of $\tilde A_P^\alpha$.
    Applying Lemma \ref{lem:lem2}, we have
    $$
    \sum_{i \in \Nn_\gamma} \EE\left[\int_0^1 \ind\{t \in A_{-i}\} f^{(i)}_0(t) dt\right]
    \leq \sum_{i \in \Nn_\gamma} m \cdot \EE\left[\mu(A_{-i})\right]
    = m_\gamma \alpha.
    $$
    For the remaining nulls $i \in \Nn_0 \setminus \Nn_\gamma$, we have by Lemma \ref{lem:universal-A} with $u = \alpha$ that if $P_i > \alpha$, then $\tilde{A}_{P}^\alpha = A_{-i}$.
    Therefore,
    \begin{align*}
    \sum_{i \in \Nn_0 \setminus \Nn_\gamma} \EE\left[\int_0^1 \ind\{t \in A_{-i}\} f^{(i)}_0(t) dt\right]
    &\leq \frac{1}{1-\gamma} \sum_{i \in \Nn_0 \setminus \Nn_\gamma} \EE\left[\ind\{P_i > \alpha\} \int_0^1 \ind\{t \in A_{-i}\} f^{(i)}_0(t) dt\right] \\
    &\leq \frac{1}{1-\gamma} \sum_{i \in \Nn_0 \setminus \Nn_\gamma} \EE\left[\int_0^1 \ind\{t \in \tilde{A}_{P}^\alpha\} f^{(i)}_0(t) dt\right] \\
    &= \frac{1}{1-\gamma}  \EE\left[\int_{\tilde{A}_{P}^\alpha} \sum_{i \in \Nn_0 \setminus \Nn_\gamma} f^{(i)}_0(t) dt\right] \\
    &\leq \frac{1}{1-\gamma}  \EE\left[m \cdot \mu(\tilde{A}_{P}^\alpha)\right] \\
    &\leq \frac{1}{1-\gamma} \cdot m \cdot \frac{\tilde{\alpha}}{m+1} \\
    &= \frac{\alpha}{1-\gamma},
    \end{align*}
    where the first inequality uses $P_i \indeps P_{-i}$ and the definition of $\Nn_\gamma$.
    The second inequality applies Lemma \ref{lem:universal-A} and the second-last line applies Lemma 
    \ref{lem:modified-lem2} with weights $w_i = 1$ and threshold level $\lambda = \alpha$ to the length $m+1$ vector of p-values $\tilde{P}$.
    Together, we have
    $$
    \mathrm{bFDR}(\Rr_\alpha)
    = \sum_{i \in \Nn_0} \EE\left[\int_0^1 \ind\{t \in A_{-i}\} f^{(i)}_0(t) dt\right]
    \leq m_\gamma \alpha + \frac{\alpha}{1-\gamma}.
    $$
    This completes the proof.
\end{proof}

\section{Additional Details for Numerical Results}\label{sec:numerical-details}

\subsection{Additional Simulation Details} \label{subsec:simulation-details}
This subsection gives the p-value formulas for the study in
Section~\ref{subsec:simulation-continuous}. Write
$$
\bar Y_i = \frac{1}{n}\sum_{j=1}^n Y_{ij},
\qquad
S_i^2 = \frac{1}{n-1}\sum_{j=1}^n (Y_{ij}-\bar Y_i)^2,
\qquad
U_i^2 = \frac{1}{n}\sum_{j=1}^n Y_{ij}^2.
$$
The individually valid two-sided t-test p-values are
$$
P_i^{t}
=
2\bar F_{t_{n-1}}\!\left(\frac{\sqrt n|\bar Y_i|}{S_i}\right),
$$
where $\bar F$ denotes a survival function. The density-compound
construction in \eqref{eq:oracle-gaussian-dcp} is
$$
P_i^{\mathrm{oracle}}
=
\frac{1}{m}\sum_{j=1}^m
2\bar\Phi\left(\frac{|\bar Y_i|}{\sigma_j/\sqrt n}\right),
$$
restated here for completeness.
The weighted and oracle compound-e-weighted p-values are
$$
\begin{aligned}
P_i^{\mathrm{w}}&=\frac{P_i^t}{w_i}\wedge1,
& w_i&=\frac{mU_i^2}{\sum_{j=1}^mU_j^2},\\
P_i^{\mathrm{ew}}&=\frac{P_i^t}{E_i^*}\wedge1,
& E_i^*&=\frac{mU_i^2}{\sum_{j=1}^m\sigma_j^2}.
\end{aligned}
$$
See \citet{ignatiadis2026empiricalbayesprepivotinggroup} for a review of
these approaches and additional compound p-value constructions.

\subsection{Additional Regression Discontinuity Results} \label{subsec:additional-plots-rd}
We supplement the RDD analysis in Section~\ref{subsec:rd} with a comparison of the p-value
distributions, a check of sensitivity to permutation sampling, and a
falsification test at artificial cutoffs.

\paragraph{Distributions of p-values.}
Figure~\ref{fig:rd-pvalue-distributions} 
compares the distributions of the individually valid and density-compound p-values.
Although the smallest permutation p-value is smaller than the smallest density-compound p-value, 
the density-compound p-values at ranks two through five are noticeably smaller than the permutation p-values at the same ranks, 
giving more SL rejections for some values of the nominal level $\alpha$.

\paragraph{Effect of randomness.} To check robustness to the random sampling of group
assignments, we compare $P_i^{\mathrm{dcperm},K}$ with the fully enumerated
compound p-values $P_i^{\mathrm{cperm}}$ of
\citet{10.1093/jrsssb/qkag116}.
This construction yields compound p-values in the sense of \eqref{eq:cp} and
averages over every group assignment within each quasi-experiment:
\begin{equation}
P_i^{\mathrm{cperm}}
=
\frac{1}{m}\sum_{j=1}^m
\left(\frac{1}{n_j!}\sum_{\pi\in\mathfrak S(n_j)}
\ind\{X_j^\pi\geq X_i\}\right),
\end{equation}
where $X_i$ denotes the (negative) one-sided Welch t-test p-value for
quasi-experiment $i$ and $X_i^\pi=T_i(Z_i^\pi)$.
At $\alpha=0.2$,
SL makes the same five rejections with both constructions, in the same p-value order.
Figure~\ref{fig:rd-pvalue-distributions_cp} shows the resulting rejection curves and Q-Q plots.
In all, $P_i^{\mathrm{dcperm},K}$ and $P_i^{\mathrm{cperm}}$ yield similar
results, suggesting that randomly sampling group assignments does not substantially
affect our findings.

\begin{figure}[!ht]
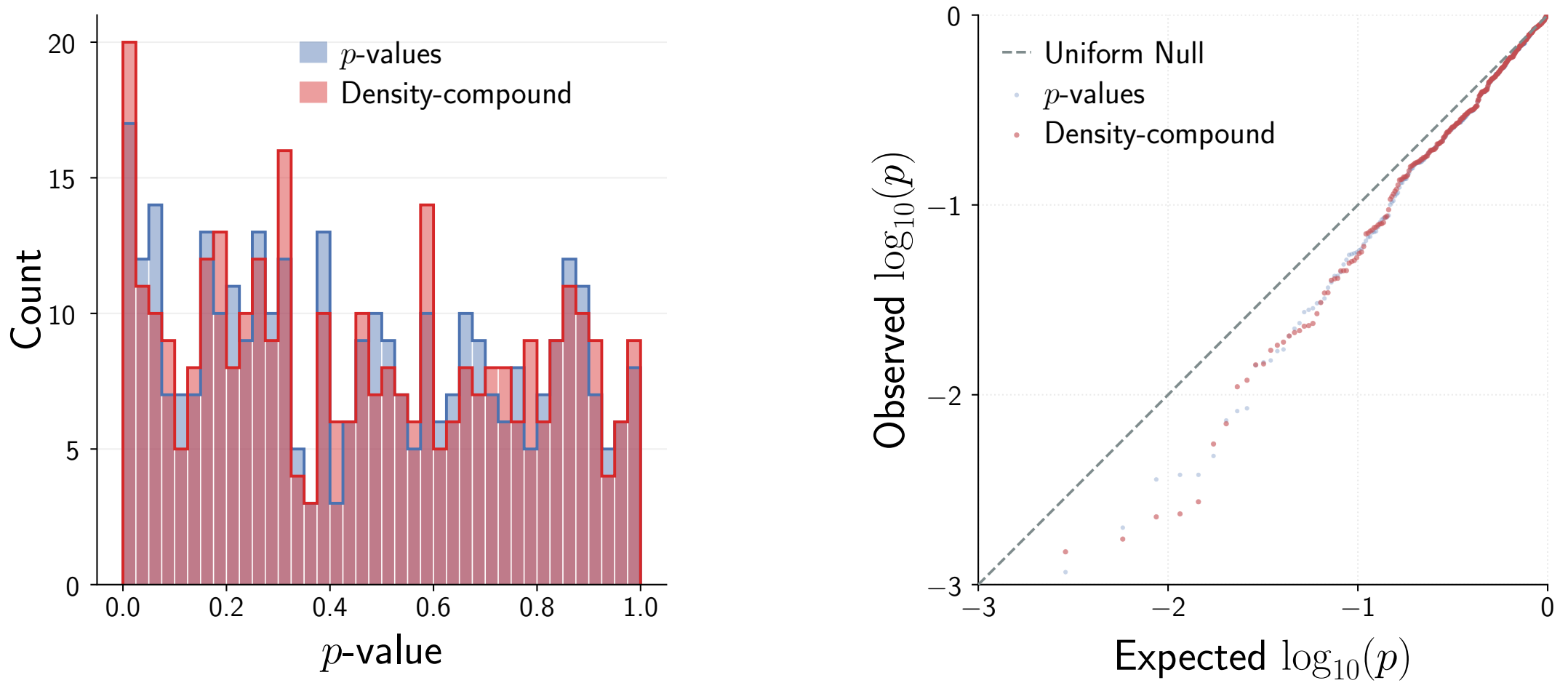

    \centering
    \begin{subfigure}[b]{0.48\textwidth}
        \centering
        \caption{}\label{fig:pvalue-histogram_rd}
        \resizebox{\linewidth}{!}{\input{images/rd_pvalue_histogram.tex}}
    \end{subfigure}
    \hfill
    \begin{subfigure}[b]{0.48\textwidth}
        \centering
        \caption{}\label{fig:pvalue-qq_rd}
        \resizebox{\linewidth}{!}{\input{images/rd_qq_tail_plot.tex}}
    \end{subfigure}
    \caption{Distribution of p-values in the regression discontinuity
    application. (a) Histograms of individually valid permutation p-values and
    density-compound p-values. (b) $\log_{10}$ Q-Q plots of both
    constructions against the uniform distribution.}
    \label{fig:rd-pvalue-distributions}
\end{figure}

\begin{figure}[!ht]
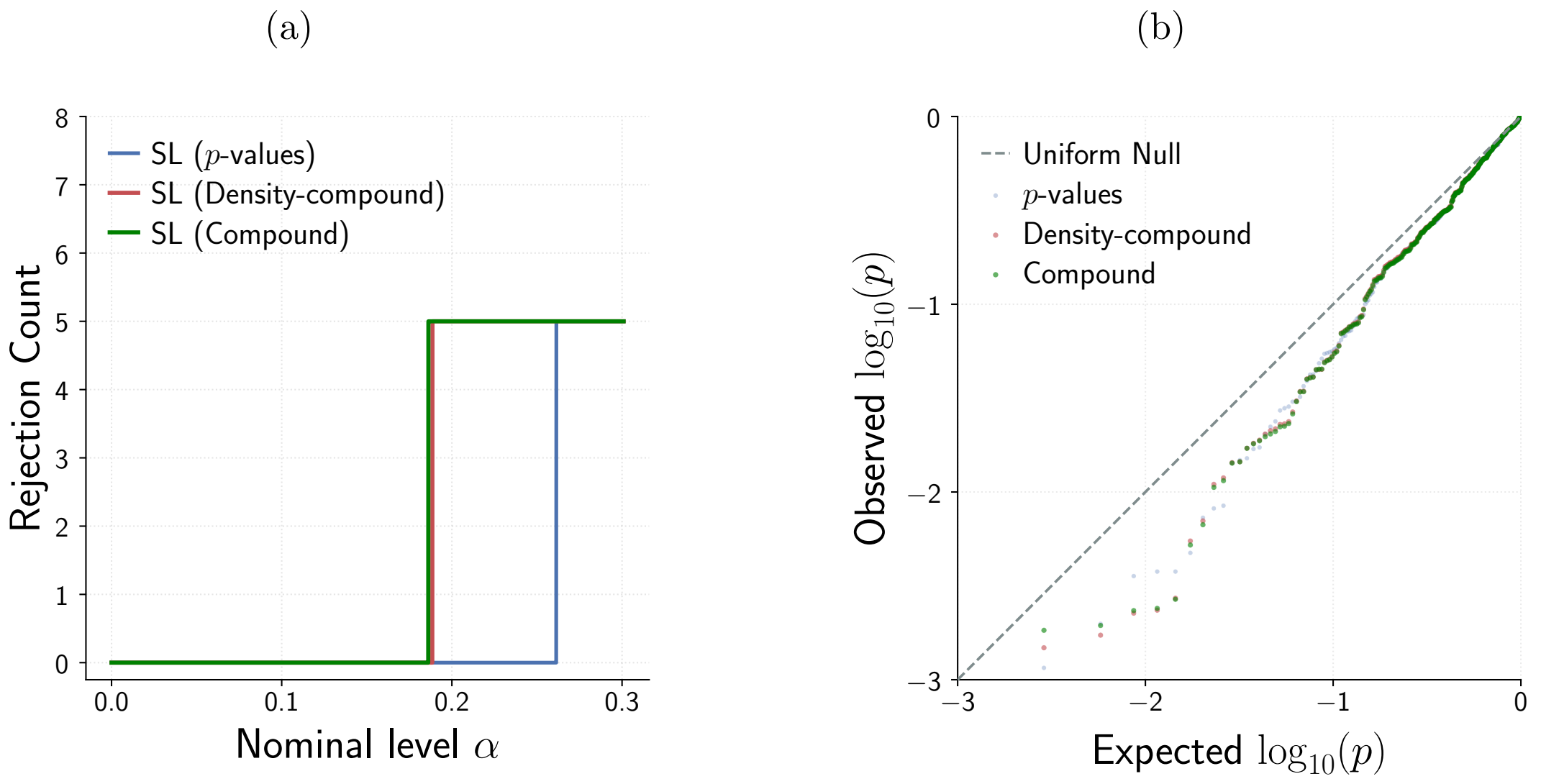

    \centering
    \begin{subfigure}[b]{0.48\textwidth}
        \centering
        \caption{}\label{fig:rejection-curves_rd_cp}
        \resizebox{\linewidth}{!}{\input{images/rd_exact_pool_comparison.tex}}
    \end{subfigure}
    \hfill
    \begin{subfigure}[b]{0.48\textwidth}
        \centering
        \caption{}\label{fig:pvalue-qq_rd_cp}
        \resizebox{\linewidth}{!}{\input{images/rd_exact_pool_qq_tail_plot.tex}}
    \end{subfigure}
    \caption{Comparison of randomized density-compound permutation p-values
    $P_i^{\mathrm{dcperm},K}$ and fully enumerated compound p-values
    $P_i^{\mathrm{cperm}}$ in the regression discontinuity application.
    (a) SL rejection counts versus nominal level $\alpha$, together with the
    individually valid permutation-p-value benchmark. (b) $\log_{10}$ Q-Q plots
    of all three constructions against the uniform distribution.}
    \label{fig:rd-pvalue-distributions_cp}
\end{figure}

\paragraph{Falsification test.}
Next, we repeat the RDD analysis at artificial thresholds that do not
correspond to real admission cutoffs. This falsification exercise examines
whether the testing procedures report discontinuities where admission
eligibility does not change.

Within each town and cohort, we place artificial cutoffs at the midpoints between adjacent real admission cutoffs, 
keeping only those whose symmetric windows of width $2h$ lie strictly between the real cutoffs.
We then follow selection criteria analogous to those in the main analysis. 
First, we pool observations across cohorts only when the same pair of school/track
cutoffs remains adjacent. Before restricting each pooled sample to $[-h,h]$,
we require at least one observation at or below $-h$ and one at or above $h$.
We exclude observations at the artificial cutoff and retain samples with at least $4$ observations on each side and at most $20$ in total.
Figure~\ref{fig:rd-placebo-construction} illustrates two such
constructions in Town 20787; colors and labels link the same construction across
cohorts, and the shaded regions show the $h=0.05$ windows.

\begin{figure}[!ht]
    \centering
    \resizebox{0.98\linewidth}{!}{\input{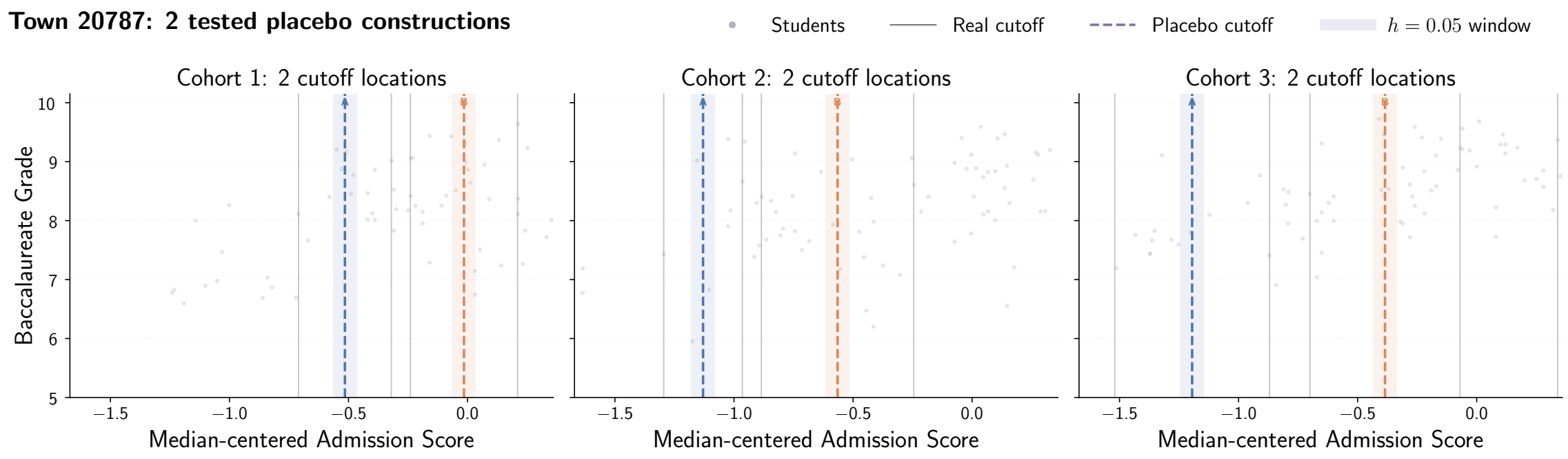}}
    \caption{Midpoint falsification constructions in Town 20787. Solid gray lines are
    real admission cutoffs, dashed colored lines are artificial cutoffs, and shaded
    regions show the corresponding $h=0.05$ windows.}
    \label{fig:rd-placebo-construction}
\end{figure}

We repeat this analysis across bandwidths $h$ to assess how the falsification
results change as the local windows widen.
Figures~\ref{fig:rd-placebo-rejection-curves} and
\ref{fig:rd-placebo-pvalue-histograms} report the resulting rejection curves and
p-value distributions. For the tested bandwidths $h\in\{0.01,0.02,0.03,0.04,0.05\}$, none of the three
p-value constructions yields a rejection over $\alpha\in[0,0.3]$ except for 
individually valid permutation p-values yielding one rejection at $h=0.01$
for $\alpha\geq0.285$ on the plotted grid.
In particular, at the bandwidth used in the main analysis, $h=0.05$, SL makes no
rejections with any of the three constructions over $\alpha\in[0,0.3]$.

Additional calculations on a bandwidth grid spaced by $0.01$ reveal one
isolated artificial construction that is rejected at $\alpha = 0.2$ by the density-compound and
fully enumerated compound p-values at $h=0.06$ and $h=0.07$. 
This rejected artificial cutoff, the same for both bandwidths, is
not testable at $h=0.05$ and ceases to satisfy the filtering rule at
$h=0.08$. 
SL makes no rejections with density-compound or fully enumerated compound p-values at
$h\in\{0.08,0.09,0.10\}$ on this grid. At wider bandwidths, rejections at artificial cutoffs begin to appear under
all three methods.

\begin{figure}[!ht]
    \centering
    \resizebox{0.92\linewidth}{!}{\input{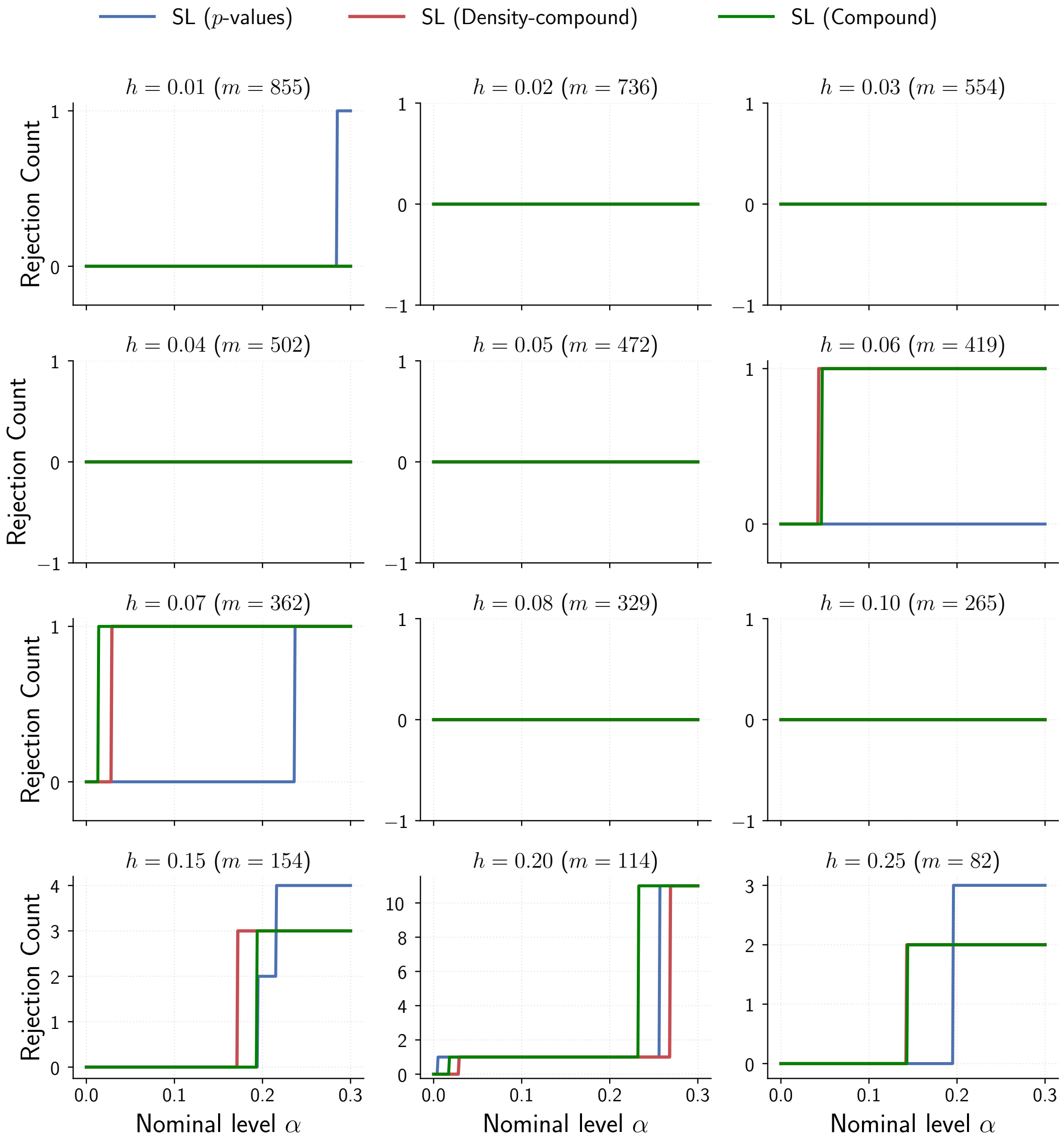}}
    \caption{SL rejection curves for the midpoint-falsification experiments across
    bandwidths. For each bandwidth, we reconstruct the eligible family and
    reapply the sample-size filters, where $m$ is the number of eligible tests.}
    \label{fig:rd-placebo-rejection-curves}
\end{figure}

\begin{figure}[!ht]
    \centering
    \resizebox{0.92\linewidth}{!}{\input{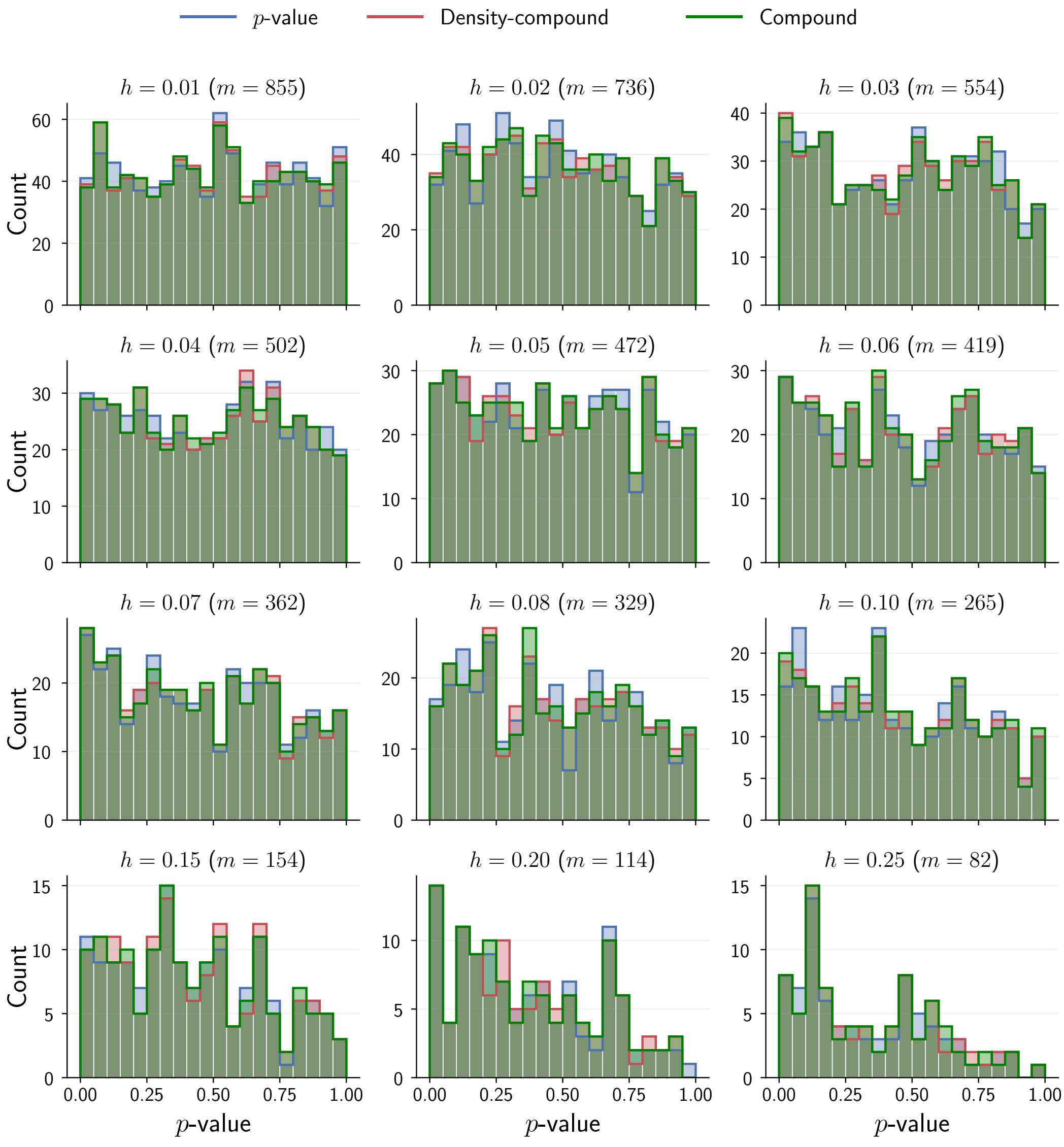}}
    \caption{Distribution of p-values for the midpoint-falsification experiments across
    bandwidths. For each bandwidth, $m$ is the number of eligible tests.}
    \label{fig:rd-placebo-pvalue-histograms}
\end{figure}

\clearpage

\end{document}